\documentclass[preprintnumbers, aps,prd,floatfix,groupedaddress,nofootinbib,superscriptaddress,twocolumn,tightenlines,10pt]{revtex4-1}
\usepackage{graphicx,amsmath,amssymb,amsfonts}
\usepackage{aas_macros}
\usepackage{hyperref}
\usepackage{enumerate} 
\usepackage{color} 
\usepackage{bm}
\usepackage[normalem]{ulem}

\newcommand{\bfv}{{\boldsymbol v}}

\newcommand{\bfx}{{\boldsymbol x}}
\newcommand{\bfk}{{\boldsymbol k}}

\newcommand{\bfu}{{\boldsymbol u}}
\newcommand{\bfp}{{\boldsymbol p}}
\newcommand{\bfq}{{\boldsymbol q}}
\newcommand{\bfs}{{\boldsymbol s}}
\newcommand{\gridspt}{{\tt GridSPT}}

\newcommand{\deltaK}{\delta^{\rm K}}
\newcommand{\deltas}{\delta^{\rm(S)}}
\begin{document}
\title{Grid-based calculations of redshift-space matter fluctuations from perturbation theory: UV sensitivity and convergence at the field level}

\author{Atsushi Taruya}
\affiliation{Center for Gravitational Physics, Yukawa Institute for Theoretical Physics, Kyoto University, Kyoto 606-8502, Japan}
\affiliation{Kavli Institute for the Physics and Mathematics of the Universe, Todai Institutes for Advanced Study, the University of Tokyo, Kashiwa, Chiba 277-8583, Japan (Kavli IPMU, WPI)}
\author{Takahiro Nishimichi}
\affiliation{Center for Gravitational Physics, Yukawa Institute for Theoretical Physics, Kyoto University, Kyoto 606-8502, Japan}
\affiliation{Kavli Institute for the Physics and Mathematics of the Universe, Todai Institutes for Advanced Study, the University of Tokyo, Kashiwa, Chiba 277-8583, Japan (Kavli IPMU, WPI)}
\author{Donghui Jeong}
\affiliation{Department of Astronomy and Astrophysics and Institute for Gravitation and the Cosmos, The Pennsylvania State University, University Park, PA 16802, USA}
\affiliation{School of Physics, Korea Institute for Advanced Study (KIAS), 85 Hoegiro, Dongdaemun-gu, Seoul, 02455, Republic of Korea}
\date{\today}
\begin{abstract}
Perturbation theory (PT) has been used to interpret the observed nonlinear large-scale structure statistics at the quasi-linear regime. To facilitate the PT-based analysis, we have presented the \gridspt\ algorithm, a grid-based method to compute the nonlinear density and velocity fields in standard perturbation theory (SPT) from a given linear power spectrum. Here, we further put forward the approach by taking the redshift-space distortions into account. With the new implementation, we have, for the first time, generated the redshift-space density field to the fifth order and computed the next-to-next-to-leading order (2 loop) power spectrum and the next-to-leading order (1 loop) bispectrum of matter clustering in redshift space. By comparing the result with corresponding analytical SPT calculation and $N$-body simulations, we find that the SPT calculation (A) suffers much more from the UV sensitivity due to the higher-derivative operators and (B) deviates from the $N$-body results from the Fourier wavenumber smaller than real space $k_{\rm max}$. Finally, we have shown that while Pad\'e approximation removes spurious features in morphology, it does not improve the modeling of power spectrum and bispectrum.
\end{abstract}

\preprint{YITP-21-95}
\maketitle

\section{Introduction}
\label{sec:introduction}

Galaxy redshift surveys \cite{Desjacques_Jeong_Schmidt2018} provide a wealth of cosmological information which enables us to probe the late-time cosmic expansion history as well as the growth of large-scale structure. They also offer a clue to probe the primordial fluctuations, from which one can address the fundamental physics questions of the early universe. In addition to several ongoing ground-based surveys such as HETDEX \cite{HETDEX}, PFS \cite{Subaru_PFS2014} and DESI \cite{DESI_ScienceBook2016}, there are space-based missions planned to probe galaxies out to higher redshifts over a large sky area, such as Euclid\footnote{{\tt https://sci.esa.int/web/euclid}} \cite{Euclid2011}, Nancy Grace Roman Space Telescope\footnote{{\tt https://roman.gsfc.nasa.gov/}} \cite{WFIRST2012}, and SPHEREx \footnote{{\tt https://spherex.caltech.edu/}} \cite{SPHEREx2018}. Those gigantic surveys aim to dramatically improve our understanding of the universe to the next level, and to  resolve puzzles such as the nature of dark matter and dark energy, and the physics of cosmic inflation. 

Surveying a larger volume with higher galaxy number density means that these surveys measure the summary statistics, such as the power spectrum and correlation function, with unprecedented precision and this can offer a tight constraint on cosmological parameters, helping us to clarify the nature of cosmic acceleration as well as to test the gravity on cosmological scales \cite{Weinberg_etal2013}. In doing so, it is indispensable to take an accurate theoretical description of the large-scale structure along with the observational systematics. In galaxy surveys, major systematics to be under control are the nonlinearities in gravitational evolution, galaxy bias, and redshift-space distortions. There have been tremendous efforts to describe these effects both from analytical treatments and numerical simulations, and it is indeed one of the major subjects in observational cosmology (e.g., \cite{Crocce:2005xy,Jeong:2006xd,Jeong:2008rj,Crocce:2007dt,Taruya:2007xy,Matsubara2008a,Bernardeau:2008fa,Nishimichi:2008ry,Lawrence:2009uk,Taruya:2010mx,Nishimichi:2011jm,2012JCAP...07..051B,Senatore2015,Mirbabayi_Schmidt_Zaldarriaga2015,Nishimichi_etal2017,Desjacques_Jeong_Schmidt2018,Nishimichi_etal2019_DQ_I}).

Among various techniques and methods, cosmological $N$-body simulations and perturbation theory calculations have established as the standard theoretical tools to accurately predict the observed large-scale structure. In particular, $N$-body simulations are powerful in describing quantitatively the clustering of dark matter and halos at nonlinear regime. Providing a real-space realization of halos, $N$-body also makes it possible to account for directly the observational systematics such as the survey window function and masks. On the other hand, perturbation theory (PT) treatment \cite{Bernardeau:2001qr} provides a faster way to predict statistical quantities at weakly nonlinear regime, and is used for a theoretical template of the measured power spectrum or correlation function. These two approaches are complementary, and a combination of them may give a more efficient theoretical tool with versatile applications (e.g., Ref.~\cite{Nishimichi_etal2017}).

To facilitate the PT-based approach, we have developed a grid-based algorithm to {\it simulate} the nonlinear density and velocity fields of large-scale structure, based on the standard perturbation theory (SPT) \cite{Taruya_Nishimichi_Jeong2018} (see  Refs.~\cite{Roth_Porciani2011,Tassev2014} for earlier works). Taking advantage of the fast Fourier Transform (FFT), its C++ implementation, called \gridspt,  enables us to quickly generate the non-linearly evolved density and velocity field at each order in SPT. Then, we can apply all analysis tools developed for the statistical analysis of the density and velocity fields on configuration-space grids, for example, for $N$-body simulations or for analysis of survey data. Furthermore, the observational systematics such as the survey window function and masks can be easily incorporated into the grid density fields. As an explicit demonstration, in Ref.~\cite{Taruya_Nishimichi_Jeong2021}, we have estimated the covariance matrix of the matter power spectrum with various shapes of survey window functions, including the higher-order corrections from the next-to-leading order (one-loop) trispectrum. 

In this paper, extending the previous grid-based algorithm to include the redshift-space distortions (RSD) \cite{Peebles:1980,Hamilton_RSD_review1998}, we present an explicit implementation of the RSD effects on the \gridspt. Previous studies, for example in Refs.~\cite{Matsubara2008a,Taruya:2010mx} and  Ref.~\cite{Scoccimarro:2004tg}, have shown that the {\it naive} SPT calculation of the matter power spectrum in redshift space does not provide as good model as that in real space, and there have been numerous works to improve the SPT predictions (e.g., \cite{Matsubara2008a,Taruya:2010mx,Vlah:2012ni,Taruya_Nishimichi_Bernardeau2013,Vlah_etal2013,Carlson_Reid_White2013,Wang:2013hwa,Matsubara2014,Hand_etal2017,Vlah_White2019,Chen_etal2021}). Making use of the grid-based treatment, we shall see how the naive SPT treatment leads to an inaccurate prediction particularly at the field level, even after including the nonlinear corrections up to the fifth order. Also, applying the Pad\'e approximations to the SPT density fields, we shall seek for the possibility of using a re-summed treatment for more accurate modeling. It is, however, to be stressed that the implementation of the RSD effect in \gridspt\, is not our final goal. In our successive work, we plan to implement the effect of galaxy bias as well as the effective-field-theory treatment (e.g., \cite{2012JCAP...07..051B,2012JHEP...09..082C,Baldauf:2015aha,Baldauf:2015aha,Nishimichi_etal2020}), the latter of which can mitigate the UV-sensitive behaviors of the SPT calculation, and we thus expect that the method has a potential to improve upon the SPT predictions. Note that the \gridspt\, algorithm has been applied to a precise calibration of the effective-field-theory counter terms for the bispectrum and trispectrum at next-to-leading order \cite{Steele_Baldauf2021a,Steele_Baldauf2021b}.

In principle, one can implement the RSD effect from the \gridspt\, output by mapping the real-space density field to the redshift-space using the line-of-sight component of the peculiar velocity field. However, a naive implementation of the mapping formula in grid space needs an interpolation, for which an accurate computation needs a non-perturbative calculation. For the perturbative treatment, we have presented a novel expression that relates the redshift-space density field in terms of the real-space density and velocity fields. We have then evaluated the expression at the redshift-space position. In this way, no interpolation technique is necessary, and one can directly {\it reconstruct} the SPT density field in redshift space from the real-space \gridspt\, calculations. With an explicit implementation of the RSD effects, we investigate the statistical and morphological properties of the redshift-space SPT density fields.

The organization of this paper is as follows. In Sec.~\ref{sec:GridSPT}, we begin by briefly reviewing the grid-based SPT calculation of large-scale structure, and comment on the aliasing effect that appears in a practical implementation. Then, in Sec.~\ref{sec:GridSPT_RSD}, we consider the RSD, and derive the expression for redshift-space density field written in terms of real-space quantities. Based on this, we present a perturbative framework to compute density fields with \gridspt. Sec.~\ref{sec:demonstration} presents explicit demonstration of the \gridspt\, calculations taking the RSD effect into account, for which we also make a detailed comparison with $N$-body simulations and analytical SPT calculations. To this end, we present for the first time the two-loop SPT power spectrum in redshift space. Sec.~\ref{sec:pade_approx} discusses the application of Pad\'e approximations to the \gridspt, and discusses a possibility to improve the SPT calculations in redshift space at field level. Finally, Sec.~\ref{sec:conclusion} is devoted to the conclusion and discussions on the future prospects. 

Throughout the paper, we use the following Fourier convention:
\begin{eqnarray}
f(\bfk)
    &=&
\int d^3x\, e^{-i\bfk\cdot\bfx}f(\bfx)
\\
f(\bfx)
    &=&
\int\frac{d^3k}{(2\pi)^3}\, e^{i\bfk\cdot\bfx}f(\bfk)
\equiv
    \int_{\bfk} e^{i\bfk\cdot\bfx} f(\bfk)\,.
\end{eqnarray}

\section{Grid-based perturbation theory}
\label{sec:GridSPT}

In this section, we present a concise review on the grid-based calculation for perturbation theory of large-scale structure named \gridspt, described in Ref.~\cite{Taruya_Nishimichi_Jeong2018}. In essence, \gridspt\, enables us to perform SPT calculations at the field-level; to generate the numerical realizations of higher-order density and velocity fields at each grid point. The heart of the algorithm is the real-space recursion relation in Eq.~(\ref{eq:recursion_formula}), on which the \gridspt\, implementation is based.

Standard perturbation theory models the gravitational evolution of matter distribution by integrating the Vlasov-Poisson equations under the assumption of the single-stream matter flow \cite{Bernardeau:2001qr}. In this framework, the large-scale matter distribution is described by the pressureless fluid equations coupled with the Poisson equation. When further combined with the irrotational flow assumption, which is also valid on large scales, the system of equations describing the nonlinear evolution of density and velocity fields is further reduced to 
\begin{align}
&\frac{d}{d\eta}\left(
\begin{array}{c}
\delta
\\
\\
\theta
\end{array}
\right)+\Omega_{ab}
\,\left(
\begin{array}{c}
\delta
\\
\\
\theta
\end{array}
\right)
=
\left(
\begin{array}{c}
{\displaystyle 
\nabla\cdot[\delta\bfu]}
\\
\\
{\displaystyle 
\nabla\cdot
\left[
\left(\bfu\cdot\nabla\right)\bfu
\right]
}
\end{array}
\right),
\label{eq:basic_PT_eqs}
\end{align}
where we introduce the time variable $\eta$ defined by $\eta\equiv\ln D_+(t)$ with $D_+$ being the linear growth factor. We denote the comoving coordinate as $\bfx$. The quantities $\delta = \delta(\bfx,\eta)$ and $\theta = \theta(\bfx,\eta)$ are the mass density and the velocity-divergence fields, respectively. The velocity-divergence field is related to the velocity field $\bfv$ through $\theta \equiv-\nabla\bfv/(f\,aH)\equiv \nabla\cdot{\bfu}$ with $f$ being the linear growth rate, defined by $f\equiv d\ln\,D_+/d\ln a$. The field $\bfu$ is the {\it reduced} velocity field given by $\bfu=\nabla[\nabla^{-2}\theta]$ for an irrotational matter flow. The matrix $\Omega_{ab} = \Omega_{ab}(\eta)$ generally depends on cosmology and time, but replacing that with the time-independent constant matrix $\Omega_{ab}^{\rm EdS}$ for the Einstein-de Sitter Universe:
\begin{align}
 \Omega_{ab}^{\rm EdS}=\left(
\begin{array}{cc}
0 & \qquad -1
\\
\\
{\displaystyle -\frac{3}{2}} & \qquad {\displaystyle \frac{1}{2}}
\end{array}
\right)\,,
\label{eq:Omega_ab_EdS}
\end{align}
provides a good approximation in a wide class of cosmology models close to the $\Lambda$CDM (e.g., Refs.~\cite{Pietroni:2008jx,Takahashi2008,Hiramatsu_Taruya2009}).  

We obtain the perturbative solutions for Eq.~(\ref{eq:basic_PT_eqs}) by expanding the density and velocity fields. For the dominant growing-mode contributions, we have 
\begin{align}
&\delta(\bfx,\eta) = \sum_n\,\delta_n(\bfx,\eta), \qquad
\theta(\bfx,\eta) = \sum_n\,\theta_n(\bfx,\eta)
\label{eq:PT_expansion}
\end{align}
with the time dependence at each order scaled as $\delta_n,\,\theta_n\,\propto e^{n\,\eta}$. Hereafter, we suppress arguments of $\eta$ for the perturbed quantities and simply write $\delta_n(\bfx)$ and $\theta_n(\bfx)$. Substituting Eq.~(\ref{eq:PT_expansion}) into Eq.~(\ref{eq:basic_PT_eqs}) and using $\Omega_{ab}$ in Eq.~(\ref{eq:Omega_ab_EdS}), the order-by-order calculation leads to the following recursion relation \cite{Taruya_Nishimichi_Jeong2018}: 
\begin{align}
\left(
\begin{array}{c}
{\displaystyle \delta_n(\bfx) }
\\
\\
{\displaystyle \theta_n(\bfx) }
\end{array}
\right) &= \frac{2}{(2n+3)(n-1)}\,\left(
\begin{array}{cc}
{\displaystyle n+\frac{1}{2}} & \qquad 1
\\
\\
{\displaystyle \frac{3}{2} } & \qquad n
\end{array}
\right) 
\nonumber
\\
&\times\,\sum_{m=1}^{n-1} \left(
\begin{array}{c}
{\displaystyle (\nabla \delta_m)\cdot\bfu_{n-m}+\delta_m \theta_{n-m}}
\\
\\
{\displaystyle \frac{1}{2}\nabla^2(\bfu_{m}\cdot\bfu_{n-m}) }
\end{array}
\right),
\label{eq:recursion_formula}
\end{align}
for $n\geq2$. Here, we have used the identity 
$\nabla\cdot\left[(\bfu\cdot\nabla)\bfu\right] = \frac12\nabla^2\left(\bfu\cdot\bfu\right)$ for an irrotational (curl-free) velocity field $\bfu$. Unlike the equivalent expression given in Ref.~\cite{Taruya_Nishimichi_Jeong2018}, Eq.~(\ref{eq:recursion_formula}) involves no tensor-field calculation, which is helpful for reducing the memory requirement in the numerical implementation. We complete the recursion relation by using the linear-order ($n=1$) growing-mode solution
\begin{align}
\left(
\begin{array}{c}
\delta_1(\bfx)
\\
\\
\theta_1(\bfx)
\end{array}
\right) = e^\eta\,
\left(
\begin{array}{c}
1
\\
\\
1
\end{array}
\right) \delta_0(\bfx),
\label{eq:recursion_n=1}
\end{align}
where $\delta_0(\bfx)$ is the linear density field given at an initial time. 

For a given linear density field $\delta_0(\bfx)$ on grids, we use them as an initial condition for the recursion (Eq.~\ref{eq:recursion_n=1}) to calculate the nonlinear source terms given at the right-hand side of Eq.~(\ref{eq:recursion_formula}). The fast Fourier transform (FFT) facilitates the calculation of the derivative operators $\nabla_j$, which simply becomes a multiplication of $i\bfk_j$ in Fourier space. We have presented details of the algorithm and implementation in Ref.~\cite{Taruya_Nishimichi_Jeong2018} (see their Sec.~II-C)\footnote{With the real-space recursion relation at Eq.~(\ref{eq:recursion_formula}), one important difference from the algorithm in Ref.~\cite{Taruya_Nishimichi_Jeong2018} is that we do not need to compute the tensor fields, $\partial_iu_j$, at every step of PT calculations. }. In Ref.~\cite{Taruya_Nishimichi_Jeong2018}, we have generated nonlinear density fields up to fifth order and studied both their morphological and statistical properties in a face-to-face comparison with $N$-body simulations that begins from {\it exactly} the same random realizations. 
One of the advantages of this method is that grid-based codes for the statistical analysis of $N$-body simulation results can be reused for the outcomes of \gridspt,
and that once the density fields are generated, the predictions can be scaled to any redshift analytically by using the fact that the time dependence of the $n$-th order fields is simply described as $\delta_n,\,\,\theta_n\propto e^{n\,\eta}$.

It is worthy noting that the operations for the \gridspt\, implementation, particularly calculating the right-hand-side of Eq.~(\ref{eq:recursion_formula}), can generate the aliasing effect, which arises when fast-Fourier-transforming the nonlinear terms evaluated in configuration space (see Appendix \ref{subsec:de-aliasing_methods}). The aliasing effect produces spurious high-wavenumber Fourier modes that affect the small-scale behaviors of the resulting nonlinear fields. Mitigating such an effect is thus critical for a practical SPT calculation at the field level.

A simple but widely used technique to mitigate the aliasing effect is to discard the high-frequency modes. In our previous papers \cite{Taruya_Nishimichi_Jeong2018,Taruya_Nishimichi_Jeong2021}, we have adopted the so-called $2/3$ rule to set Fourier modes in the high frequency range of $k>(2/3)k_{\rm Nyq}$ to zero at each step of the \gridspt\, calculation. Here, the wavenumber $k_{\rm Nyq}$ is the Nyquist frequency defined by $k_{\rm Nyq}\equiv\pi/L_{\rm p}$, where $L_{\rm p}\equiv (L_{\rm box}/N_{\rm grid}^{1/3})$ is the grid separation, $L_{\rm box}$ and $N_{\rm grid}$ are respectively the side length and the total number of grids for the comoving cubic box inside which the fields $\delta_n$ and $\theta_n$ are defined. Strictly speaking, however, the $2/3$ rule is valid only for the aliasing effect arising from the quadratic operations of the fields. For the nonlinear terms with the $N$-th power of the fields, instead, the $2/3$ rule has to be generalized to the $2/(N+1)$-rule. That is, the modes with wavenumber $k>2/(N+1)\,k_{\rm Nyq}$ are to be discarded before the calculation of nonlinear terms. Applying the $2/(N+1)$ rule has been essential in computing the redshift-space density field with \gridspt, since the redshift-space density field is constructed perturbatively with higher powers of the density and velocity fields. In Appendix \ref{sec:aliasing_correction}, we discuss this point in greater detail and present a comparison among results of the \gridspt\, calculations with various de-aliasing treatments.

Finally, a cautionary remark is in order; the single-stream PT treatment ceases to be adequate in the nonlinear regime where the multi-stream flow is generated, and recent studies show that the multi-stream effect on the matter distribution is manifest even on large scales and becomes more significant at higher order (e.g., \cite{Blas:2013aba,Bernardeau:2012ux,Nishimichi:2014rra,Nishimichi_etal2017}). The effective-field-theory treatment can remedy the situation by introducing counter terms that absorb the UV sensitivity. We shall leave a grid-based implementation of the effective-field-theory treatment for our future work, and focus on modeling RSD in the \gridspt\, framework.

\section{Implementing redshift-space distortions on \gridspt}
\label{sec:GridSPT_RSD}


In this section, based on the standard PT treatment, we present an algorithm to compute perturbatively the redshift-space density fields on grids. 

First, recall that the observed position of a galaxy in redshift space, $\bfs$, is related to the real-space position $\bfx$ through 
\begin{align}
 \bfs =\bfx - f\,u_z(\bfx)\,\hat{z}, 
\label{eq:mapping_s_and_x}
\end{align}
where $u_z$ is the line-of-sight component of the field $\bfu$, defined earlier $\bfu\equiv -\bfv/(f\,a H)$, with $\bfv$ being the peculiar velocity. Throughout the paper, we work with the distant-observer limit and take the $z$-axis as the line-of-sight direction. With the mapping relation in Eq.~(\ref{eq:mapping_s_and_x}),  one finds an expression for the density field in redshift space, denoted by $\deltas$, in terms of the real-space quantities as (e.g., Refs.~\cite{Scoccimarro:2004tg,Desjacques_Jeong_Schmidt2018}, see also Refs.~\cite{Kaiser1987,Cole:1993kh,Raccanelli_etal2018} for the expression without taking the distant-observer limit)
\begin{align}
 \deltas(\bfs)&=\Bigl|\frac{\partial\bfs}{\partial \bfx} \Bigr|^{-1}\,\Bigl\{1+\delta(\bfx)\Bigr\}-1
\nonumber
\\
&=\frac{\delta(\bfx)+f\,\nabla_z u_z(\bfx)}{1-f\,\nabla_zu_z(\bfx)},
\label{eq:delta_in_s-space}
\end{align}
where the operator $\nabla_z$ stands for the line-of-sight derivative, $\hat{z}\cdot\nabla_x$. The above expression is exact in the distant-observer limit, and using {\tt GridSPT}, the quantities on the right-hand side can be computed up to an arbitrary order without expanding the denominator. Note, however, that the right hand side of Eq.~(\ref{eq:delta_in_s-space}) is still to be evaluated at the real-space position. In order to obtain the density field in redshift space, therefore, we have to transform the
quantities at the real space position $\bfx$ to the redshift-space position $\bfs$ through Eq.~(\ref{eq:mapping_s_and_x}). Although such a transformation can be implemented rigorously up to an arbitrary order in PT calculations, the resultant redshift-space density fields no longer reside at the original grids. To obtain a regularly-spaced density field, we have to interpolate among the
resultant density fields. Such an operation obscures the counting of PT order, so it is incompatible with a PT calculation in a strict sense. We shall leave this implementation as a future work.

To circumvent the situation, we derive an alternative expression for the redshift-space density field. To do so, consider the Fourier transform of the redshift-space density field:  
\begin{align}
 \deltas(\bfk)&=\int d^3\bfs \,e^{-i\bfk\cdot\bfs}\,\deltas(\bfs)
\nonumber
\\
& = \int d^3\bfs\, e^{-i\bfk\cdot\bfs}\,\Biggl[\Bigl|\frac{\partial\bfs}{\partial \bfx} \Bigr|^{-1}\,\Bigl\{1+\delta(\bfx)\Bigr\}-1\Biggr]
\nonumber
\\
& = \int d^3\bfx\, e^{-i\bfk\cdot(\bfx-f\,u_z(\bfx)\hat{z})}\,
\Bigl\{\delta(\bfx)+f\,\nabla_z\,u_z(\bfx)\Bigr\}.
\end{align}
In the last line, we changed the variable of integral from $\bfs$ to $\bfx$, using Eq.~(\ref{eq:mapping_s_and_x}) and the Jacobian $|\partial\bfs/\partial\bfx|=1-f\,\nabla_zu_z(\bfx)$. Taylor-expanding the velocity field in the exponent and substituting the Fourier transform of the quantities $\delta$ and $u_z$, we have obtained
\begin{widetext}
\begin{align}
 \deltas(\bfk)&=\int d^3\bfx\, e^{-i\bfk\cdot\bfx} \sum_{n=0}\frac{i^n}{n!}\,(fk_z)^n\Bigl\{\delta(\bfx)+f\,\nabla_z\,u_z(\bfx)\Bigr\}\{u_z(\bfx)\}^n
\nonumber
\\
&=\sum_{n=0} 
    \frac{(fk_z)^n}{n!}\,\int d^3\bfx e^{-i\bfk\cdot\bfx}
    \int_{\bfp} e^{i\bfp\cdot\bfx}
    \int_{\bfq_1} e^{i\bfq_1\cdot\bfx}\cdots\int_{\bfq_n}e^{i\bfq_n\cdot\bfx}
\,\Bigl\{\delta(\bfp)+f\,\frac{p_z^2}{p^2}\theta(\bfp)\Bigr\}
\frac{q_{1,z}}{q_1^2}\theta(\bfq_1)\cdots \frac{q_{n,z}}{q_n^2}\theta(\bfq_n).
\label{eq:deltas_Fourier} 
\end{align}
\end{widetext}
Here, we consider the irrotational velocity flow\footnote{To be precise, in deriving Eq.~(\ref{eq:deltas_expansion}), we do not necessarily assume the irrotationality.}, and used the velocity-divergence field $\theta$ 
[see Eq.~(\ref{eq:basic_PT_eqs}) above], with which 
$u_z(\bfk)=(-i\,k_z/k^2)\theta(\bfk)$.

Going back to the configuration space, the inverse Fourier transform of Eq.~(\ref{eq:deltas_Fourier}) gives
\begin{widetext}
\begin{align}
\deltas(\bfs) &=\int\frac{d^3\bfk}{(2\pi)^3}e^{i\bfk\cdot\bfs}\deltas(\bfk)
\nonumber
\\
&=\sum_{n=0}
\int_{\bfp}\int_{\bfq_1}\cdots\int_{\bfq_n}
e^{i(\bfp+\bfq_1+\cdots+\bfq_n)\cdot\bfs}
    \frac{f^n(p_z+\sum_{i=1}^n q_{i,z})^n}{n!}\,\Bigl\{\delta(\bfp)+f\frac{p_z^2}{p^2}\theta(\bfp)\Bigr\}\frac{q_{1,z}}{q_1^2}\theta(\bfq_1)\cdots\frac{q_{n,z}}{q_n^2}\theta(\bfq_n).
\label{eq:deltas_Fourier_quantities}
\end{align}
\end{widetext}
Finally, the above expression can be recast as
\begin{align}
\deltas(\bfs)=\sum_{n=0} \frac{f^n}{n!} \tilde{\nabla}_z^{n}\Bigl[\Bigl\{\delta(\bfs)+f\tilde{\nabla}_zu_z(\bfs)\Bigr\}\,\{u_z(\bfs)\}^{n}\Bigr]
\label{eq:deltas_expansion}
\end{align}
with the operator $\tilde{\nabla}_z$ defined by $\tilde{\nabla}_z\equiv\hat{z}\cdot\nabla_s$. Note that the $\hat{z}$ direction
is well-defined both in the real space and the redshift space.

Eq.~(\ref{eq:deltas_expansion}) is the key equation to perform a grid-based PT calculation in redshift space. In contrast to Eq.~(\ref{eq:delta_in_s-space}), 
the right hand side is now expressed as function of the redshift-space position $\bfs$. Hence, we use Eq.~(\ref{eq:deltas_expansion}) as a basis to directly compute the redshift-space density field from the real-space quantities without any interpolation. To be explicit, let us apply the SPT expansion given in Eq.~(\ref{eq:PT_expansion}), and substitute these expansions in real space into Eq.~(\ref{eq:deltas_expansion}). Computing perturbatively the redshift-space density field,  the order-by-order calculation leads to
\begin{align}
 \deltas=\sum_{n=1}\,\deltas_n
\end{align}
with the explicit expression of $\deltas_n$ given below up to the fifth order:
\begin{align}
 \deltas_1&= D_1
\label{eq:deltas_1}
\\
 \deltas_2&= D_2+ f\tilde{\nabla}_z\Bigl(D_1 u_{z,1}\Bigr),
\label{eq:deltas_2}
\\
 \deltas_3&= D_3+ f\tilde{\nabla}_z\Bigl(D_1u_{z,2}+D_2u_{z,1}\Bigr)+ \frac{f^2}{2!}\tilde{\nabla}_z^2\Bigl(D_1u_{z,1}^2\Bigr),
\label{eq:deltas_3}
\\
 \deltas_4&= D_4+ f\tilde{\nabla}_z\Bigl(D_1u_{z,3}+D_2u_{z,2}+D_3u_{z,1}\Bigr)
\nonumber
\\
&+ \frac{f^2}{2!}\tilde{\nabla}_z^2\Bigl(2D_1u_{z,1}u_{z,2}+D_2u_{z,1}^2\Bigr) + \frac{f^3}{3!}\tilde{\nabla}_z^3\Bigl(D_1u_{z,1}^3\Bigr),
\label{eq:deltas_4}
\\
 \deltas_5&= D_5+ f\tilde{\nabla}_z\Bigl(D_1u_{z,4}+D_2u_{z,3}+D_3u_{z,2}+D_4u_{z,1}\Bigr) 
\nonumber
\\
&+ \frac{f^2}{2!}\tilde{\nabla}_z^2\Bigl\{D_1(2u_{z,1}u_{z,3}+u_{z,2}^2)+2D_2u_{z,1}u_{z,2}+D_3u_{z,1}^2\Bigr) 
\nonumber
\\
&+ \frac{f^3}{3!}\tilde{\nabla}_z^3\Bigl(3D_1u_{z,1}^2u_{z,2}+D_2u_{z,1}^3\Bigr)+\frac{f^4}{4!}\tilde{\nabla}_z^4\Bigl(D_1u_{z,1}^4\Bigr),
\label{eq:deltas_5}
\end{align}
where we introduce the perturbed quantity $D_n$ defined by
\begin{align}
 D_n\equiv \delta_n+f\tilde{\nabla}_zu_{z,n}.
\end{align}
Now, the recipe to compute $\deltas$ with {\tt GridSPT} is to first evaluate the real-space density and velocity fields, $\delta$ and $u_z$, up to an arbitrary order, and then to plug them into the above expressions. All the calculation is done in the same grid space as we obtain the real-space quantities. Note that as a matter of course, the Fourier transform of the density field at each order, $\delta_n^{\rm(S)}(\bfk)$, yields the expression identical to the one with the redshift-space kernel $Z_n$ in literature [see Eq.~(\ref{eq:deltaS_kernelZ})]. 

\section{Results}
\label{sec:demonstration}

Using the prescription in Sec.~\ref{sec:GridSPT_RSD}, we are in position to present the results of \gridspt\, calculations in redshift space, and to compare them with the results from $N$-body simulations. Here, for the sake of comprehensive study parallel to our previous works, we adopt the same cosmological parameters as used in Ref.~\cite{Taruya_Nishimichi_Jeong2018}, assuming the flat-$\Lambda$CDM model:  $\Omega_{\rm m}=0.279$ for matter density, $\Omega_{\Lambda}=0.721$ for dark energy
with equation-of-state parameter $w=-1$, $\Omega_{\rm b}/\Omega_{\rm m}=0.165$ for baryon fraction, $h=0.701$ for Hubble parameter, $n_s=0.96$ for scalar spectral index, and finally, $\sigma_8=0.8159$ for the normalization of the fluctuation amplitude at $8\,h^{-1}$Mpc. We have then used the results of the cosmological $N$-body simulation done in Ref.~\cite{Taruya_Nishimichi_Jeong2018}. The simulation has been carried out by a publicly available code, GADGET-2 \cite{Springel:2005mi}, with
$N_{\rm particle}=1,024^3$ particles in comoving periodic cubes of $L_{\rm box}=1,000\,h^{-1}$Mpc, with the initial density field calculated from the {\tt 2LPT} code \cite{Crocce:2006ve}. Specifically, we use the output data at $z=0$ and $1$ to create the redshift-space density field as well as to measure the statistical quantities. With the same initial seed and cubic box, we perform the \gridspt\, calculations up to the fifth order.  Unless otherwise stated,  the number of grids is set to
$N_{\rm grid}=1,200^3$ as a default setup. To mitigate the aliasing effect, based on the discussion earlier and in Appendix \ref{sec:aliasing_correction}, we adopt the $2/(1+5)=1/3$ rule (instead of the $2/3$ rule that we have adopted in Ref.~\cite{Taruya_Nishimichi_Jeong2018}) with an isotropic sharp-$k$ filter, which is applied only once to the initial density field.

\begin{figure*}[tb]
\vspace*{-0.8cm}
\begin{center}
\hspace*{-1.0cm}
\includegraphics[width=10.3cm,angle=0]{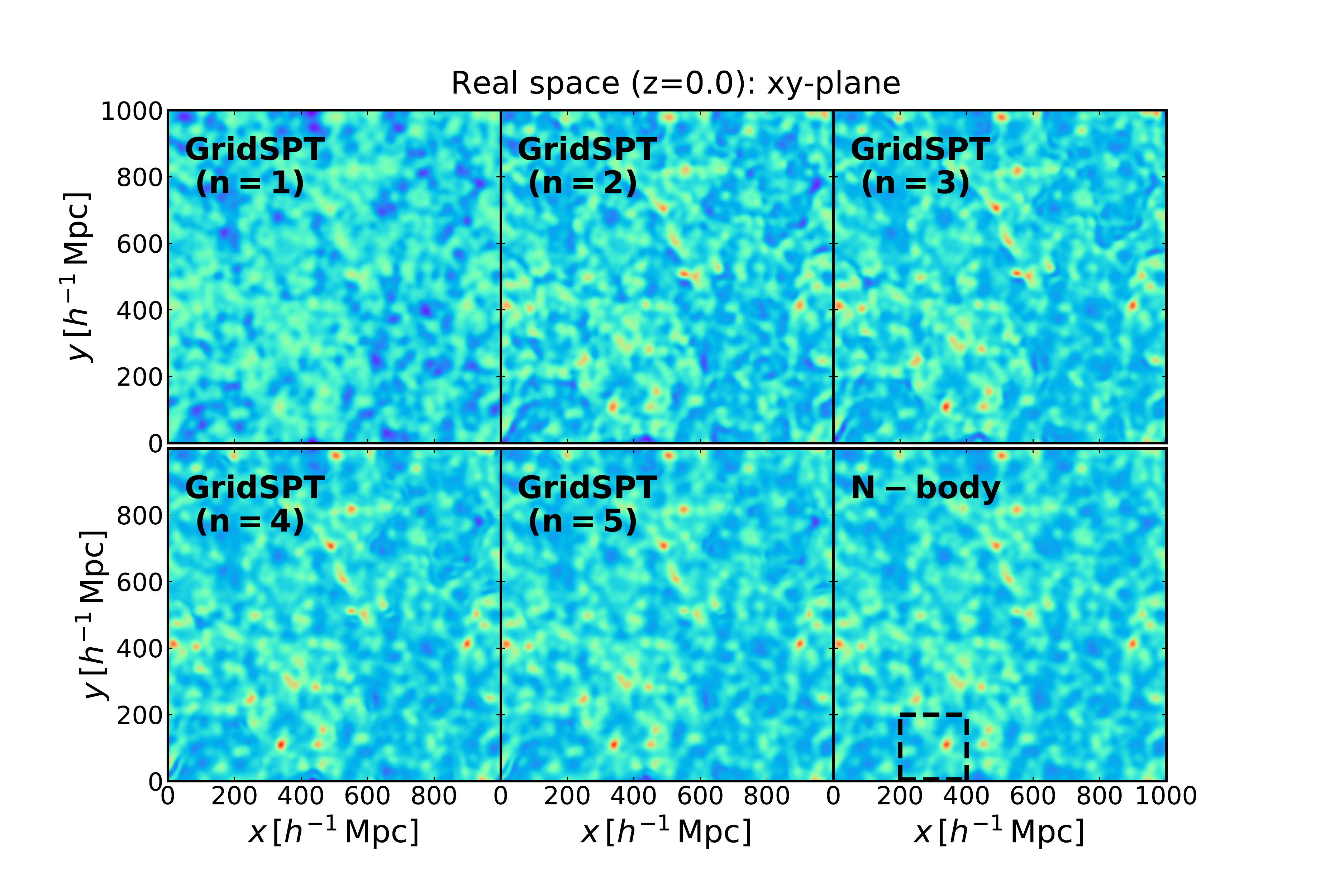}
\hspace*{-2.0cm}
\includegraphics[width=10.3cm,angle=0]{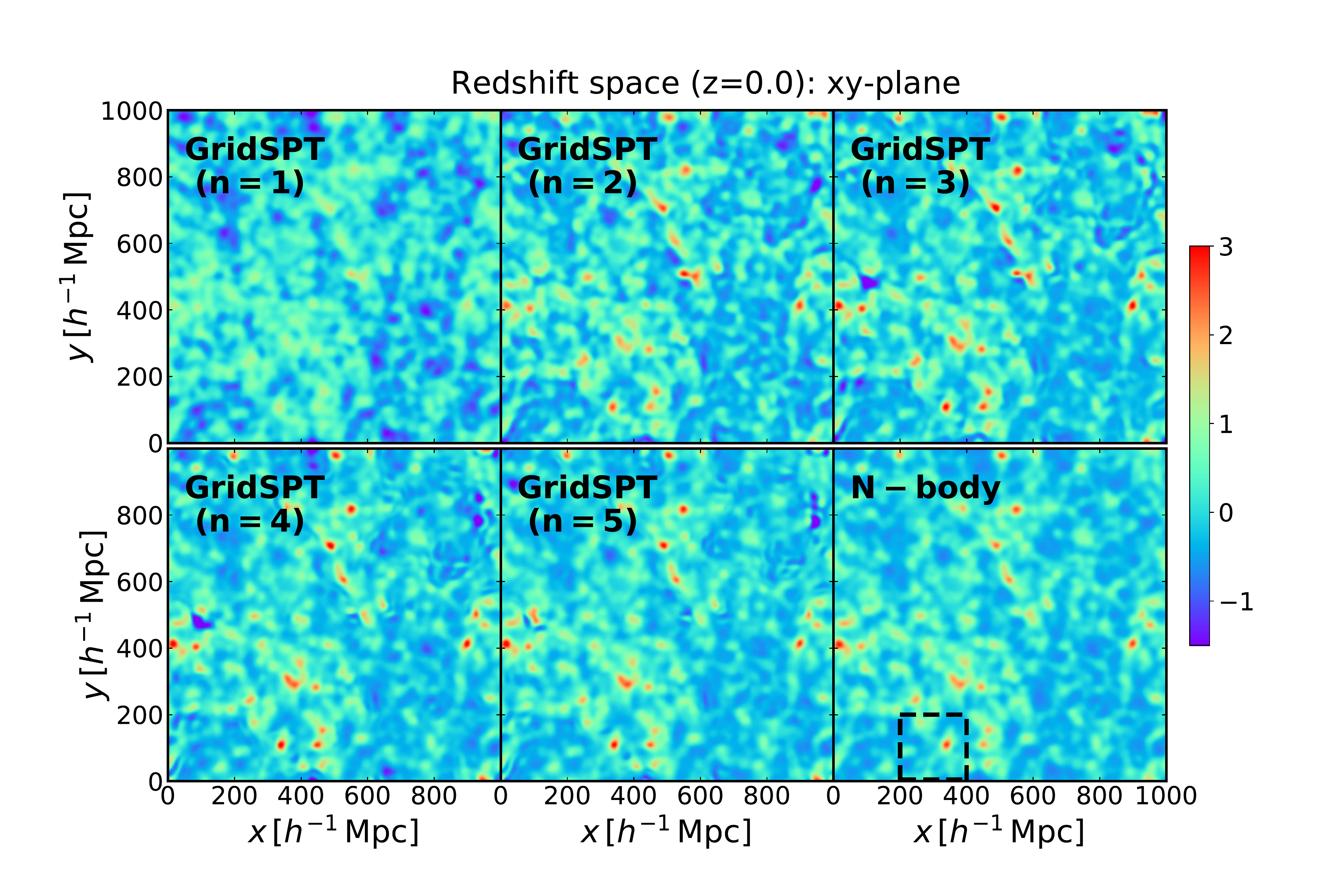}
\end{center}
\vspace*{-0.6cm}
\caption{2D density field at $z=0$ smoothed with a Gaussian filter of $R=10\,h^{-1}$Mpc. A slice of $xy$ plane is taken, and the density field averaged over $10\,h^{-1}$Mpc depth is shown. Left and right panel represent the results in real and redshift space, respectively. In each panel, the results generated with {\tt GridSPT} code are shown (from top left to bottom middle). Here, the color scale represents the amplitude of the density field, $\delta_{\rm SPT}=\sum_{j=1}^n\,\delta_j$ or $\delta_{\rm SPT}^{\rm(S)}=\sum_{j=1}^n\,\delta_j^{\rm(S)}$ with the number $n$ indicated in each panel. For comparison, the bottom right panel shows the density field from $N$-body simulation, evolved with the same initial condition as used in {\tt GridSPT} calculations.   
\label{fig:Slice_real_red_xy}
}
\vspace*{-0.5cm}
\begin{center}
\hspace*{-1.0cm}
\includegraphics[width=10.3cm,angle=0]{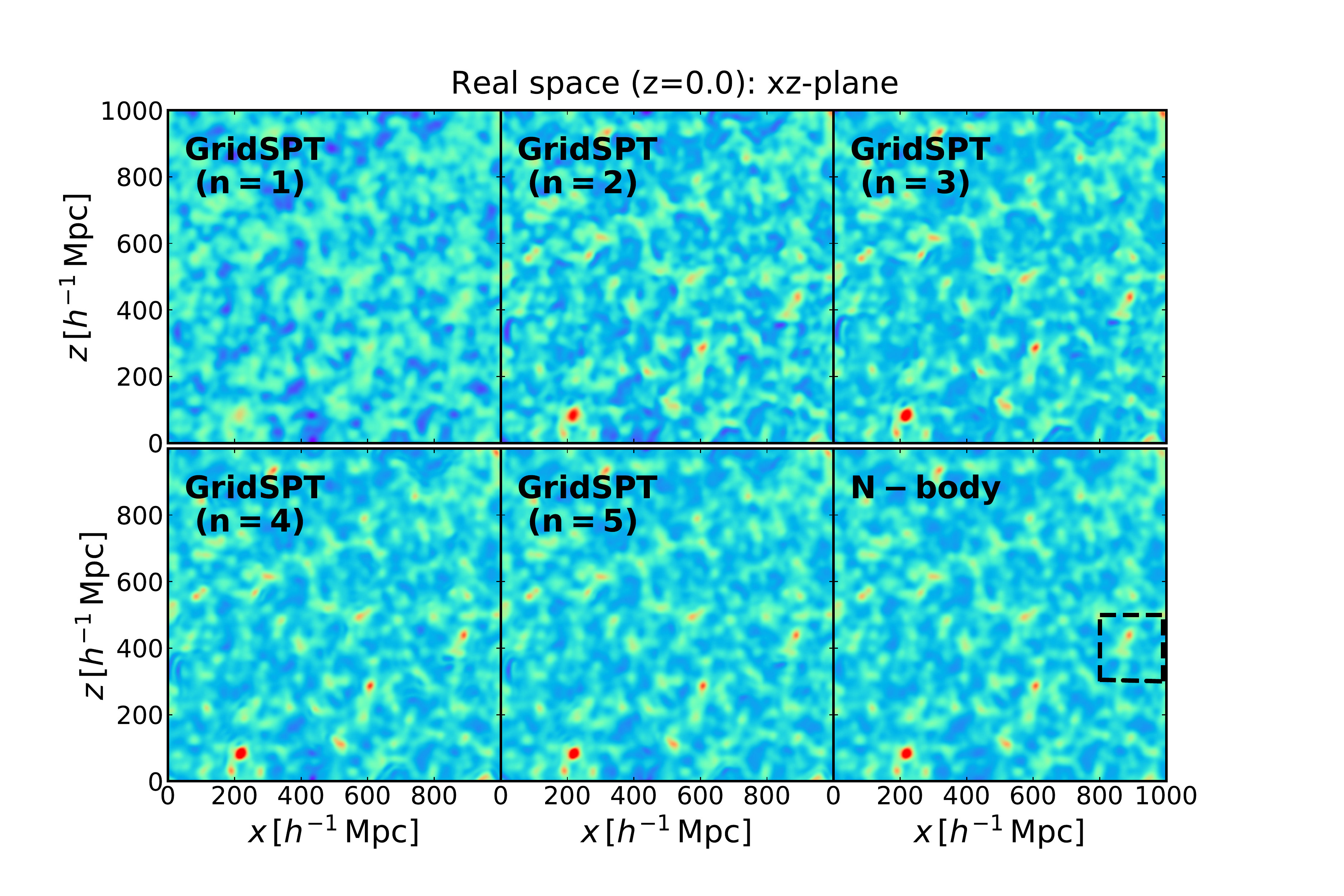}
\hspace*{-2.0cm}
\includegraphics[width=10.3cm,angle=0]{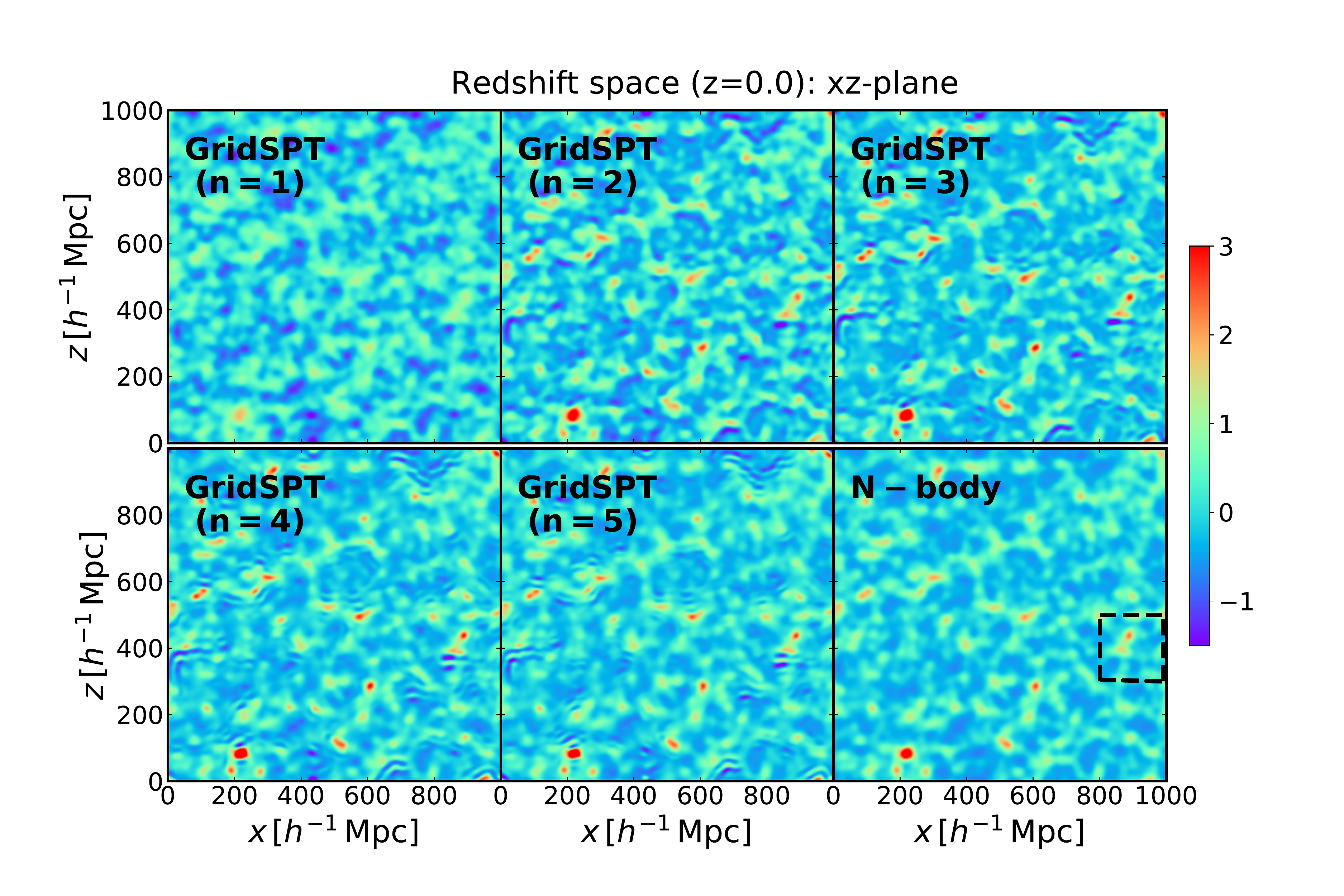}
\end{center}
\vspace*{-0.6cm}
\caption{Same as Fig.~\ref{fig:Slice_real_red_xy}, but the results for a slice of $10\,h^{-1}$\,Mpc depth in $x$-$z$ plane is shown. 
\label{fig:Slice_real_red_xz}
}
\end{figure*}

\subsection{Properties of SPT density fields}

Let us begin by looking at the generated density fields in real and redshift space. 

\begin{figure*}[tb]
\vspace*{-0.8cm}
\begin{center}
\hspace*{-1.0cm}
\includegraphics[width=10.3cm,angle=0]{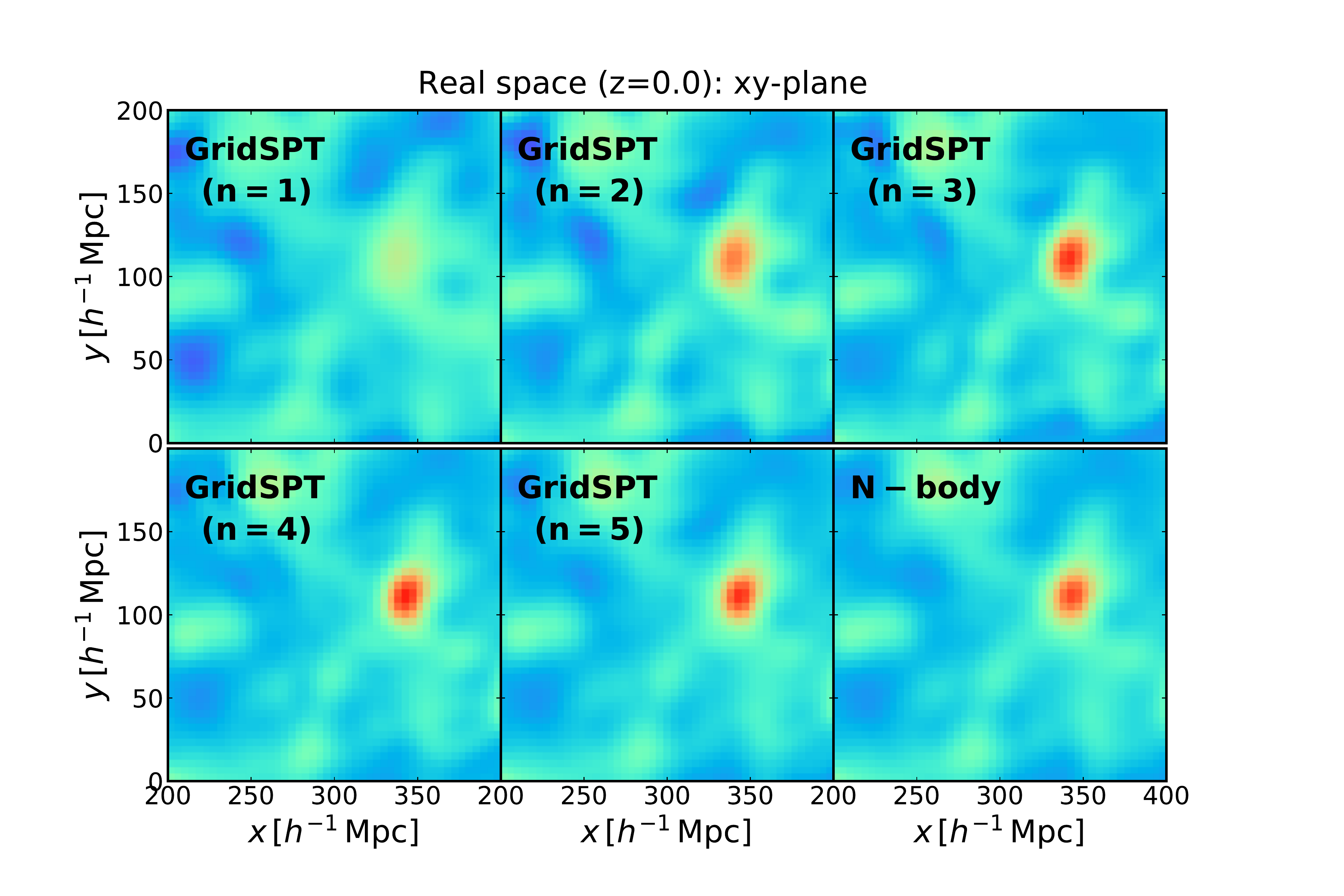}
\hspace*{-2.0cm}
\includegraphics[width=10.3cm,angle=0]{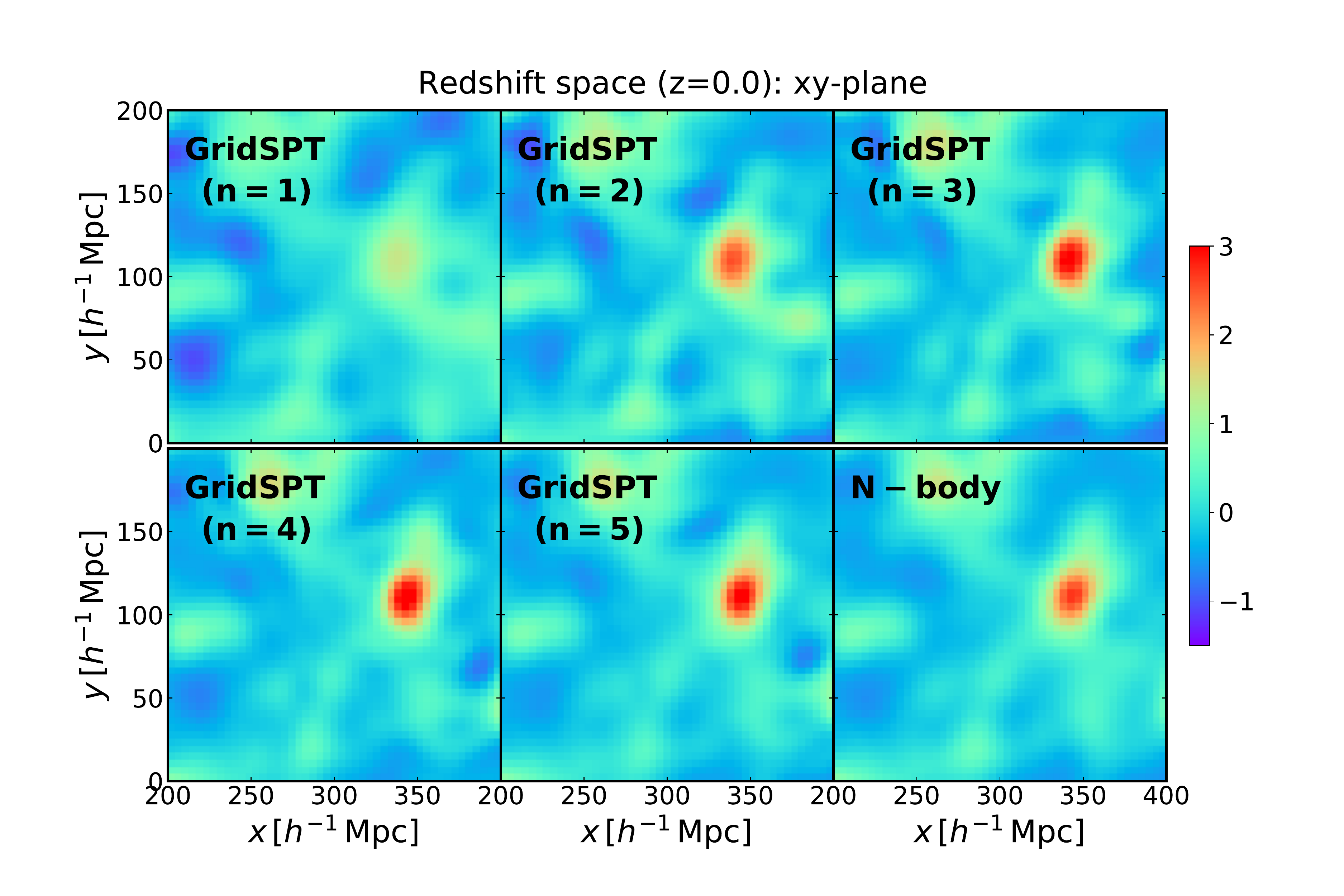}
\end{center}
\vspace*{-0.6cm}
\caption{Same as Fig.~\ref{fig:Slice_real_red_xy}, but enlarged plot of the 2D density field over $200 \times 200\, h^{-1}$\,Mpc size is shown for the region enclosed by the dashed line in the bottom right panel of Fig.~\ref{fig:Slice_real_red_xy}. 
\label{fig:Slice_zoom_real_red_xy}
}
\vspace*{-0.5cm}
\begin{center}
\hspace*{-1.0cm}
\includegraphics[width=10.3cm,angle=0]{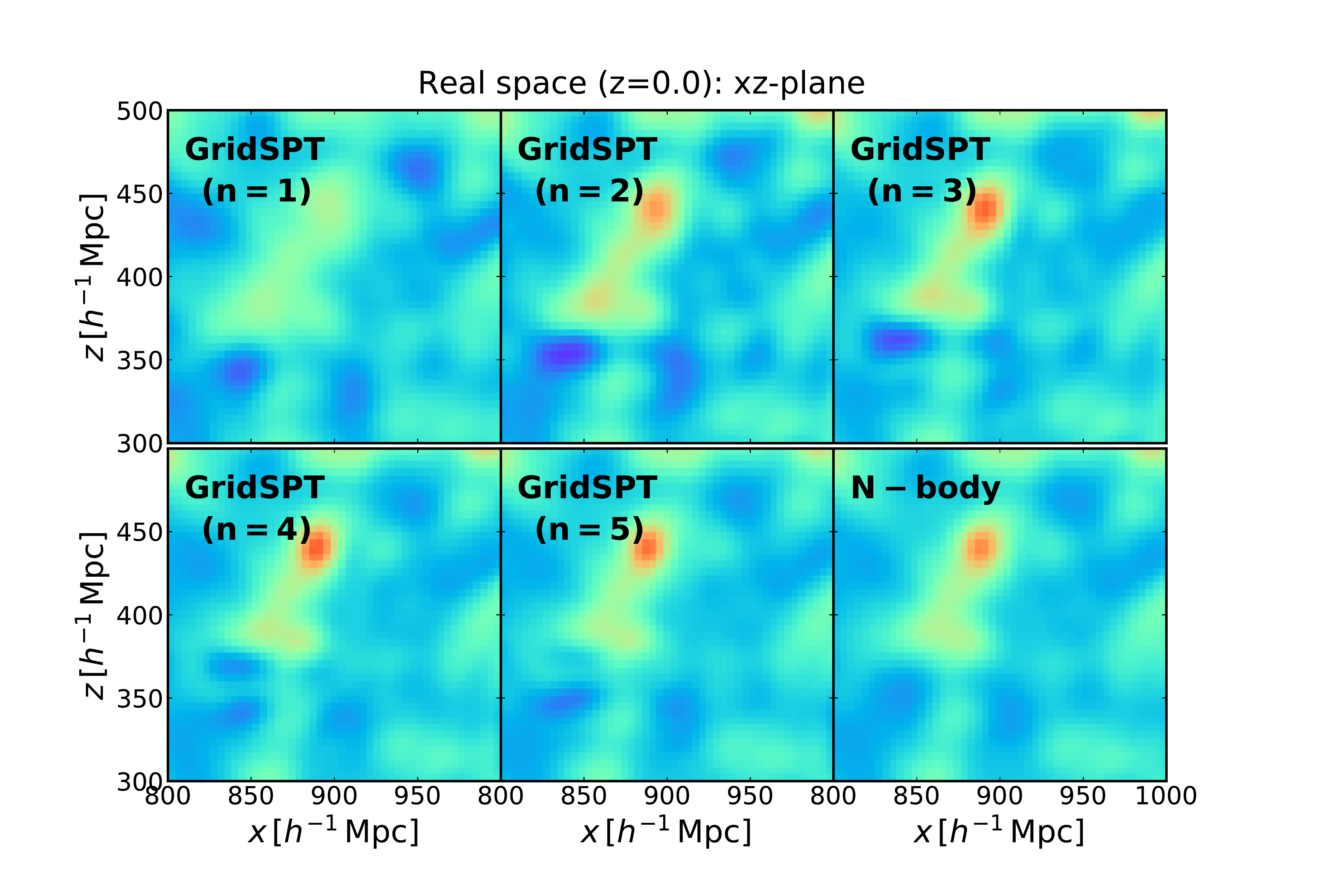}
\hspace*{-2.0cm}
\includegraphics[width=10.3cm,angle=0]{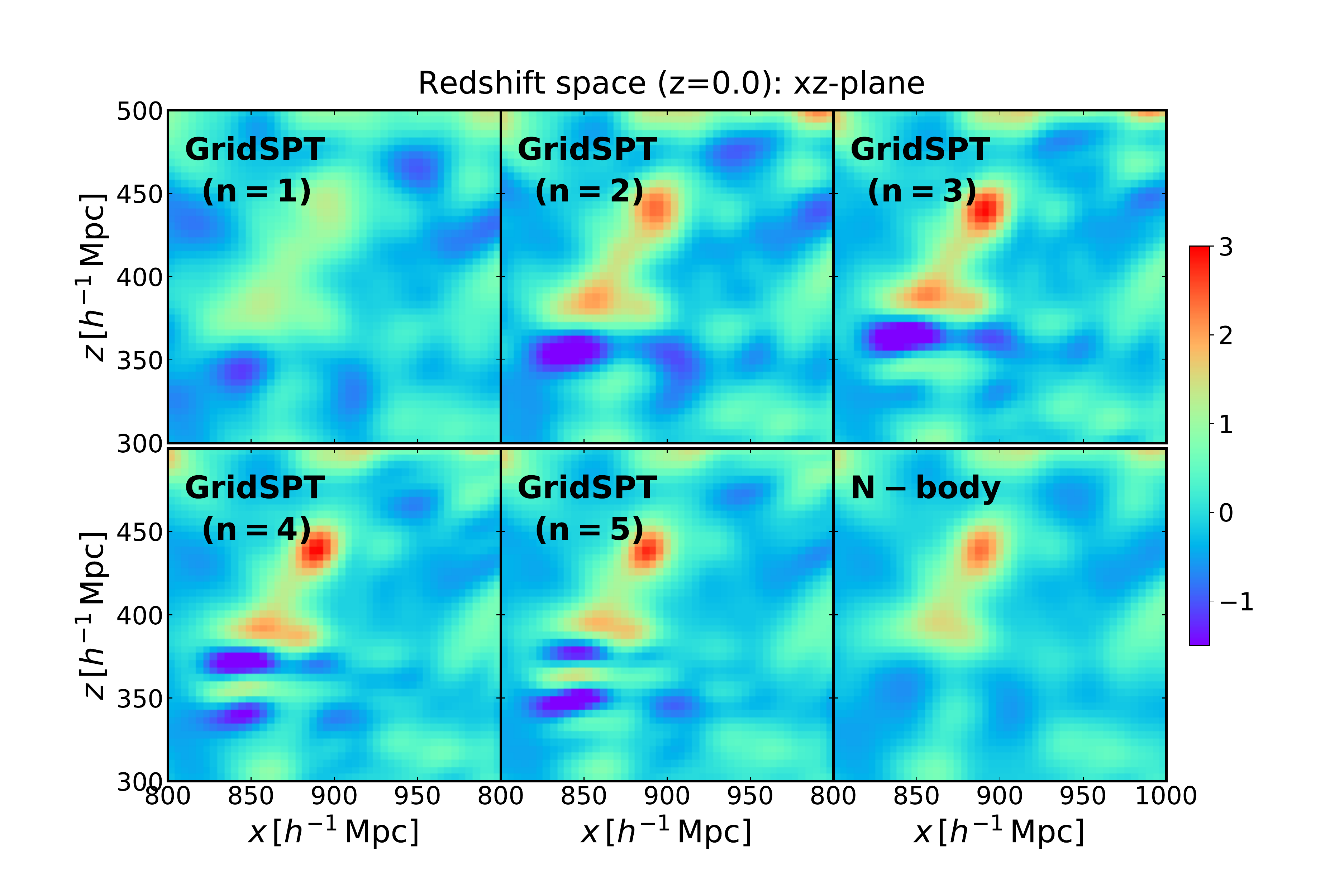}
\end{center}
\vspace*{-0.6cm}
\caption{Same as Fig.~\ref{fig:Slice_real_red_xz}, but enlarged plot of the 2D density field over $200 \times 200\,h^{-1}$\,Mpc size is shown for the region enclosed by the dashed line in the bottom right panel of Fig.~\ref{fig:Slice_real_red_xz}. 
\label{fig:Slice_zoom_real_red_xz}
}
\end{figure*}

Figs.~\ref{fig:Slice_real_red_xy} and \ref{fig:Slice_real_red_xz} present the 
2D slices of the real- (left) and redshift-space (right) density fields at $z=0$ obtained from \gridspt\, and $N$-body results, taking the $z$-axis to be the line-of-sight direction. Applying the Gaussian filter of radius $R=10\,h^{-1}$\,Mpc, a slice of $xy$- (Fig.~\ref{fig:Slice_real_red_xy}) and $xz$-plane (Fig.~\ref{fig:Slice_real_red_xz}) is taken, and is averaged over $10\,h^{-1}$\,Mpc depth on each plane.  In Figs.~\ref{fig:Slice_real_red_xy} and \ref{fig:Slice_real_red_xz}, the density fields over the entire box are shown. On the other hand, Figs.~\ref{fig:Slice_zoom_real_red_xy} and \ref{fig:Slice_zoom_real_red_xz} plot a zoom-in view over the $200\times200\,h^{-1}$\,Mpc-sized region, which are taken from Figs.~\ref{fig:Slice_real_red_xy} and \ref{fig:Slice_real_red_xz} enclosed by the dashed line in the bottom right panel. In all figures, the amplitudes of density fields, plotted in linear scale, are indicated by the same color scale. 

In each panel, five successive sub-panels from top left to bottom middle are the \gridspt\, results summing up higher-order corrections one by one, i.e., $\sum_{j=1}^n\,\delta_j$ or $\sum_{j=1}^n\,\delta_j^{\rm (S)}$, with the number $n$ indicated in each sub-panel. These are compared with the $N$-body results shown in the bottom right sub-panel. Note that the real-space results in the left panels of Figs.~\ref{fig:Slice_real_red_xy} and \ref{fig:Slice_zoom_real_red_xy} are exactly the same as Figs. 1 and 2 of Ref.~\cite{Taruya_Nishimichi_Jeong2018}, but with a different color scheme. Adding higher-order PT corrections, the real-space density fields obtained from \gridspt\, get closer to the $N$-body result, and at the fifth order, the PT density field smoothed over $10\,h^{-1}$\,Mpc agrees well with the $N$-body result.

Similarly, the $xy$-plane density fields (Figs.~\ref{fig:Slice_real_red_xy} and \ref{fig:Slice_zoom_real_red_xy}) in redshift space show a good agreement between the $5$-th order PT result and the $N$-body result. A closer look at the amplitude reveals that the contrast between under- and over-dense regions becomes more pronounced in the redshift-space than in the real-space. This would be partly ascribed to the Kaiser effect \cite{Kaiser1987,Jeong/etal:2015}, but the fact that the effect looks more significant in higher-order \gridspt\, and $N$-body density fields implies that there is a certain amount of nonlinear contribution, boosting the linear-order enhancement. 

On the other hand, in the $xz$-plane (Figs.~\ref{fig:Slice_real_red_xz} and \ref{fig:Slice_zoom_real_red_xy}), the \gridspt\, density fields exhibit wobbly structures with successive under- and over-dense regions, which appear most significant along the line-of-sight direction (e.g., see the region around $(x,z)=(850,350)\, h^{-1}$\,Mpc in right panels of Fig.~\ref{fig:Slice_real_red_xz} or \ref{fig:Slice_zoom_real_red_xz}). We have found that those structures are typically found around the underdense regions in the $N$-body results. We ascribe the feature to the higher-derivative terms in the higher-order SPT density field [see Eqs.~(\ref{eq:deltas_3})-(\ref{eq:deltas_5})], based upon the fact that such a structure is not seen in the real-space results, and that the feature becomes more prominent as we increase the PT order in redshift space.
In particular, the \gridspt\, implementation requires evaluating the higher-order derivative operator $\nabla_z^n$, and we have calculated them in Fourier space by multiplying the factor $(i\,k_z)^n$, which might enhance the aliasing effect beyond the level remedied by $2/(N+1)$-rule.
We have also checked that even implementing the higher-order differential scheme (e.g., see Appendix C of Ref.~\cite{Tanaka_Satoshi_etal2017}), results are hardly changed. Thus, fake wobbly structures in the $xz$-plane are a direct outcome inherent in our implementation of SPT involving higher-order derivatives. As a result, the overall agreement between \gridspt\, and $N$-body simulation in redshift space is not as good as that in real space even at the fifth order, indicating a slower
convergence of the SPT expansion in redshift space. We shall discuss this point in more detail from the statistical point of view in the next subsection.

\begin{figure*}[tp]
\begin{center}
\hspace*{-1cm}
 \includegraphics[width=20cm,angle=0]{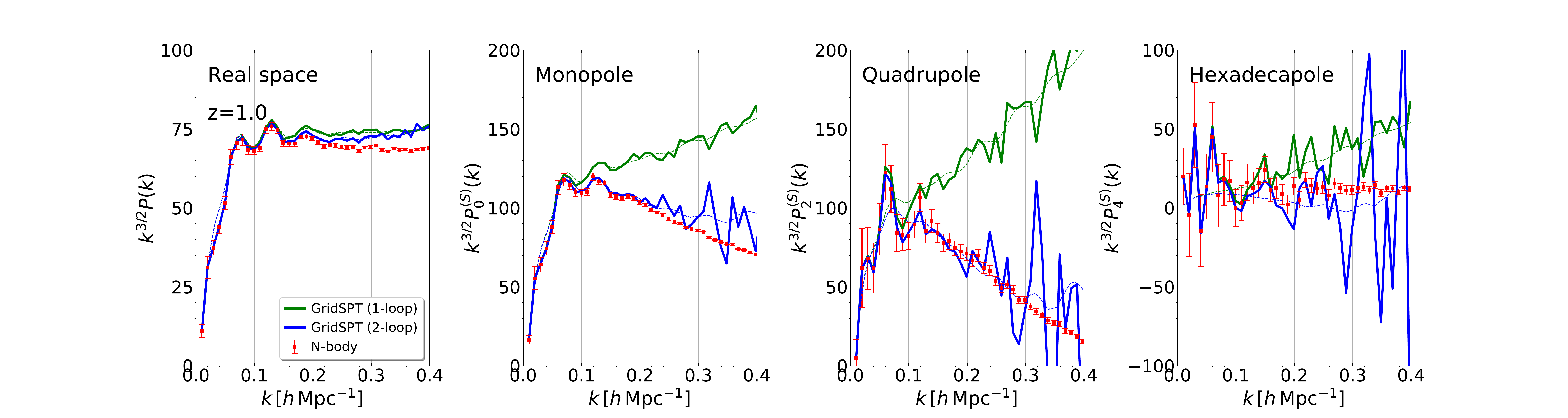}
\end{center}

\vspace*{-0.5cm}

\caption{Power spectrum in real (left) and redshift space (from second left to right) at $z=1$. Solid lines are the results from the \gridspt\, calculation with the number of grids $N_{\rm grid}=1200^3$. The analytical SPT results are also shown for reference, depicted as dotted lines. Note that the cutoff scale of $k_{\rm cut}=1.4\,h$\,Mpc$^{-1}$ is introduced in the analytical SPT calculations. In both cases, the green and blue curves respectively indicate the results at one- and two-loop order. On the other hand, the red symbols represent the measured result from $N$-body simulation with the same initial seed as used in \gridspt. Note that the errorbars shown in $N$-body result are the sampling noise estimated from the number of Fourier modes. 
\label{fig:pk_real_red_zred1_single_run}
}
\vspace*{-0.5cm}
\begin{center}
\hspace*{-1cm}
 \includegraphics[width=16cm,angle=0]{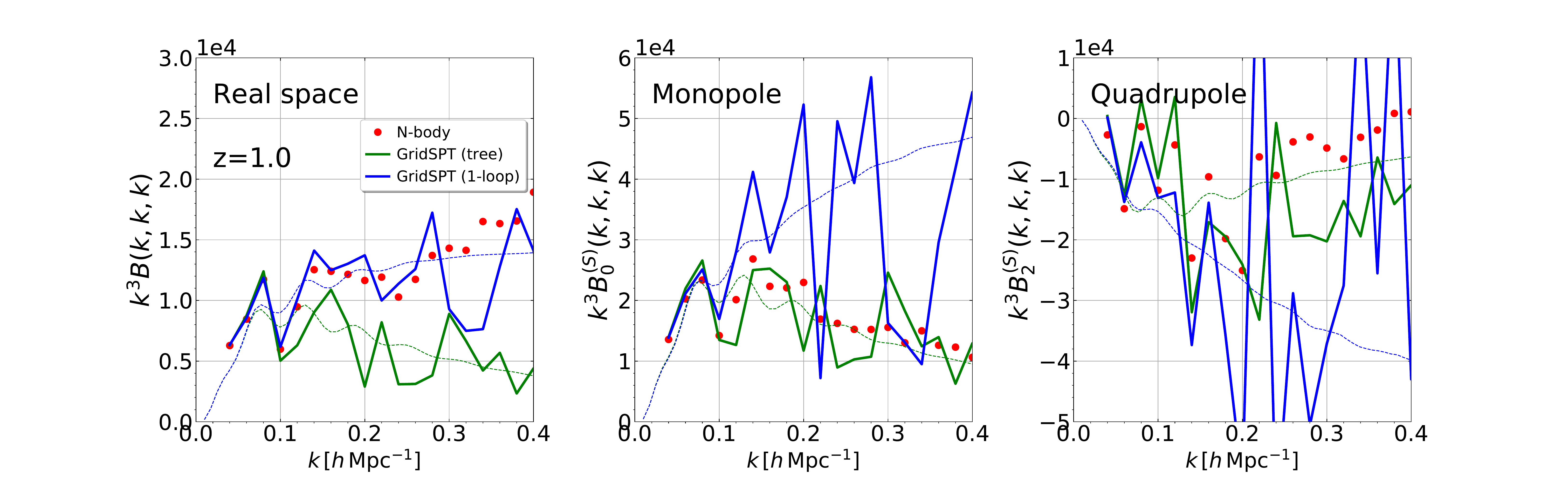}
\end{center}

\vspace*{-0.5cm}

\caption{Bispectrum in real (left) and redshift (middle and right) space at $z=1$, with the number of grids $N=1200^3$ in \gridspt\, calculations. The \gridspt\, results are shown in thick solid lines. The analytical SPT results are also plotted for reference in dotted lines. In both cases, tree-level and one-loop results are depicted as green and blue curves, respectively. The red filled circles are the measured results of the bispectrum obtained from $N$-body simulations. 
\label{fig:bk_real_red_zred1_single_run}
}
\end{figure*}

\subsection{Power spectrum and bispectrum}
\label{subsec:powerspec_bispec}

Inspecting the density fields on grids, we next consider the statistical quantities, focusing particularly on the power spectrum and the bispectrum of matter field. In both \gridspt\, and $N$-body simulations, we measure them with the same grid-based codes using FFT\footnote{To be precise, in the case of $N$-body simulations, we first assign $N$-body particles on grids to generate the density fields. We here adopt the cloud-in-cell (CIC) interpolation to do this. The interlacing de-aliasing correction is made based on Ref.~\cite{sefusatti2016} before we divide by the CIC window function to obtain our final estimate of the density field on grids.}. In redshift space, the statistical isotropy is known to be manifestly broken, and measured results of the power spectrum and bispectrum, which we respectively denote by $P^{\rm(S)}$ and $B^{\rm(S)}$, exhibit anisotropies along the line-of-sight direction ($z$-axis in our case). To characterize their anisotropic nature, we apply the multipole expansion and define the multipole moments as follows: 
\begin{align}
& P_\ell^{\rm(S)}(k) \equiv \frac{2\ell+1}{2}\int_{-1}^1 d\mu \,P^{\rm(S)}(\bfk)\,\mathcal{P}_\ell(\mu)
\label{eq:pk_multipoles}
\end{align}
for the power spectrum.  The function $\mathcal{P}_\ell$ is the Legendre polynomials, and the quantity $\mu$ is the directional cosine given by $\mu\equiv \hat{k}\cdot\hat{z}$, or equivalently, $k_z/k$ in our setup. For the bispectrum, we adopt the definition used in Ref.~\cite{Hashimoto_Rasera_Taruya2017}: 
\begin{align}
&  B_\ell^{\rm(S)}(k_1,k_2,k_3) 
\nonumber
\\
&\,\,\,
\equiv \frac{2\ell+1}{2} \int_{-1}^1 d\mu \,\int_0^{2\pi}\,\frac{d\phi}{2\pi}\, B^{\rm(S)}(\bfk_1,\bfk_2,\bfk_3)\,\mathcal{P}_\ell(\mu),
\label{eq:bk_multipoles}
\end{align}
where the directional cosine $\mu$ is defined with the orientation angle between the line-of-sight direction and the vector normal to the triangle formed with three wave vectors. The angle $\phi$ represents the azimuthal angle characterizing the rotation of the triangle on the plane. To be specific, we set  
\begin{align}
& \mu=\cos\omega=\frac{(\hat{k}_1\times\hat{k}_2)\cdot\hat{z}}{\sin\theta_{12}},\label{eq:def_of_mu_for_bk}
\\
& \cos\phi=\frac{\{\hat{z}\times(\hat{k}_1\times\hat{k}_2)\}\cdot\hat{k}_1}{\sin\omega}.
\label{eq:def_of_phi_for_bk}
\end{align}
Note that the bispectrum multipoles $B^{\rm(S)}$ defined above differ from those used in the literature (e.g., Refs.~\cite{Scoccimarro_etal1999,Scoccimarro2015,Gagrani_Samushia2017,Yamamoto_Nan_Hikage2017}, see also Ref.~\cite{Sugiyama_etal2019} for a comparison between different coordinate choices), but a nice property of this definition is that they are symmetric under the permutation of the order of $k_1$, $k_2$ and $k_3$.

\begin{figure*}[tp]
\begin{center}
 \includegraphics[width=20cm,angle=0]{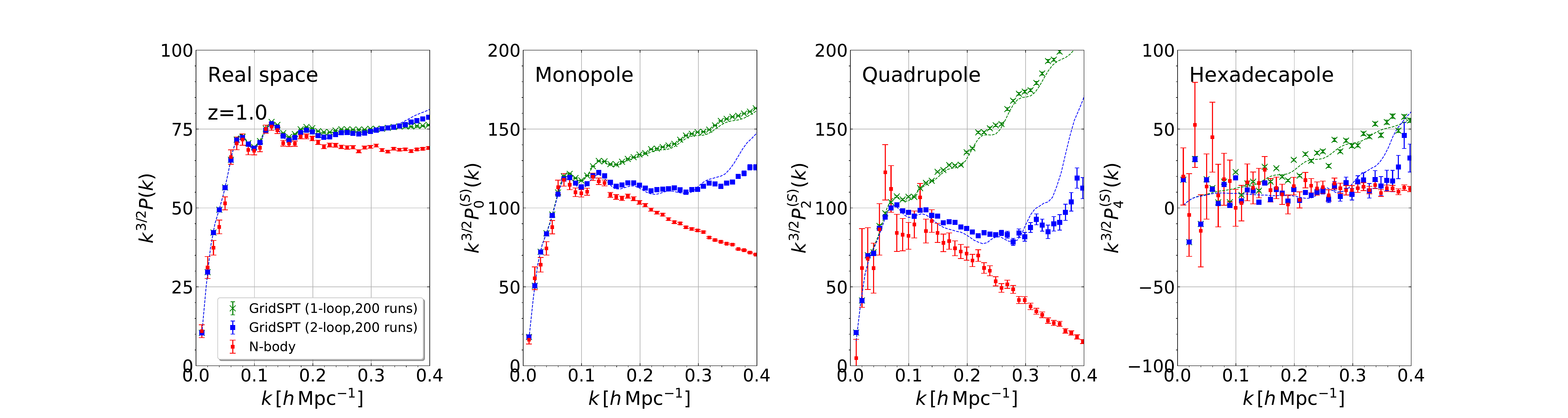}
\end{center}

\vspace*{-0.5cm}

\caption{Same as Fig.~\ref{fig:pk_real_red_zred1_single_run}, but the \gridspt\, results at one- and two-loop order, respectively shown in green and blue symbols, are averaged over $200$ realizations, adopting the number of grids $N_{\rm grid}=600^3$. The analytical SPT results (dotted lines) are computed with the cutoff scale $k_{\rm cut}=0.8~h{\rm Mpc}^{-1}$.
    The errorbars on the \gridspt\,results indicate the standard error of the mean over the $200$ realizations. The $N$-body results, depicted by the symbols in red, are identical to those shown in Fig.~\ref{fig:pk_real_red_zred1_single_run}, and their errorbars indicate the sampling error estimated from the number of Fourier modes for a single realization data. Note that the analytical SPT results shown here (dotted) adopt a different cutoff wavenumber (see footnote), and thus differs from those in  Fig.~\ref{fig:pk_real_red_zred1_single_run}. 
\label{fig:pk_real_red_zred1_200_runs}
}
\vspace*{-0.5cm}
\begin{center}
\hspace*{-1cm}
 \includegraphics[width=16cm,angle=0]{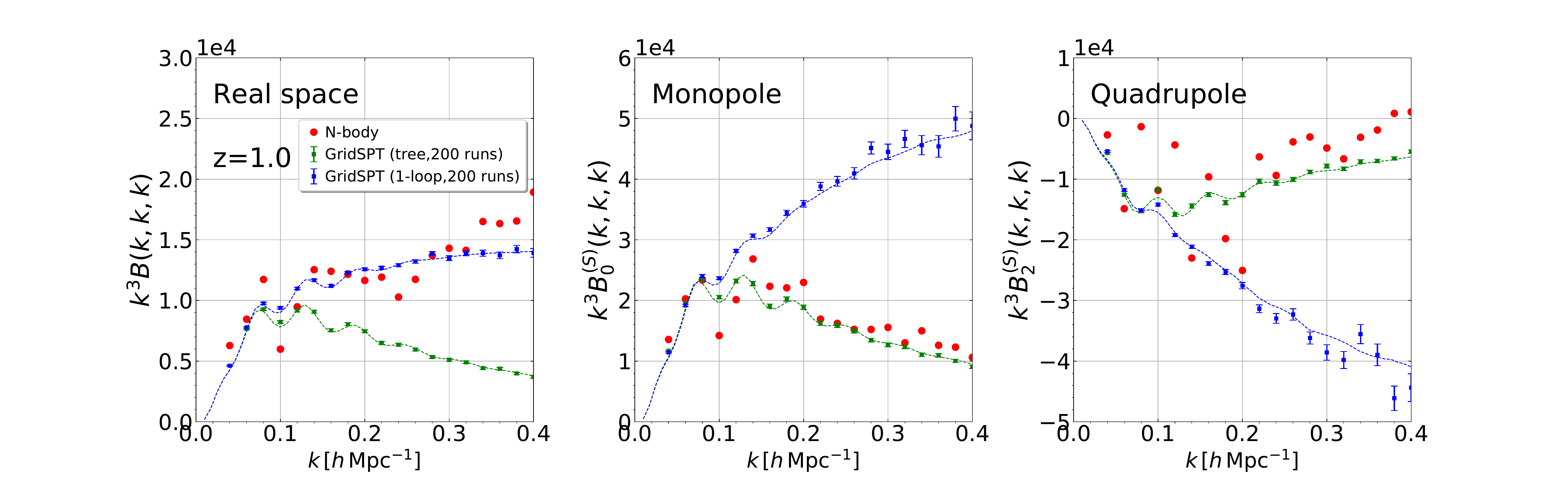}
\end{center}

\vspace*{-0.5cm}

\caption{
Same as Fig.~\ref{fig:bk_real_red_zred1_single_run}, but the \gridspt\, results of tree-level and one-loop calculations, 
respectively depicted as green and blue symbols, are averaged over $200$ realizations, adopting the number of grids $N_{\rm grid}=600^3$. 
The errorbars for the \gridspt\, results represent the standard error of the mean over the $200$ realizations. 
\label{fig:bk_real_red_zred1_200_runs}
}
\end{figure*}

Figs.~\ref{fig:pk_real_red_zred1_single_run} and \ref{fig:bk_real_red_zred1_single_run} show the results for the matter power spectrum and the matter bispectrum from a single-realization at $z=1$. Here, the bispectrum is measured in the equilateral configuration, taking the three wavenumbers to be the same $(k_1=k_2=k_3\equiv k)$, and is plotted as function of $k$. The \gridspt\, results depicted as solid lines, are respectively constructed up to the two-loop and one-loop order, through 
\begin{align}
& P^{\rm(S)}(\bfk)=P_{\rm lin}^{\rm(S)}(\bfk)+
P_{\rm 1\mbox{-}loop}^{\rm(S)}(\bfk)+
P_{\rm 2\mbox{-}loop}^{\rm(S)}(\bfk);
\label{eq:pkSPT}
\\
 &\quad P_{\rm lin}^{\rm(S)}(\bfk)=P_{11}^{\rm(S)}(\bfk),
\label{eq:pk_linear}
\\
 &\quad P_{\rm 1\mbox{-}loop}^{\rm(S)}(\bfk)=2\,P_{13}^{\rm(S)}(\bfk)+P_{22}^{\rm(S)}(\bfk),
\label{eq:pk_1loop}
\\
 &\quad P_{\rm 2\mbox{-}loop}^{\rm(S)}(\bfk)=2\,P_{15}^{\rm(S)}(\bfk)+2\,P_{24}^{\rm(S)}(k)+P_{33}^{\rm(S)}(\bfk)
\label{eq:pk_2loop}
\end{align}
for the power spectrum, and 
\begin{align}
&B^{\rm(S)}(\bfk_1,\,\bfk_2,\,\bfk_3) 
\nonumber
\\
&\quad=B_{\rm tree}^{\rm(S)}(\bfk_1,\,\bfk_2,\,\bfk_3)+B_{\rm 1\mbox{-}loop}^{\rm(S)}(\bfk_1,\,\bfk_2,\,\bfk_3);
\label{eq:bkSPT_multipoles}
\\
 &\quad B_{\rm tree}^{\rm(S)}(\bfk_1,\,\bfk_2,\,\bfk_3)=B_{112}^{\rm(S)}(\bfk_1,\,\bfk_2,\,\bfk_3) 
\nonumber
\\
&\qquad
+ \mbox{2\,\,perms}\,\, (\bfk_1\leftrightarrow \bfk_2\leftrightarrow \bfk_3),
\label{eq:bk_tree}
\\
 &\quad B_{\rm 1\mbox{-}loop}^{\rm(S)}(\bfk_1,\,\bfk_2,\,\bfk_3) =\Bigl\{B_{123}^{\rm(S)}(\bfk_1,\,\bfk_2,\,\bfk_3) 
\nonumber
\\
&\qquad + \mbox{5\,\,perms}\,\, (\bfk_1\leftrightarrow \bfk_2\leftrightarrow \bfk_3)\Bigr\} + \Bigl\{B_{114}^{\rm(S)}(\bfk_1,\,\bfk_2,\,\bfk_3)
\nonumber
\\
&\qquad  + \mbox{2\,\,perms}\,\, (\bfk_1\leftrightarrow \bfk_2\leftrightarrow \bfk_3)\Bigr\} 
+ B_{222}^{\rm(S)}(\bfk_1,\,\bfk_2,\,\bfk_3)
\label{eq:bk_1loop}
\end{align}
for the bispectrum. In the above, building blocks of the power spectrum and bispectrum, $P_{ab}$ and $B_{abc}$, are defined respectively by
\begin{align}
& \langle\delta_a^{\rm(S)}(\bfk)\delta_b^{\rm(S)}(\bfk')\rangle=
(2\pi)^3\delta_{\rm D}(\bfk+\bfk')\,P_{ab}^{\rm(S)}(\bfk),
\label{eq:def_pkred_SPT}
\\
& \langle\delta_a^{\rm(S)}(\bfk_1)\delta_b^{\rm(S)}(\bfk_2)\delta_c^{\rm(S)}(\bfk_3)\rangle
\nonumber
\\
&\quad=(2\pi)^3\delta_{\rm D}(\bfk_1+\bfk_2+\bfk_3)\,B_{abc}^{\rm(S)}(\bfk_1,\,\bfk_2,\,\bfk_3).
\label{eq:def_bkred_SPT}
\end{align}
Applying the multipole expansion to each term, the multipole moments of the redshift-space power spectrum and bispectrum are respectively evaluated up to $\ell=4$  and $\ell=2$, together with the real-space power spectrum and bispectrum\footnote{In practice, measurements from the density fields on grids are made with discrete Fourier modes, and  we use the FFT-based algorithm to directly evaluate the power spectrum and bispectrum multipoles (e.g.,
Ref.~\cite{Scoccimarro2015,Baldauf_etal2015,Bianchi_etal2015}). }.

In Figs.~\ref{fig:pk_real_red_zred1_single_run} and \ref{fig:bk_real_red_zred1_single_run}, we plot the measurements from the $N$-body simulation in red symbols. The errorbars particularly shown for the power spectra indicate the sampling noise estimated from the number of Fourier modes in each bin. In addition, we plot the analytical SPT predictions, which we obtain by directly performing the relevant loop integrals numerically, at both next-to-leading (one-loop) and next-to-next-to-leading
(two-loop) orders, depicted as dotted lines. In Appendix~\ref{sec:analytical_SPT}, for the sake of completeness, we present the analytical expressions for the SPT power spectrum and bispectrum in redshift space. Note that in both \gridspt\, and analytical SPT calculations, the two-loop redshift-space power spectra are the results presented for the first time in this paper.

Overall, the \gridspt\, power spectra consistently reproduce the analytical SPT calculations. Note here that for analytical SPT calculations, we introduce the cutoff scales in the linear power spectrum so as to accommodate with \gridspt\, calculations\footnote{To be precise, we introduce the low-$k$ cutoff $k_{\rm min}$ set to the fundamental mode determined by the box size (i.e., $k_{\rm min}=2\pi/L$). Further, the high-$k$ cutoff is introduced, setting $k_{\rm max}$ to $1.4\,h$\,Mpc$^{-1}$, which is close to the de-aliasing filter scale in \gridspt, $k_{\rm crit}=k_{\rm Nyq}/3\simeq1.26\,h$\,Mpc$^{-1}$.}. Compared to the real-space results, adding the two-loop corrections to the one-loop spectra largely suppresses the amplitude of the power spectra. As a result, the predictions at $z=1$ get closer to the $N$-body results at $k\lesssim0.2\,h$\,Mpc$^{-1}$, above which the \gridspt\, results become 
slightly noisier. 

On the other hand,  adding the one-loop order, the SPT predictions of the bispectrum positively (negatively) increase its amplitude for the monopole (quadrupole) moment. While the one-loop prediction seems to reasonably match the real-space results in $N$-body simulations, a quick look at the redshift-space results indicates that rather than the one-loop SPT, the tree-level predictions better explain the $N$-body results. Although these are qualitatively similar to what have been found in previous works (e.g., Ref.~\cite{Hashimoto_Rasera_Taruya2017}), the bispectrum measured from the \gridspt\, fields 
are rather noisy, difficult to judge whether it is quantitatively consistent or not.

\begin{figure*}[tp]
\begin{center}
 \includegraphics[width=8.2cm,angle=0]{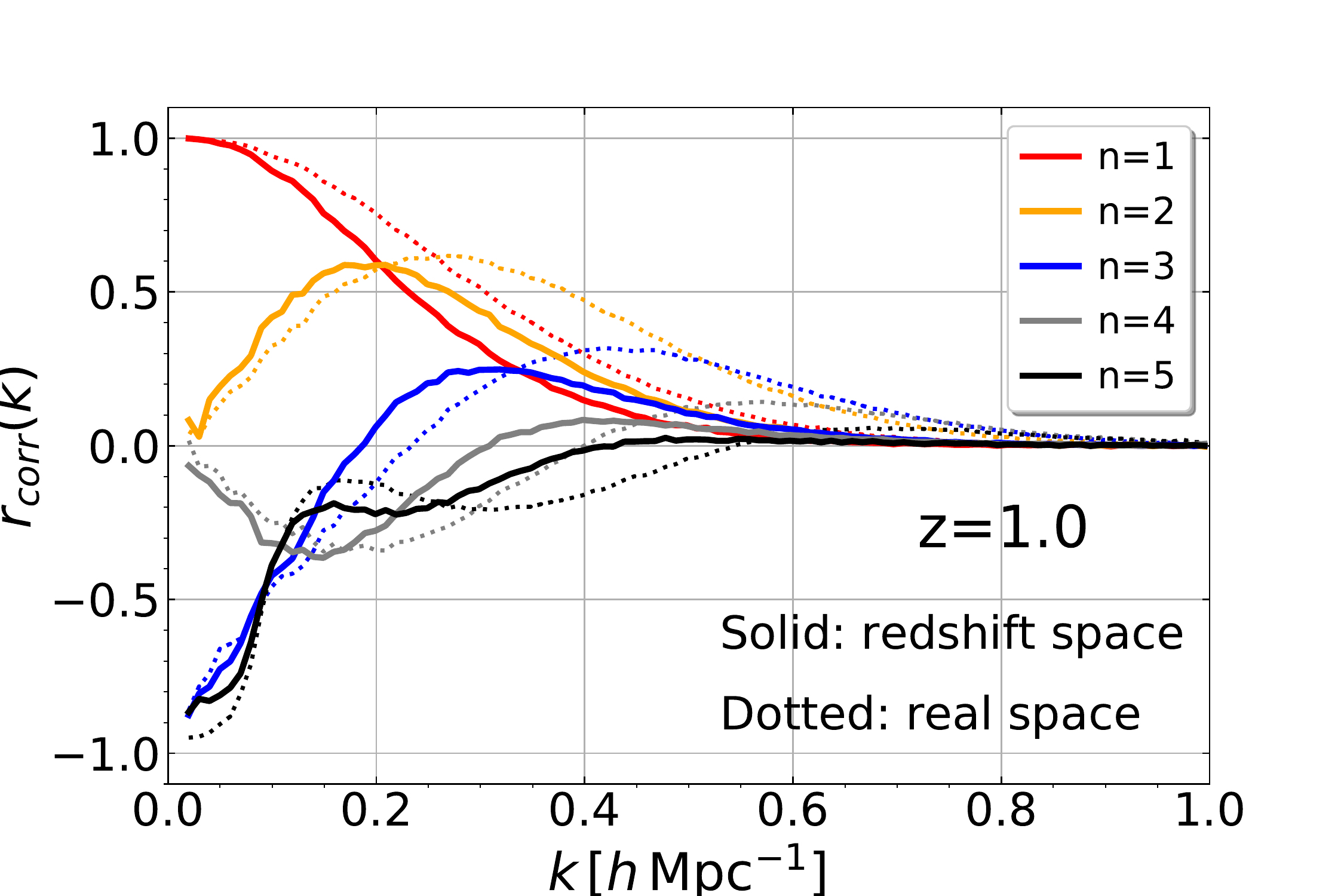}
 \includegraphics[width=8.2cm,angle=0]{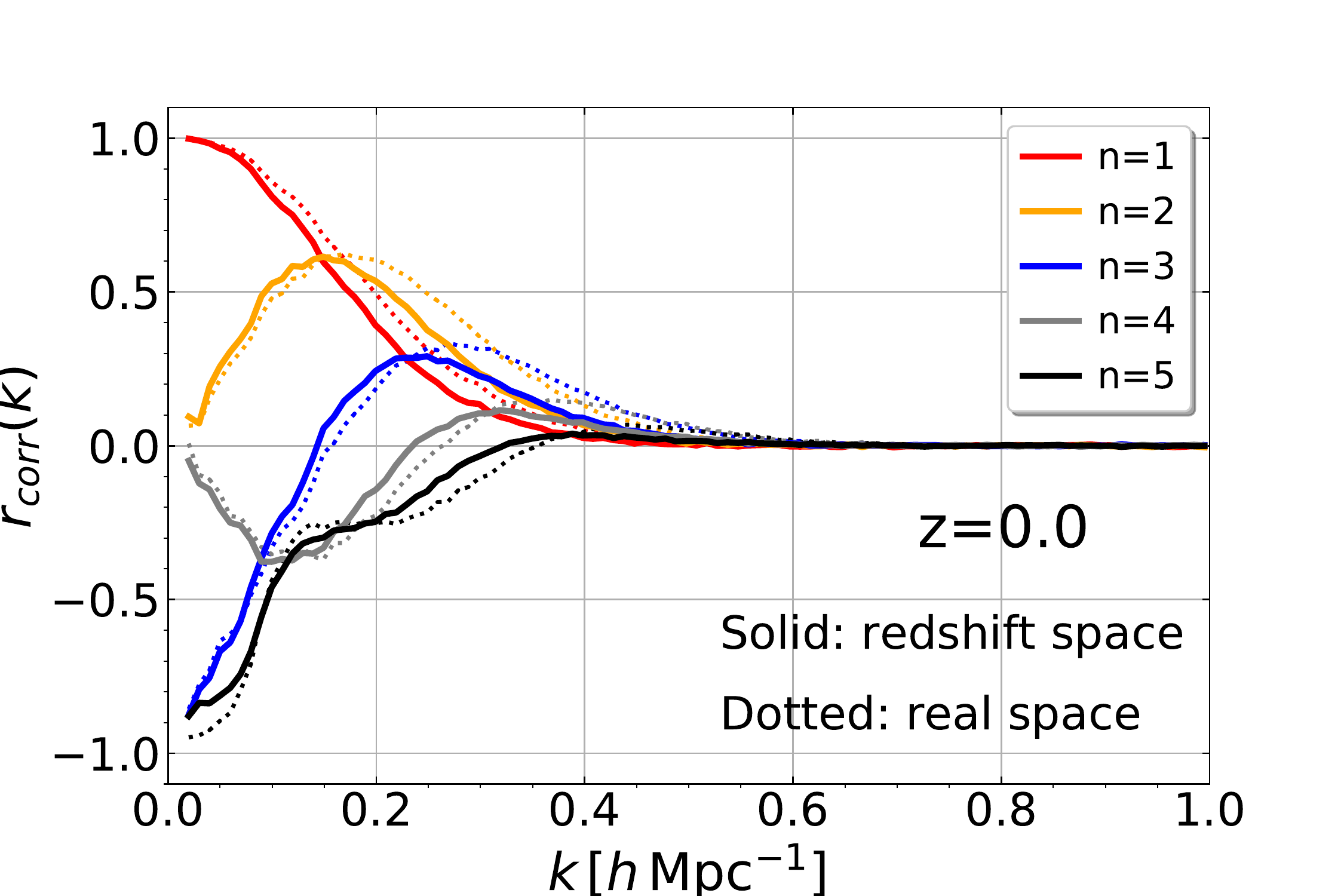}
\end{center}

\vspace*{-0.5cm}

\caption{Cross-correlation coefficient for \gridspt\, and $N$-body density fields, $r_{\rm corr}^{(n)}$, defined 
in Eq.~(\ref{eq:def_r_corr}). Results at $z=1$ (left) and $0$ (right) are shown. The solid and dotted lines are the results in redshift and real space, respectively. 
\label{fig:cross_corr_real_red}
}
\vspace*{-0.5cm}
\begin{center}
 \includegraphics[width=8.2cm,angle=0]{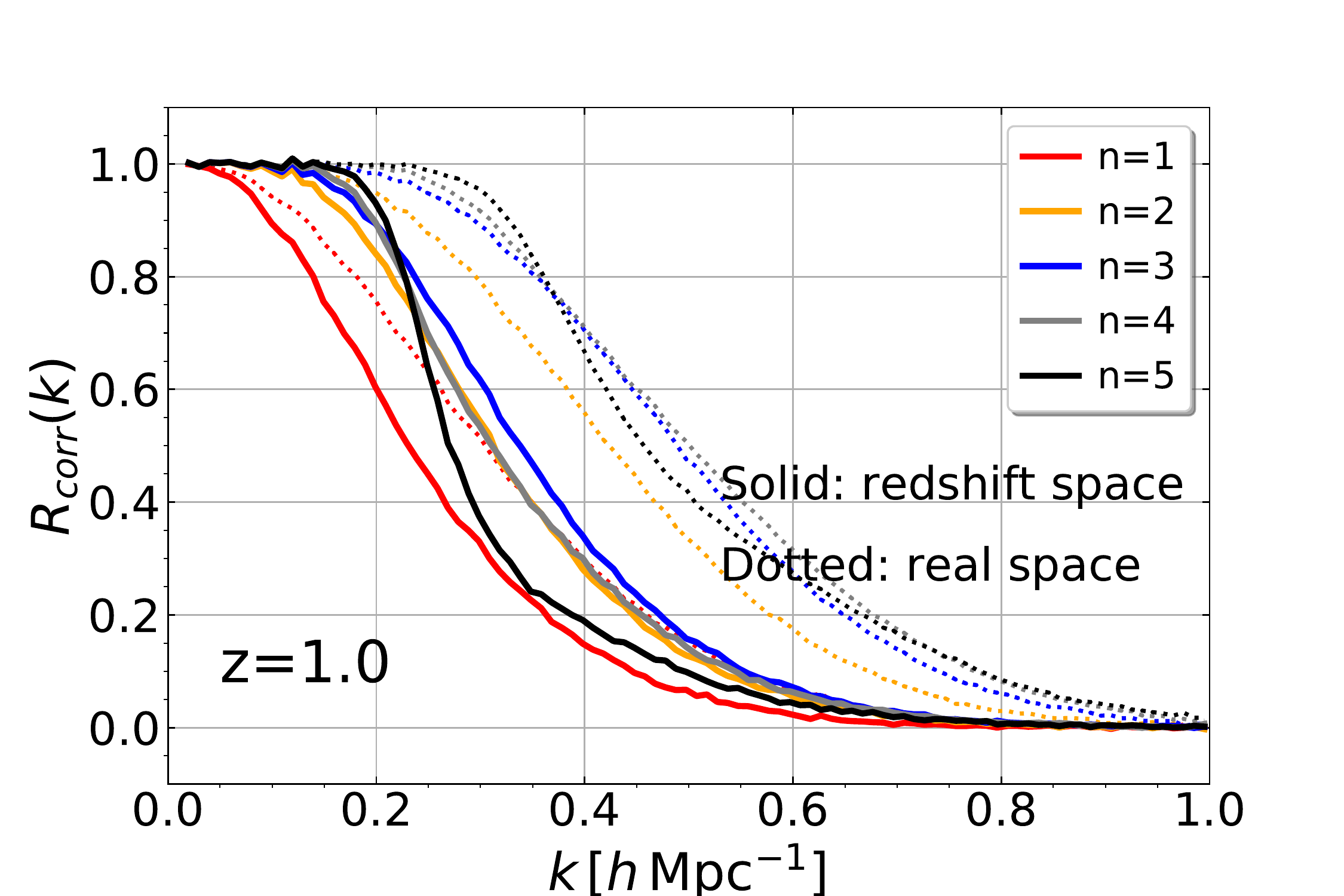}
 \includegraphics[width=8.2cm,angle=0]{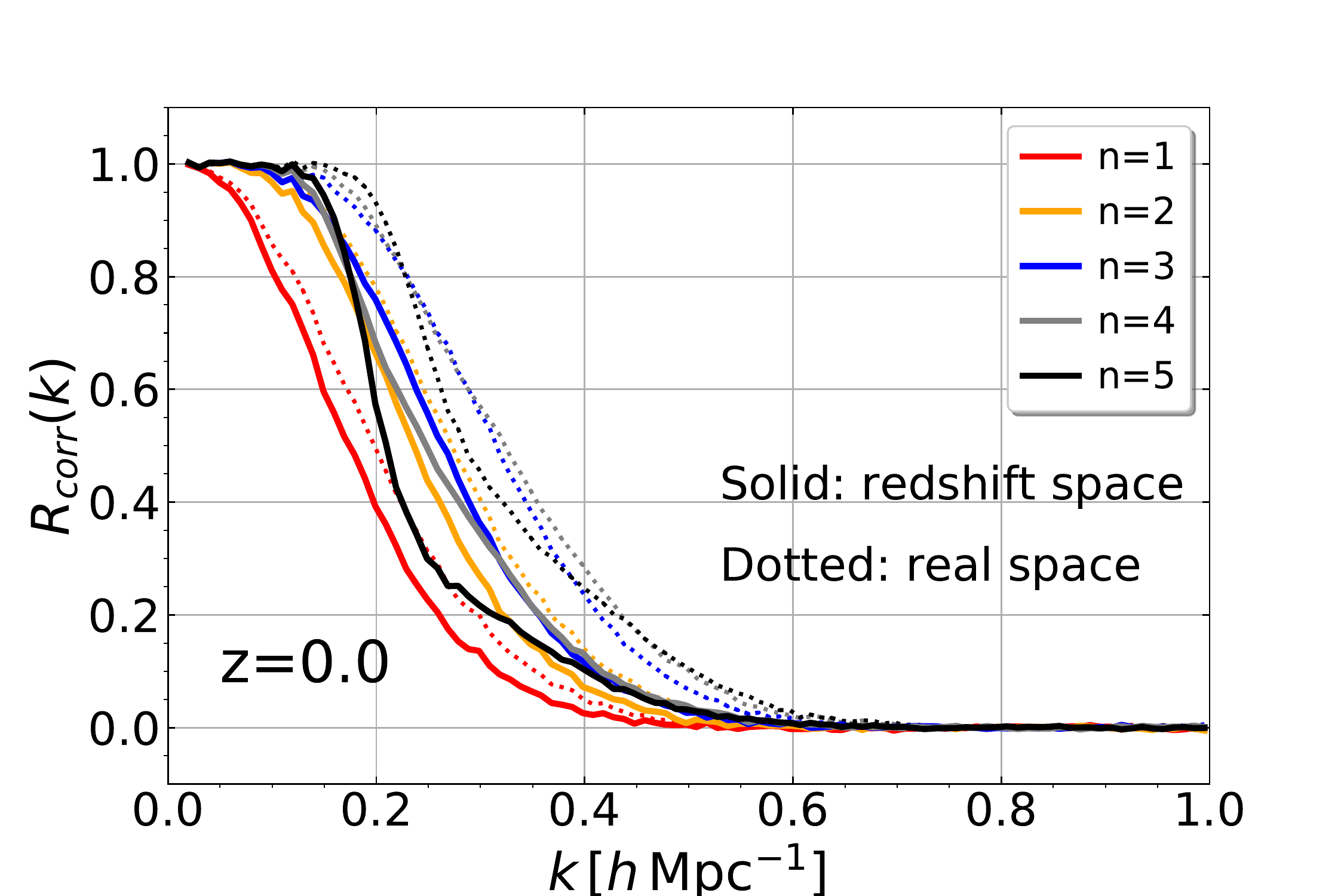}
\end{center}

\vspace*{-0.5cm}

\caption{Cross-correlation coefficient for \gridspt\, and $N$-body density fields, $R_{\rm corr}^{(n)}$, defined 
in Eq.~(\ref{eq:def_R_corr}). The results at $z=1$ (left) and $0$ (right) are shown in the case of real (dotted) and redshift (solid) space.  
\label{fig:R_corr_real_red}
}
\end{figure*}

For a more quantitative assessment of the statistical predictions, we increase the number of realizations in \gridspt\, calculations up to $200$,  and in Figs.~\ref{fig:pk_real_red_zred1_200_runs} and \ref{fig:bk_real_red_zred1_200_runs}, the average over the realizations are shown, with the error bars of the \gridspt\, results indicating the standard error of the mean over $200$ realizations. 
To speed up the calculations, we here adopt a smaller number of grid points, $N_{\rm grid}=600^3$.
It is now clear that the \gridspt\, results agree well with analytical SPT predictions not only for the power spectrum but also for the bispectrum. Note that the cutoff scales of the analytical SPT calculations were adjusted again close to the one introduced in the \gridspt\,calculations\footnote{In this case, while the low-$k$ cutoff is kept fixed to the one used in Figs.~\ref{fig:pk_real_red_zred1_single_run} and \ref{fig:bk_real_red_zred1_single_run}, the high-$k$ cutoff is changed to $k_{\rm max}=0.8\,h$\,Mpc$^{-1}$. Note that adopting $N_{\rm grid}=600^3$, the de-aliasing filter scale of the \gridspt\, calculations is $k_{\rm crit}\simeq 0.62\,h$\,Mpc$^{-1}$.}. Then, in redshift space, the one-loop bispectra are shown to largely deviate from $N$-body simulations, and the tree-level bispectra rather match the $N$-body results (Fig.~\ref{fig:bk_real_red_zred1_200_runs}), in marked contrast to the real-space bispectrum. These are fully consistent with previous results.

Finally, going back to the results of the power spectra in Fig.~\ref{fig:pk_real_red_zred1_200_runs}, we find that the agreement between the SPT predictions and $N$-body simulations gets worse, compared to the single-realization results in Fig.~\ref{fig:pk_real_red_zred1_single_run}. The discrepancy is particularly manifest and significant at small scales for the monopole and quadrupole moments. The major reason of this comes from the resolution of the \gridspt\, calculations, originating from the UV sensitivity inherent in the SPT. Indeed, as discussed in detail in Appendix \ref{sec:UV_sensitivity}, the SPT predictions of redshift-space power spectra sensitively depend on the small-scale cutoff. Reducing the high-$k$ cutoff significantly enhances the power spectrum amplitude on small scales.

\subsection{Cross correlation}
\label{subsec:cross_correlation}

So far, comparisons between the \gridspt\, calculations and $N$-body simulations have been made by presenting their respective predictions. In this subsection, we evaluate the cross correlation between their density fields, and investigate statistically the (dis)similarity of the fields evolved by these different techniques starting from the same initial seed. 

Consider first the density field at each PT order of \gridspt\, and compute its cross-correlation with the density field obtained from the $N$-body simulation. 
Following Ref.~\cite{Taruya_Nishimichi_Jeong2018}, we define the cross-correlation coefficient, $r_{\rm corr}^{(n)}$, given by 
\begin{align}
 r_{\rm corr}^{(n)}(k)\equiv \frac{P_{0,n\times{\rm N\mbox{-}body}}^{\rm(S)}(k)}{\sqrt{P_{0,nn}^{\rm(S)}(k)P_{0,{\rm N\mbox{-}body}}^{\rm(S)}(k)}}.
\label{eq:def_r_corr}
\end{align}
Here, the quantity in the numerator, $P_{0,n\times{\rm N\mbox{-}body}}^{\rm(S)}$, represents the monopole moment of the cross power spectrum between the $n$-th order SPT density field and the measurement from the $N$-body simulation, defined by 
\begin{align}
 \langle \delta_n^{\rm (S)}(\bfk) \delta_{\rm N\mbox{-}body}^{\rm (S)}(\bfk')\rangle =(2\pi^3)\delta_{\rm D}(\bfk+\bfk')\,P_{n\times{\rm N\mbox{-}body}}^{\rm(S)}(\bfk).
\end{align}
In the presence of the RSD effect, the above spectrum exhibits anisotropies, for which we take only the monopole moment to evaluate Eq.~(\ref{eq:def_r_corr}), i.e., averaged over the wavevectors in spherical bins. 
In Fig.~\ref{fig:R_corr_real_red}, the results in redshift and real space, depicted respectively as solid and dotted lines, are shown up to the fifth order $(n=5)$ at redshifts $z=1$ (left) and $0$ (right). Note that the real-space results are identical to those obtained in Ref.~\cite{Taruya_Nishimichi_Jeong2018}. While the low-$k$ behaviors exhibit a non-monotonic scale dependence having a positive or negative value depending on the perturbative order, all the cross-correlation coefficients asymptotically go to zero at high $k$. In real space, it has been suggested by Ref.~\cite{Taruya_Nishimichi_Jeong2018} that the asymptotic convergence at high $k$ comes from the randomness of the linear displacement field, and this is quantitatively predicted by the analytical treatment with re-summed PT calculations \cite{Taruya:2012ut}. Qualitatively, the results in redshift space show similar trends, but a closer look at small scales reveals that the asymptotic convergence to zero seems faster than that in real space, implying that the convergence of SPT expansion gets worse in redshift space, as we  expected.

To elucidate this point more clearly, we next compute the cross-correlation coefficient summing up each PT correction up to $n$-th order, $R_{\rm corr}^{(n)}$, defined by 
\begin{align}
 R_{\rm corr}^{(n)}(k)= \frac{\sum_{a=1}^n \,P_{0,a\times{\rm N\mbox{-}body}}^{\rm(S)}(k)}{\sqrt{\bigl\{\sum_{a,b=1}^{n} \,P_{0,ab}^{\rm(S)}(k)\bigr\}P_{0,{\rm N\mbox{-}body}}^{\rm(S)}(k)}}
\label{eq:def_R_corr}
\end{align}
Here, the summation in the denominator is taken only for even numbers of $a+b$ for which the expectation values are non-vanishing. The measured results of $R_{\rm corr}^{(n)}$ are plotted up to $n=5$ in Fig.~\ref{fig:R_corr_real_red}, adopting the same color scheme and line types as in Fig.~\ref{fig:cross_corr_real_red}. As anticipated, the correlation coefficient in redshift space starts to be suppressed at larger scales, and the suppression gets rather faster, compared to the results in real space. These features are more prominent at $z=0$. It is also to be noted that adding higher-order SPT corrections does not always improve the cross correlation. At $n>3$, the correlation with $N$-body simulation is rather worsen at $k\gtrsim0.2\,h$\,Mpc$^{-1}$. Although this is also seen in real space, and would be ascribed to the UV-sensitive features of higher-order SPT expansion, a more prominent feature seen in the redshift-space results suggests that 
the perturbative description of the redshift-space density field in Eq.~(\ref{eq:deltas_expansion}) further worsens the convergence of PT expansion. Physically, in redshift space, the velocity fields around and inside virialized objects are known to give a significant impact on the density fields even at large scales, referred to as the Fingers-of-God effect \cite{1983ApJ...267..465D,1972MNRAS.156P...1J}. This is partly deduced from the exact expressions given in Eq.~(\ref{eq:delta_in_s-space}) or (\ref{eq:deltas_expansion}), where the terms involving the line-of-sight velocity field makes the density field non-perturbative. Since the SPT treatment naively Taylor expands all the contributions, it would be difficult for calculations at finite order to capture the Fingers-of-God effect, and any improvement on the PT prediction would need a non-perturbative treatment or phenomenological description (e.g., Refs.~\cite{Matsubara2008a,Taruya:2010mx,Okumura:2011pb,Vlah:2012ni}). In the next section, we shall examine one such approach, and discuss its usability by looking at the morphological and statistical properties of redshift-space density fields. 

\begin{figure*}[tb]
\begin{center}
 \includegraphics[width=20cm,angle=0]{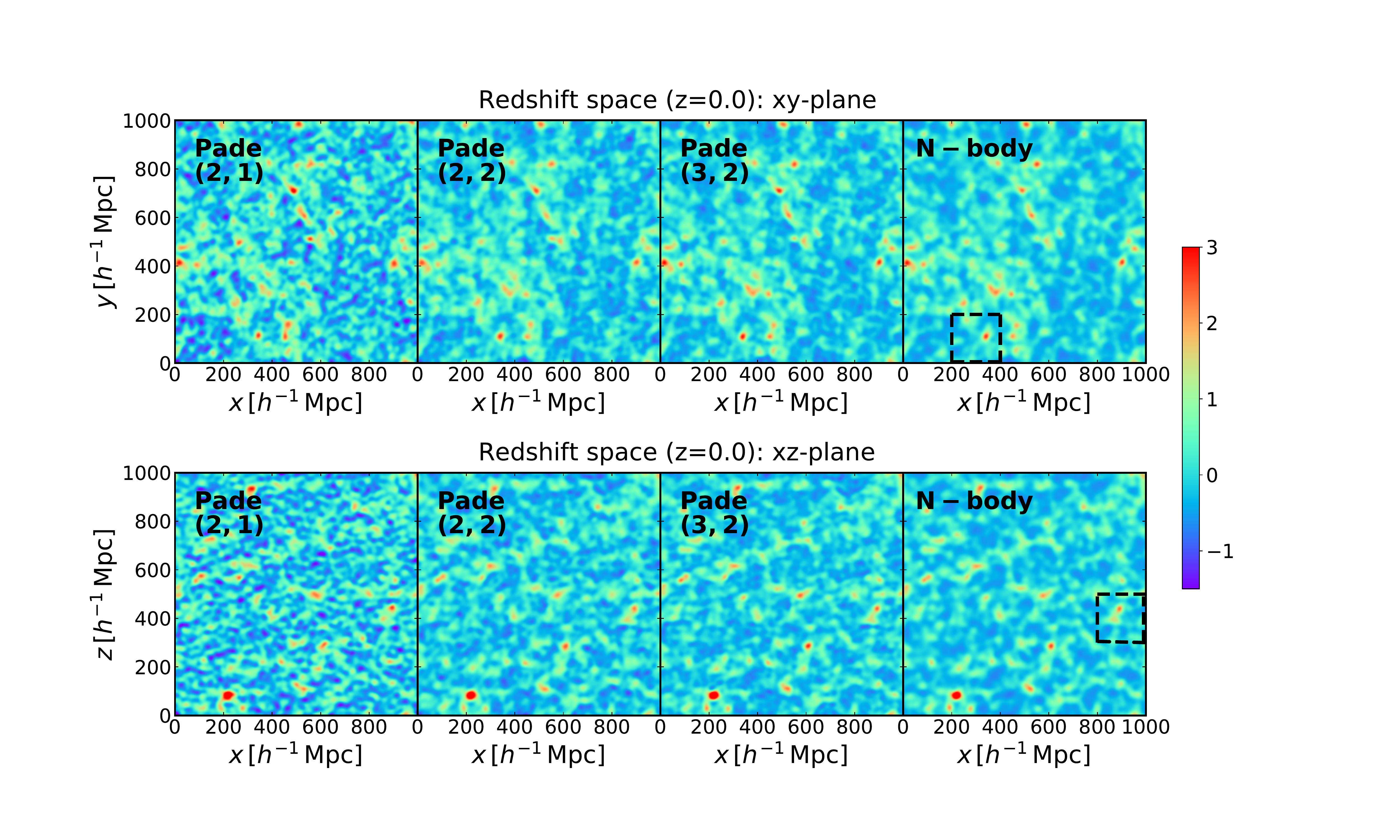}
\end{center}

\vspace*{-1.2cm}

\caption{2D density field at $z=0$ smoothed with a Gaussian filter of $R=10\,h^{-1}$Mpc. The redshift-space density fields obtained from Pad\'e approximations are plotted, together with the $N$-body results. Similar to Figs.~\ref{fig:Slice_real_red_xy} and \ref{fig:Slice_real_red_xz}, a slice of $xy$- (upper) and $xz$- (lower) planes is taken, and the density fields averaged over $10\,h^{-1}$Mpc depth are shown. 
\label{fig:Slice_red_pade}
}
\vspace*{-1.4cm}
\begin{center}
 \includegraphics[width=20cm,angle=0]{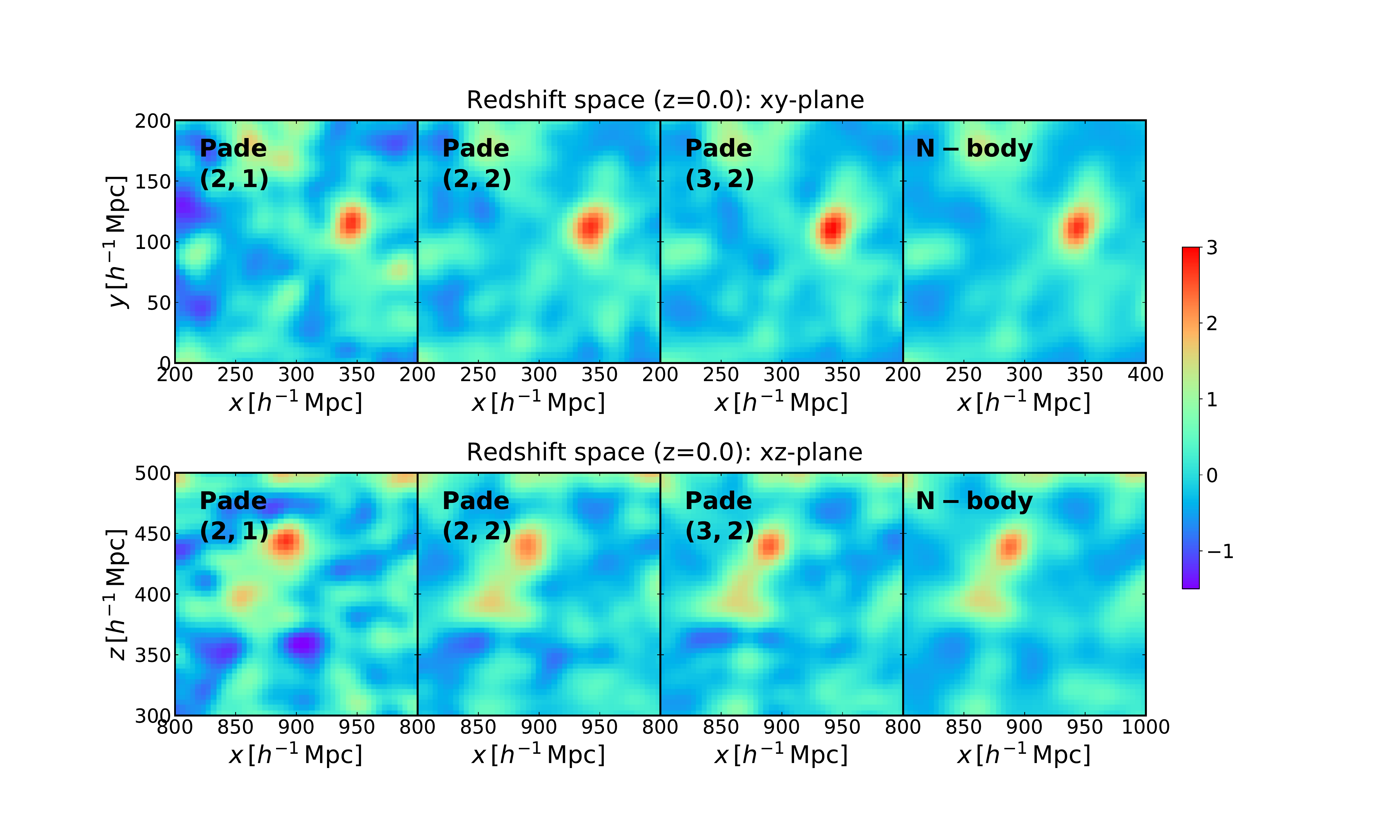}
\end{center}

\vspace*{-1.2cm}

\caption{Same as Fig.~\ref{fig:Slice_red_pade}, but enlarged plot of the 2D density field over $200 \times 200\, h^{-1}$\,Mpc size is shown for the region enclosed by the dashed line in the right panels of Fig.~\ref{fig:Slice_red_pade}. 
\label{fig:Slice_zoom_red_pade}
}
\end{figure*}

\section{Pad\'e approximation}
\label{sec:pade_approx}

\begin{figure*}[tb]
\begin{center}
 \includegraphics[width=16cm,angle=0]{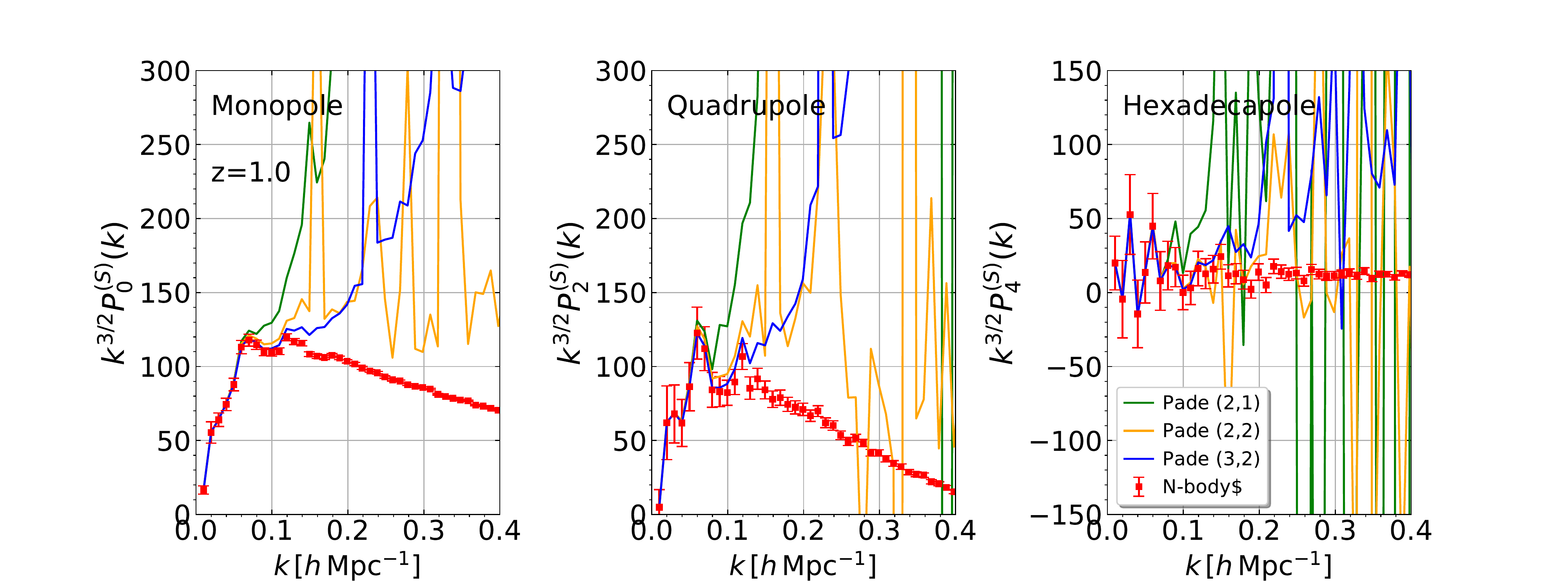}
\end{center}

\vspace*{-0.5cm}

\caption{Redshift-space power spectrum at $z=1$, obtained from the Pad\'e approximations of \gridspt\, calculations. 
\label{fig:pk_red_pade_zred1_single_run}
}
\vspace*{-0.5cm}
\begin{center}
 \includegraphics[width=8.2cm,angle=0]{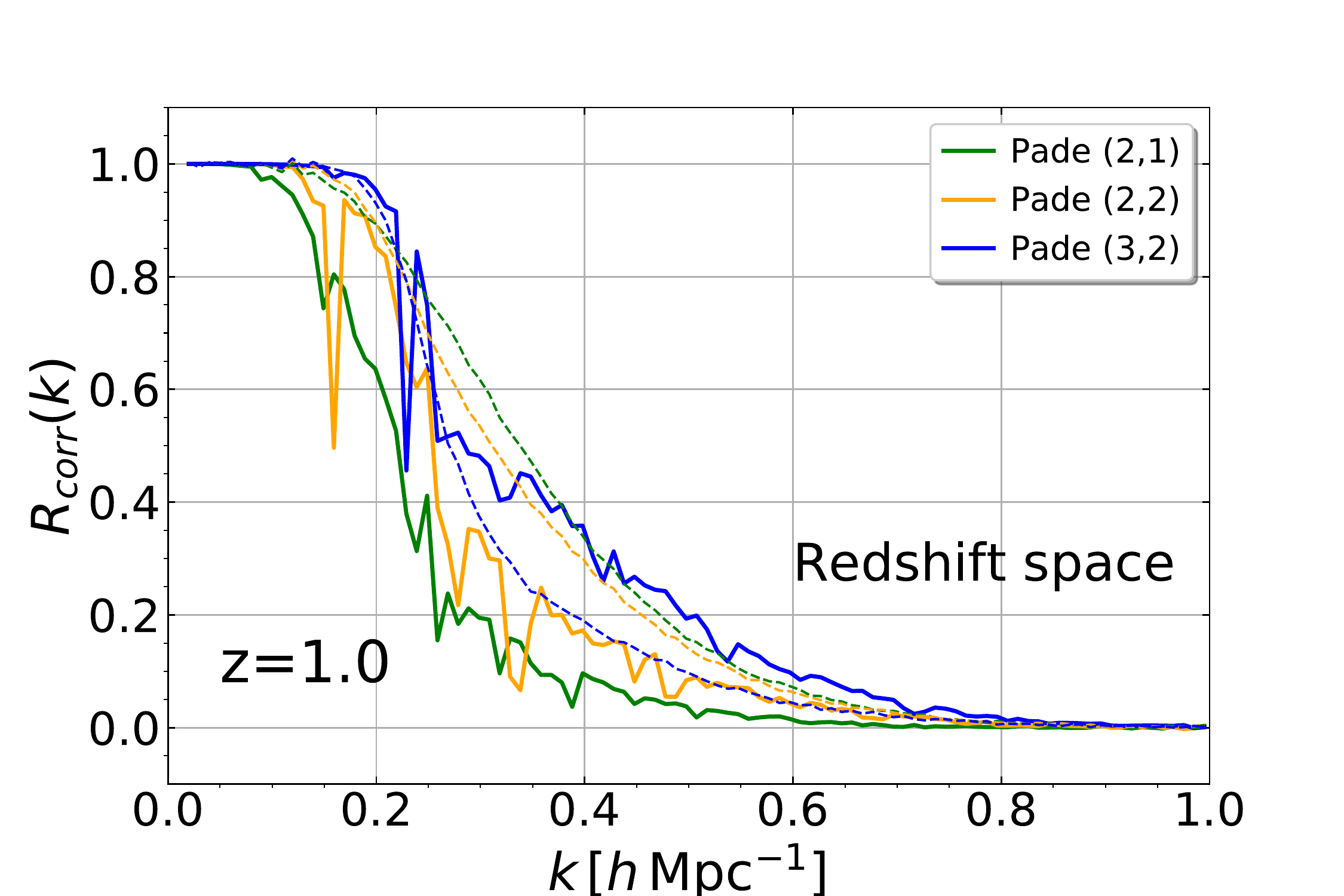}
 \includegraphics[width=8.2cm,angle=0]{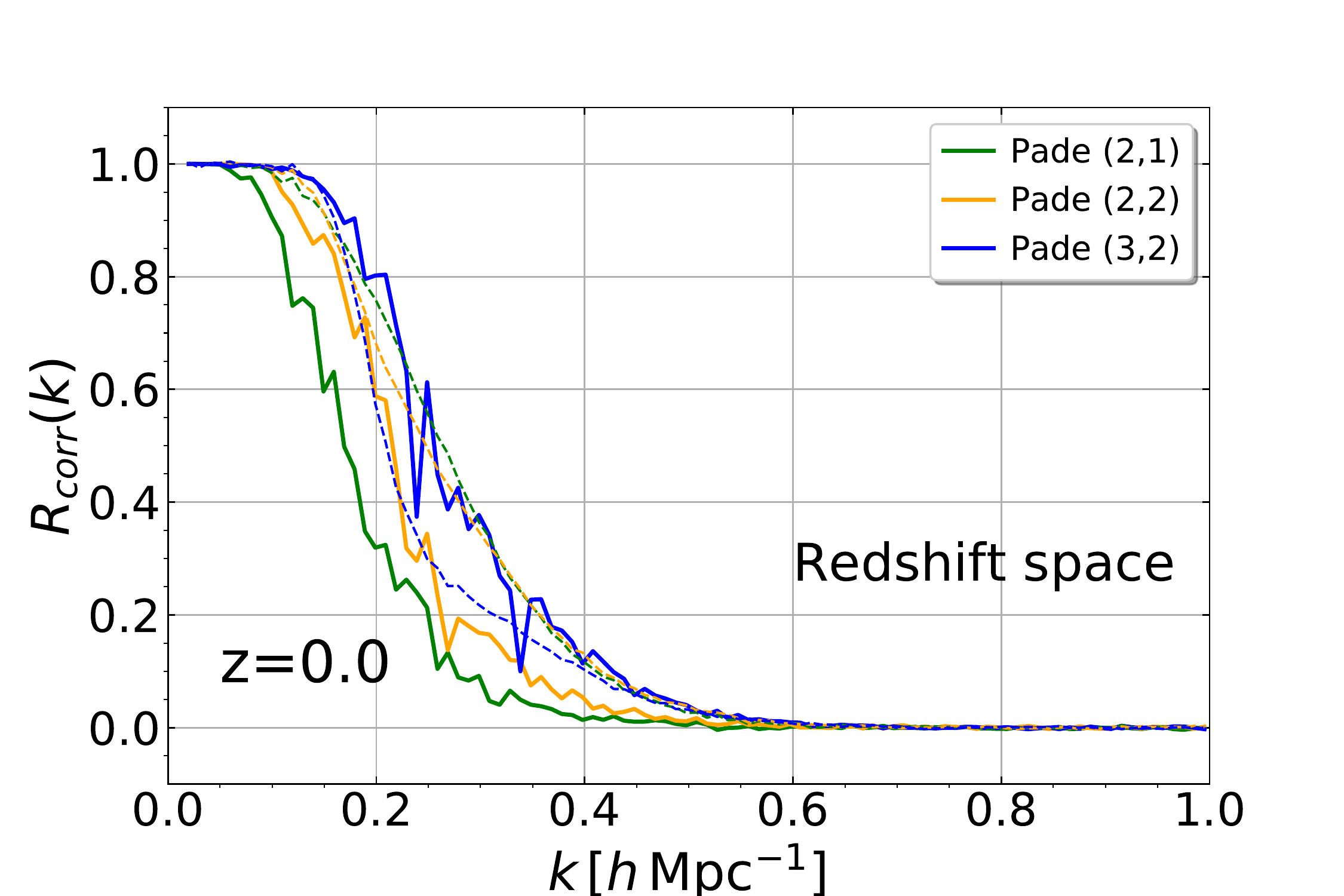}
\end{center}

\vspace*{-0.5cm}

\caption{Cross-correlation coefficient for density fields obtained from the 
Pad\'e approximations of the \gridspt\, calculations and $N$-body simulations (solid lines). The results in redshift space are presented at at $z=1$ (left) and $0$ (right).
The dashed lines show the \gridspt\, results at $3$, $4$ and $5$-th order, which are the same as the solid lines in Fig.~\ref{fig:R_corr_real_red}.
\label{fig:R_corr_pade_red}
}
\end{figure*}

In this section, as one of the non-perturbative resummation methods, we consider the Pad\'e approximation, and compute the {\it re-summed} density field using the \gridspt\, results up to the fifth order. The Pad\'e approximation re-organizes the original power-series expansion by considering its rational form. Pad\'e approximations are known to be superior to Taylor series when functions contain poles. There have been several works on the application of Pad\'e approximations in the context of the perturbation theory of large-scale structure \cite{Yoshisato_Matsubara_Morikawa1998,Matsubara_Yoshisato_Moriakwa1998, Blas:2013aba} (see also Ref.~\cite{Tatekawa2007} for the application of Shanks transformation). Here, we particularly focus on the redshift-space density field, and apply the Pad\'e approximations specifically to the \gridspt\, calculations.

Provided the PT expansion up to the $(M+N)$-th order, Pad\'e approximations provide a way to construct a rational expansion form involving the series expansion up to the $M$-th and $N$-th orders in the numerator and denominator, respectively, which we denote by Pad\'e $(M,N)$:
\begin{align}
 \delta_{\rm SPT}^{\rm(S)}=\sum_{n=1}^{M+N}c_n\quad\longrightarrow\quad
 \delta_{\rm Pade}^{\rm(S)}(\bfk)=\frac{\sum_{m=1}^M a_m\,}{1+\sum_{n=1}^N b_n},
\label{eq:pade_approx}
\end{align}
where the coefficient $c_n$ is given by $c_n=\delta_n^{\rm (S)}$, with the quantity $\delta_n^{\rm (S)}$ being the Fourier-space density field computed from \gridspt\, based on Eqs.~(\ref{eq:deltas_1})-(\ref{eq:deltas_5}). Given the positive integers $M$ and $N$, the coefficients $a_n$ and $b_n$ are expressed in terms of $\{c_n\}$. In general, $M=N$ is the best choice (e.g., Ref.~\cite{Hinch1991}). Here, we consider the Pad\'e $(2,1)$, $(2,2)$ and $(3,2)$, which are respectively computed with the SPT density fields up to third, forth and fifth order. In Appendix \ref{sec:Pade_coefficients}, we summarize the explicit form of the coefficients $a_n$ and $b_n$ for each case.

In Fig.~\ref{fig:Slice_red_pade}, the projected density fields in redshift space smoothed with the Gaussian filter of the radius $R=10\,h^{-1}$\,Mpc, as similarly shown in Figs.~\ref{fig:Slice_real_red_xy} and \ref{fig:Slice_real_red_xz}, are plotted.  Also, in Fig.~\ref{fig:Slice_zoom_red_pade}, we have enlarged the plot of the redshift-space density fields, taken from the regions enclosed by the dashed lines in Fig.~\ref{fig:Slice_red_pade}. These should be compared with the \gridspt\, results
for $n=3$, $4$, and $5$ in Figs.~\ref{fig:Slice_real_red_xy}-\ref{fig:Slice_zoom_real_red_xz}. We then find that the fake wobbly structures seen in the $xz$ plane, which exhibit successive low- and high-density regions along the line of sight, fade in the re-summed results with Pad\'e approximations. As a result, the morphology and structure of density fields from the Pad\'e $(2,2)$ and $(3,2)$ get much closer to the $N$-body results, visually regarded as an improvement. 

The improvement, however, does not hold true for the statistical measures. Fig.~\ref{fig:pk_red_pade_zred1_single_run} shows the redshift-space power spectra at $z=1$ measured from the density fields constructed with the Pad\'e approximation for a single realization data. The resulting monopole and quadrupole spectra exhibit a rather large enhancement on small scales. This is presumably due to the UV-sensitive behaviors inherent in the SPT calculation. Unlike in the naive PT treatment that evaluates the power spectrum perturbatively from several loop corrections [see Eqs.~(\ref{eq:pkSPT})-(\ref{eq:pk_2loop})], no cancellation of the higher-order corrections is expected in the Pad\'e approximation. Accordingly, the measured power spectra significantly deviate from those obtained from the $N$-body simulations. 

In Fig.~\ref{fig:R_corr_pade_red}, the cross-correlation coefficient, $R_{\rm corr}$, is computed for the Pad\'e approximation, and the results at $z=1$ (left) and $0$ (right) are compared with those obtained in Sec.~\ref{subsec:cross_correlation} especially for $n=3$, $4$ and $5$, depicted as dotted lines. Note again that these are obtained from the same single realization data as used in Figs.~\ref{fig:Slice_red_pade}-\ref{fig:pk_red_pade_zred1_single_run}. We then find that the correlation coefficients from the Pad\'e approximations are prone to be more suppressed than those of the naive SPT calculations. A closer look at the results of Pad\'e $(3,2)$ reveals that the suppression at intermediate scales around $k\sim0.3-0.4\,h$\,Mpc$^{-1}$ becomes milder compared to the SPT results at $n=5$, but the  improvement of the cross-correlation coefficient is moderate. In all cases, the results of the Pad\'e approximations show a rather noisy behavior, accompanying spikes and dips, which are also seen in the power spectra at small scales, $k\gtrsim0.2\,h$\,Mpc$^{-1}$. Note that applying the Pad\'e approximation to the real-space density fields, we have also seen similar noisy behaviors. They are possibly caused by artificial singularities coming from the rational function at Eq.~(\ref{eq:pade_approx})\footnote{In evaluating Eq.~(\ref{eq:pade_approx}) numerically, we added a small positive number to the denominator to prevent the divergence.}. Although those singular behaviors can be apparently eliminated by applying the smoothing function and hence we do not see such a spiky structure in Figs.~\ref{fig:Slice_red_pade} and \ref{fig:Slice_zoom_red_pade}, these could severely affect the statistical quantities measured from the un-filtered density fields. Since the singular points eventually appear at the regions where the higher-order density fields receive a large correction, the application of the Pad\'e approximation may not be generally suited to improve the convergence of SPT expansion at field level. We conclude that re-writing simply the SPT expansion in a rational form does not improve the predictions in redshift space. Rather, mitigating the UV sensitivity in the SPT calculations would be essential, and implementing a regularization scheme including the effective-field-theory treatment would be thus important.

\section{Conclusion and discussions}
\label{sec:conclusion}

In this paper, we have extended our previous works on a grid-based SPT algorithm, called \gridspt, to implement the redshift-space distortions (RSD) on grids. The key expression is given in Eq.~(\ref{eq:deltas_expansion}), in which the redshift-space density field is expressed in terms of the real-space quantities (density and velocity fields) given at a redshift-space position.
This expression, thus, allows us to apply the real-space results of \gridspt\, calculations directly for a perturbative evaluation of the redshift-space density field on grids. With this new implementation, we have demonstrated the \gridspt\, calculations in redshift space up to the fifth order, and investigated the morphological and statistical properties of the SPT density fields, which we have also compared with cosmological $N$-body simulations.

We found that the redshift-space power spectrum and bispectrum obtained from the \gridspt\, calculations agree well with analytical SPT results up to the two-loop and one-loop order, respectively. Note that the two-loop SPT power spectra are numerically evaluated and presented for the first time in this paper. In redshift space, adding the higher-loop corrections is shown to give a significant change in the amplitudes of both the power spectrum and bispectrum. In particular, we found that the power spectrum sensitively depends on the small-scale cutoff. This implies that the convergence of SPT expansion [Eq.~(\ref{eq:deltas_expansion})] in redshift space is rather worse than that in real space due to
the higher-derivative operators $\tilde{\nabla}_z^n$. In fact, comparing the generated density fields from the \gridspt\, calculations with those obtained from the $N$-body simulations, we see rather prominently that the SPT is prone to produce fake structures in redshift space, and even at large scales, un-physical wobbly structures appear manifest along the line of sight. Accordingly, the statistical correlation of the \gridspt\, density field with $N$-body results is rather poor, and as we go to higher $k$, the resultant cross correlation becomes suppressed more rapidly than that in real space.

To remedy the poor convergence of the SPT expansion in redshift space, we have considered the Pad\'e approximation, and applied it to the Fourier-space density fields. Rewriting the SPT expansion with a rational expansion form, the morphological properties of the smoothed density fields get visually better, and the wobbly structures mostly disappear. However, the resultant power spectra exhibit a large enhancement at small scales, accompanying spikes and dips, which are also seen in the cross-correlation coefficients. These are presumably originated from the UV-sensitive behaviors inherent in the SPT calculation, and higher-order density fields get a rather large correction at small scales. With the re-organized expansion in a rational form, no cancellation occurs unlike in the SPT calculation and the singularities eventually happen. We thus conclude that simply re-organizing the SPT expansion does not improve the predictions, and mitigating the UV-sensitivity would be rather crucial.

Finally, as we mentioned in Sec.~\ref{sec:introduction}, the implementation of the RSD effect in \gridspt\, is not our final goal, but rather an important and necessary step toward a practical application of the method to observations. In Ref.~\cite{Taruya_Nishimichi_Jeong2021}, we have demonstrated that the grid-based algorithm for SPT calculations allows us to easily incorporate the observational systematics such as the survey window function and masks. 
In addition, it is rather straightforward to implement a general expansion scheme to deal with the galaxy bias that has been actually exploited on the basis of SPT (e.g., Refs.~\cite{McDonald_Roy2009,Chan_Scoccimarro_Sheth2012,Saito_etal2014,Mirbabayi_Schmidt_Zaldarriaga2015,Senatore2015,Fujita_Vlah2020}, see Ref.~\cite{Desjacques_Jeong_Schmidt2018} for review). With an effective-field-theory treatment at the field level, we anticipate that the UV-sensitive behaviors can be mitigated, and an efficient and stable PT prediction would become possible in redshift space. Consistently incorporating all observational effects to the theoretical calculations with \gridspt, the grid-based method may provide an efficient framework to maximize the cosmological information obtained from the galaxy survey data (e.g., Refs.~\cite{Kitaura_Enslin2008,Jasche_Wandelt2013,Schmidt_etal2019}). An investigation along this direction is interesting and important toward a practical application, and we will continue to work on these.

\acknowledgments
AT thank Satoshi Tanaka for his comments and discussion on the higher-order differential scheme, and Takahiko Matsubara for his suggestions on the Pad\'e approximation. This work was supported in part by MEXT/JSPS KAKENHI Grant Number JP17H06359 (AT), JP17K14273, JP19H00677 (TN), JP20H05861 and JP21H01081 (AT and TN). We also acknowledge financial support from Japan Science and Technology Agency (JST) AIP Acceleration Research Grant Number JP20317829 (AT and TN).
DJ acknowledges support from NASA 80NSSC18K1103. Numerical computation was partly carried out at the Yukawa Institute Computer Facility.

\appendix
\section{On the aliasing correction in \gridspt\, calculations}
\label{sec:aliasing_correction}

In this Appendix, we discuss the de-aliasing treatment to mitigate the spurious high-frequency modes arising from the nonlinear calculations of fields on grids. After describing de-aliasing methods in Appendix \ref{subsec:de-aliasing_methods}, we compare the results of \gridspt\, calculations between several de-aliasing treatments in Appendix \ref{subsec:comparison_de-alising}.

\begin{figure*}[tb]
\begin{center}
\includegraphics[width=18cm,angle=0]{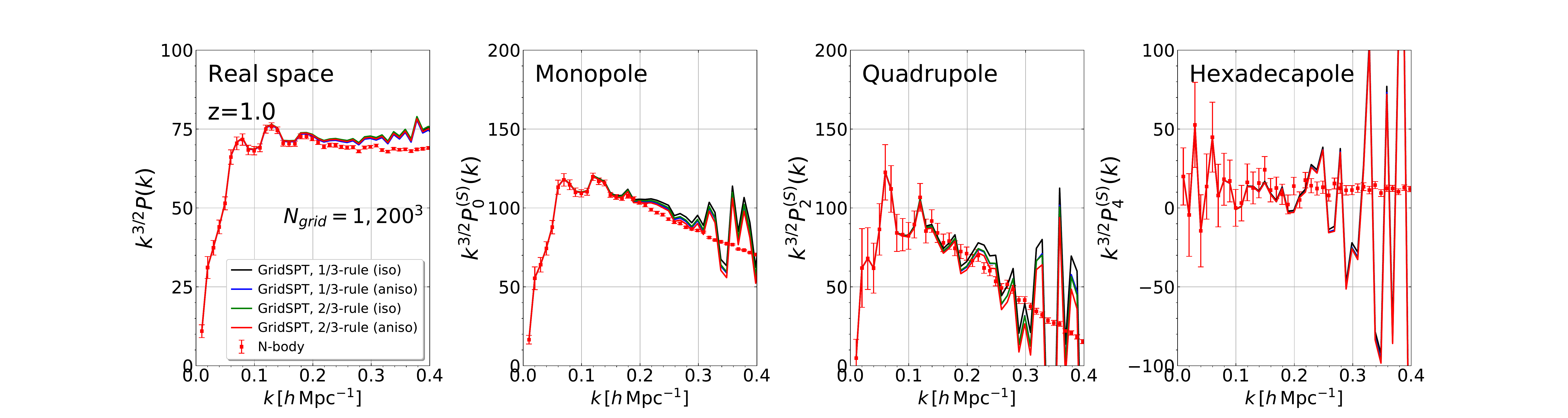}

\vspace*{-0.5cm}

\includegraphics[width=18cm,angle=0]{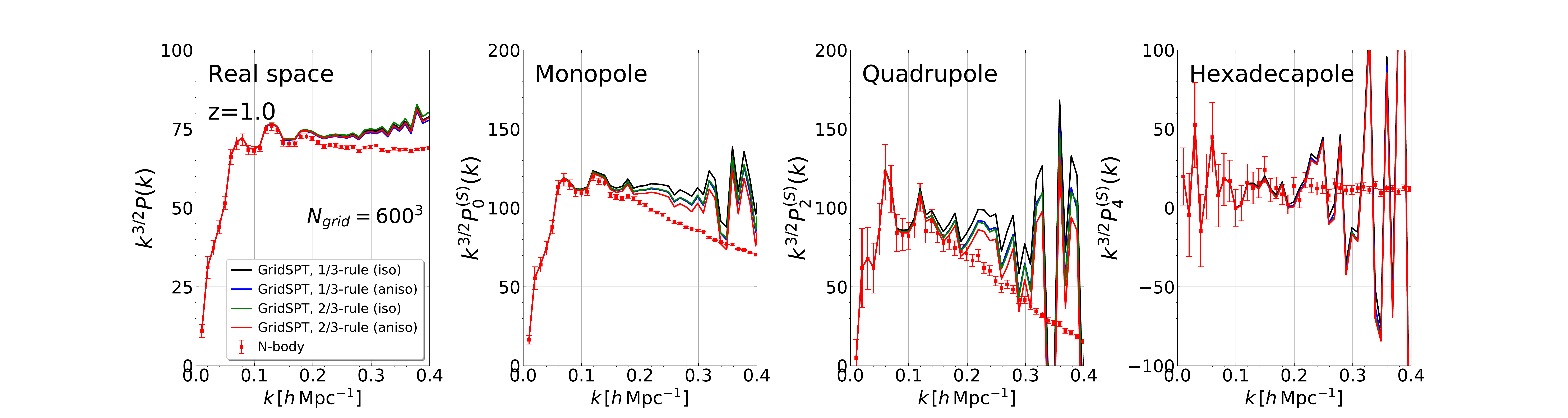}
\end{center}

\vspace*{-0.5cm}
\caption{Impacts of the de-aliasing treatments on the real- and redshift-space power spectra at two-loop order in \gridspt\, calculations, depicted as solid lines. The results at $z=1$ are shown. The upper and lower panels respectively represent the \gridspt\, results adopting the number of grids $N_{\rm grid}=1,200^3$ and $600^3$. For reference, $N$-body results are also shown in each panel, depicted as red crosses, with errorbars indicating the sampling noise estimated from the number of Fourier modes in each bin.    
\label{fig:pk_gridSPT_dealiasing}
}
\vspace*{-0.4cm}
\begin{center}
\includegraphics[width=17cm,angle=0]{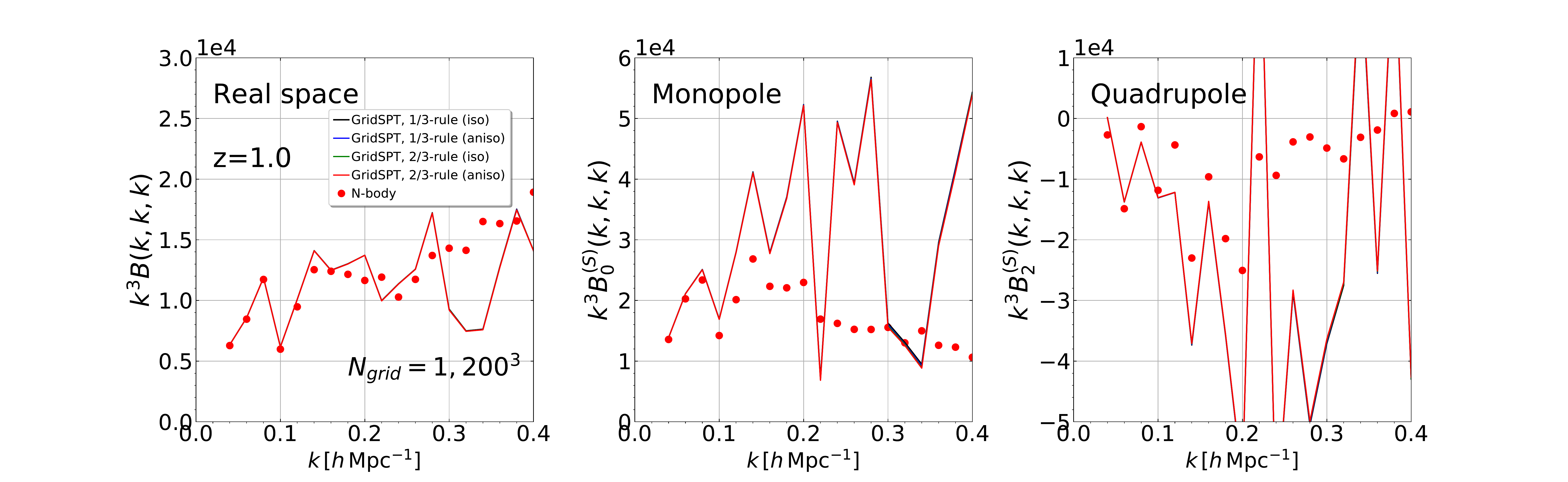}

\vspace*{-0.2cm}

\includegraphics[width=17cm,angle=0]{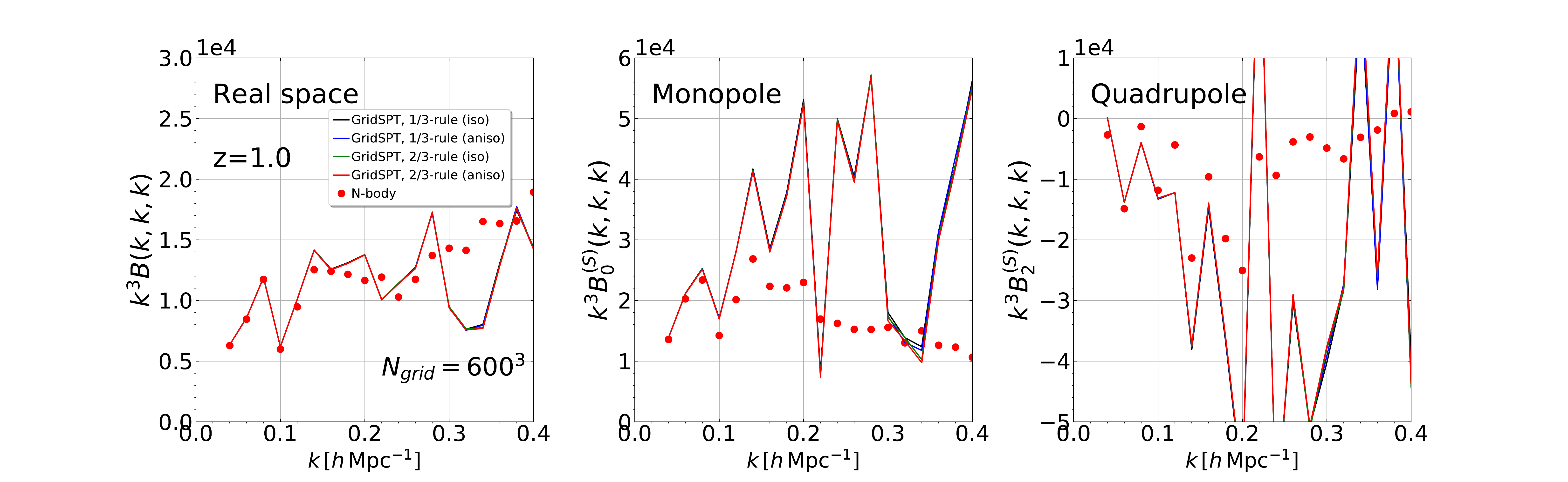}
\end{center}

\vspace*{-0.5cm}
\caption{Same as Fig.~\ref{fig:pk_gridSPT_dealiasing}, but shown for the \gridspt\, results of the one-loop bispectrum in equilateral configuration. The upper and lower panels plot the \gridspt\, results adopting the number of grids $N_{\rm grid}=1,200^3$ and $600^3$, together with the measured results from $N$-body simulations (red crosses). 
\label{fig:bk_gridSPT_dealiasing}
}
\end{figure*}

\subsection{Aliasing corrections}
\label{subsec:de-aliasing_methods}

Let us first recall how the aliasing effect affects the \gridspt\,calculations. For simplicity, we consider the one-dimensional grid space with a side length of $L$, and compute the product of the two fields $A_1(x)$ and $A_2(x)$, where the position $x$ is defined over the range $0\leq x\leq L$. For a grid number $N$, the discrete Fourier transform of the fields $A_k$ is described by  
\begin{align}
 A_k(x_j)&=\sum_{n=-N/2}^{N/2-1}A_k(k_n)\,e^{i\,k_n\,x_j},\quad (k=1,2),
\label{eq:Fourier_field_A_j}
\end{align}
where the discrete Fourier mode $k_n$ is given by $k_n=2n\pi/L$, and the position in grid space, $x_j$, is discretised as $x_j=(j/N)L$ for $j=0,\cdots,N-1$. In \gridspt,\, the product of two fields, $A_1$ and $A_2$, is computed in configuration space, and then the derivative operations are applied in Fourier space. Using Eq.~(\ref{eq:Fourier_field_A_j}), the Fourier coefficient of the product $A_1(x)A_2(x)$ for the mode $k_n$ becomes
\begin{widetext}
\begin{align}
& \frac{1}{N}\sum_{j=-N/2}^{N/2-1}\,A_{1}(x_j)A_{2}(x_j)\,e^{-i\,k_n x_j}
=\sum_{\ell,m=-N/2}^{N/2-1}\deltaK_{\ell+m,n}\,A_1(k_\ell)A_2(k_m)+
\sum_{\ell,m=-N/2}^{N/2-1}\deltaK_{\ell+m,n\pm 
N}A_1(k_\ell)A_2(k_m),
\label{eq:Fourier_product}
\end{align}
\end{widetext}
where we used the fact that
\begin{align}
 \frac{1}{N}\sum_{j=0}^{N-1}e^{i\,p\,x_j}=\left\{
\begin{array}{ll}
 1& {\displaystyle (p=\frac{2\pi}{L}N\,m,\quad m=0,\pm1,\pm2,\cdots)}
\\
\\
 0& \mbox{otherwise}
\end{array}
\right. .
\end{align}
In Eq.~(\ref{eq:Fourier_product}), the first term  
on the right hand side represents the contribution that we want to calculate. On the other hand, the second term is the aliasing contribution originating from the discreteness of the grid space. To eliminate this spurious contributions, a simple way is to discard the high-frequency modes that can produce the aliasing effect. To be precise, if we set the fields $A_1(k_n)$ and $A_2(k_n)$ to zero for $|n|>N/3$, the non-vanishing modes in Eq.~(\ref{eq:Fourier_product}) are restricted to $|\ell+m-n|<N$, and thus the aliasing contribution does not appear. Since the mode $k_n$ at $|n|=N/3$ corresponds to $2/3$ times the Nyquist frequency, this zero-padding method is called the $2/3$ rule \cite{Orszag1971a}. Note that the prescription given here can be generalized to the case for the higher-order products. That is, in order to avoid the aliasing effect for a product of the $M$ fields, $\Pi_{k=1}^M A_k(x_j)$, modes of the fields $A_k(k_n)$ 
for $|n|>N/(M+1)$ should be set to zero, corresponding to the modes larger than $2/(M+1)$ times the Nyquist frequency.

Generalizing further the discussion above to the three-dimensional grid space, Refs.~\cite{Taruya_Nishimichi_Jeong2018,Taruya_Nishimichi_Jeong2021} adopted the isotropic low-pass filter (called sharp-$k$) 
with the critical wavenumber $k_{\rm crit}=(2/3)k_{\rm Nyq}$, by which the Fourier modes in $|\bfk|>k_{\rm crit}$ are set to zero. The filter was applied at each step when we proceed to  
higher-order \gridspt\, calculations. Then, the generated PT fields up to the fifth order reproduce the desired properties known in the analytical calculations. However, the procedure used in previous works is not a unique choice. Instead of using isotropic filter, we may introduce the anisotropic filter, in which the zero-padding is applied to the modes having $|k_{x,y,z}|>k_{\rm crit}$. This also eliminates the spurious aliasing contributions. Furthermore, recalling that the $n$-th order PT fields 
are expressed as the $n$-th order product of the linear density fields, an alternative way of aliasing correction for the \gridspt\, calculation at $n$-th order is to adopt the $2/(n+1)$-rule only once. That is, the low-pass filter with $k_{\rm crit}=2/(n+1)\,k_{\rm Nyq}$ is applied only to the (initial) linear density field, and the subsequent higher-order PT calculations are performed up to $n$-th order, without taking any filter.

\subsection{Comparison of de-aliasing treatments}
\label{subsec:comparison_de-alising}

Let us quantitatively study the possible impact of the de-aliasing treatment on the \gridspt\, calculations, focusing on the statistical quantities obtained from the density fields up to the 5th order in real and redshift space. 

Based on the discussions in Appendix \ref{subsec:de-aliasing_methods}, de-aliasing prescriptions one can think of are summarized as follows:
\begin{description}
 \item[$2/3$-rule (iso)] an isotropic sharp-$k$ filter 
 with $k_{\rm crit}=(2/3)k_{\rm Nyq}$ is applied to the PT fields at every order of PT calculations.
 \item[$2/3$-rule (aniso)] an anisotropic sharp-$k$ filter 
 with $k_{\rm crit}=(2/3)k_{\rm Nyq}$, by which the modes having $|k_{x,y,z}|>k_{\rm crit}$ are set to zero, is applied to the PT fields at every order of PT calculations.
 \item[$1/3$-rule (iso)] an isotropic sharp-$k$ filter 
with $k_{\rm crit}=(1/3)k_{\rm Nyq}$ is applied only to the initial density fields before PT calculations.
 \item[$1/3$-rule (aniso)] an anisotropic sharp-$k$ filter 
 with $k_{\rm crit}=(1/3)k_{\rm Nyq}$, by which the modes having $|k_{x,y,z}|>k_{\rm crit}$ are set to zero, is applied only to the initial density fields before PT calculations.
\end{description}

In Figs.~\ref{fig:pk_gridSPT_dealiasing} and \ref{fig:bk_gridSPT_dealiasing}, using the above de-aliasing treatments, \gridspt\, results of the power spectra and bispectra are respectively shown at $z=1$, adopting respectively the number of grids $N_{\rm grid}=1,200^3$ and $600$ in the upper and lower panels. Here, the power spectra computed with \gridspt\, are at two-loop order, while the bispectra are at one-loop order. Both results are obtained from the same initial seed as used in the $N$-body simulation, whose results are also depicted as filled circles for reference. With a single realization data, the number of available Fourier modes is limited at large scales, and the measured results from the $N$-body simulation suffer from the effect of finite-mode sampling, which is known to be significant at low $k$ modes \cite{Takahashi:2008wn}. Hence, to make a fair comparison, we added corrections due to the finite-mode sampling to the \gridspt\,results. That is, the contributions having the odd powers of the (Gaussian) linear density field, $(P_{12},\, P_{23},\, P_{14})$ and $(B_{111},\, B_{113}, \,B_{122})$, are added to the power spectrum and bispectrum, respectively [see Eqs.~(\ref{eq:def_pkred_SPT}) and (\ref{eq:def_bkred_SPT}) for definitions of $P_{ab}$ and $B_{abc}$]. Although 
the odd-power contributions usually vanish in the limit of the infinite number of Fourier modes, these contributions do exist in the $N$-body realization. Indeed,  taking them into account in the \gridspt\, calculations makes the agreement with $N$-body results better especially at $k\lesssim0.1\,h$\,Mpc$^{-1}$. 

In Fig.~\ref{fig:pk_gridSPT_dealiasing}, apart from a bumpy scale-dependent feature at $k\gtrsim0.2\,h$\,Mpc$^{-1}$, a prominent difference arising from the de-aliasing treatments appears manifest if we adopt a smaller number of grids, $N_{\rm grid}=600$ (lower). Typically, the impact gets large for the redshift-space monopole and quadruple spectra, and adopting the anisotropic sharp-$k$ filter tends to suppress the power spectrum amplitude compared to the isotropic counterpart. The $1/3$ rule applied only to the initial condition also suppresses the power, compared to the $2/3$ rule at every PT order. These behaviors are originated from the change of the mode transfer due to different cutoff strategies and the cutoff scales imposed, leading to a visible change in the power spectrum amplitude. The effect would become more significant as decreasing redshifts. On the other hand, looking at the one-loop bispectrum shown in Fig.~\ref{fig:bk_gridSPT_dealiasing}, we hardly see a clear difference. 

Based on the discussion and the results in Figs.~\ref{fig:pk_gridSPT_dealiasing} and \ref{fig:bk_gridSPT_dealiasing}, in the main text, we adopt the $1/3$ rule for the de-aliasing treatment, since it seems less affecting the mode-coupling structure. Using a simple isotropic sharp-$k$ filter, the \gridspt\, calculation is performed mainly with $N_{\rm grid}=1,200^3$, and the results are presented in Sec.~\ref{sec:demonstration}.

\section{Analytical expressions for SPT power spectrum and bispectrum in redshift space}
\label{sec:analytical_SPT}

In this Appendix, we present the analytical expressions of the SPT power spectrum and bispectrum in redshift space. 

Let us first recall that in the SPT treatment, the redshift-space density field, $\delta^{\rm(S)}$, is expanded in powers of the linear density field $\delta_1$ [see Eq.~(\ref{eq:recursion_n=1})], and in Fourier space, we have
\begin{align}
\delta^{\rm(S)}(\bfk) &= \sum_{n=1}\int\frac{d^3\bfp_1 \cdots d^3\bfp_n}{(2\pi)^{3n}}\delta_{\rm D}(\bfk-\bfp_{1\cdots n}) 
\nonumber
\\
&\times Z_n(\bfp_1,\,\cdots,\,\bfp_n) \delta_1(\bfp_1)\cdots\delta_1(\bfp_n)
\label{eq:deltaS_kernelZ}
\end{align}
with the wavevector $\bfp_{1\cdots n}$ defined by $\bfp_{1\cdots n}\equiv \bfp_1+\cdots+\bfp_n$. 
Here, the kernels $Z_n$ characterize the mode coupling in redshift space between Fourier modes, and they are symmetric with respect to the exchange of their arguments. These kernels are analytically constructed, and are expressed in terms of the real-space PT kernels, $F_n$ and $G_n$, for the $n$-th order density and velocity-divergence fields, given by   
\begin{align}
    \delta_n(\bfk) &= \int \frac{d^3\bfp\cdots d^3\bfp_1}{(2\pi)^{3n}}\delta_{\rm D}(\bfk-\bfp_{1\cdots n})
\nonumber    
\\
&\quad \times F_n(\bfp_1,\cdots,\,\bfp_n) \delta_1(\bfp_1)\cdots\delta_1(\bfp_n),
\label{eq:delta_SPT}
\\
    \theta_n(\bfk) &= \sum_{n=1} \int \frac{d^3\bfp\cdots d^3\bfp_1}{(2\pi)^{3n}}\delta_{\rm D}(\bfk-\bfp_{1\cdots n}) 
\nonumber    
\\
&\quad \times 
    G_n(\bfp_1,\cdots,\,\bfp_n) \delta_1(\bfp_1)\cdots\delta_1(\bfp_n),
\label{eq:theta_SPT}    
\end{align}
with $F_1=1=G_1$. The explicit forms of these kernels are constructed through the recurrence relation (see e.g., Refs.~\cite{Goroff:1986ep,Bernardeau:2001qr,Jain_Betschinger1994}), which corresponds to the Fourier transform of the formula given by Eq.~(\ref{eq:recursion_formula}). Using the expressions in Eqs.~(\ref{eq:delta_SPT}) and (\ref{eq:theta_SPT}), the expansion form of $\delta^{\rm(S)}$, given in Eq.~(\ref{eq:deltas_Fourier}), is re-organized with respect to the powers of $\delta_1$, leading to the form given by Eq.~(\ref{eq:deltaS_kernelZ}), from which we can read off the analytical expressions for the kernel $Z_n$ recursively. At the last step, the kernel $Z_n$ has to be symmetrized by summing up the expressions with all possible permutations of their arguments. The explicit forms of $Z_n$ can be found in the literature, e.g., in Refs.\cite{Scoccimarro1999,Hashimoto_Rasera_Taruya2017} up to third and fourth order, respectively. Note that the kernel $Z_n$ includes the terms having an explicit dependence on the linear growth factor $f$, and by setting $f$ to zero, it is reduced to the real-space PT kernel $F_n$.

Provided the kernel $Z_n$, the analytical expressions for the redshift-space power spectrum and bispectrum are derived
 based on the definitions in Sec.~\ref{subsec:powerspec_bispec}. The SPT power spectrum at two-loop order, given in Eq.~(\ref{eq:pkSPT}), consists of the six contributions, summarized in Eqs.~(\ref{eq:pk_linear})--(\ref{eq:pk_2loop}). 
 With a help of the Wick theorem, their analytical expressions are obtained from Eq.~(\ref{eq:def_pkred_SPT}):
\begin{align}
 P^{\rm(S)}_{11}(\bfk) & =\bigl\{Z_1(\bfk)\bigr\}^2\,P_{\rm L}(k),
\label{eq:pkred_11}
\\
P^{\rm(S)}_{13}(\bfk) & =3\,Z_1(\bfk)\,P_{\rm L}\, \int\frac{d^3\bfp}{(2\pi)^3}\,Z_3(\bfp,\,-\bfp,\,\bfk)\,P_{\rm L}(p),
\label{eq:pkred_13}
\\
P^{\rm(S)}_{22}(\bfk) & =2\,\int\frac{d^3\bfp}{(2\pi)^3}\,\bigl\{Z_2(\bfp,\,\bfk-\bfp)\bigr\}^2\,P_{\rm L}(p)P_{11}(|\bfk-\bfp|),
\label{eq:pkred_22}
\end{align}
\begin{align}
P^{\rm(S)}_{15}(\bfk) & = 15\,Z_1(\bfk)P_{\rm L}(k)\,
\nonumber
\\
&\times \int\frac{d^3\bfp d^3\bfq}{(2\pi)^6} 
\bigl\{Z_5(\bfp,\,\bfq,\,-\bfp,\,-\bfq,\,\bfk\bigr\}P_{\rm L}(p)P_{\rm L}(q),
\label{eq:pkred_15}
\\
P^{\rm(S)}_{24}(\bfk) & = 12\,\int\frac{d^3\bfp d^3\bfq}{(2\pi)^6}\,Z_2(\bfp,\,\bfk-\bfp)\,
\nonumber
\\
&\times Z_4(\bfp,\,\bfq,\,-\bfq,\,\bfk-\bfp)
P_{\rm L}(p)P_{\rm L}(q)P_{\rm L}(|\bfk-\bfp|)
\label{eq:pkred_24}
\\
P^{\rm(S)}_{33}(\bfk) & = 9\,Z_1(\bfk)\,P_{\rm L}(k) \Bigl\{\int\frac{d^3\bfp}{(2\pi)^3}\,Z_3(\bfp,\,-\bfp,\,\bfk)\,P_{\rm L}(p)\Bigr\}^2
\nonumber
\\
& + 6\,\int\frac{d^3\bfp d^3\bfq}{(2\pi)^6} \bigl\{Z_3(\bfp,\,\bfq,\,\bfk-\bfp-\bfq)\bigr\}^2
\nonumber
\\
&\qquad\qquad \times P_{\rm L}(p)P_{\rm L}(q)P_{\rm L}(|\bfk-\bfp-\bfq|),
\label{eq:pkred_33}
\end{align}
where the function $P_{\rm L}$ is the linear power spectrum in real space, i.e., $P_{\rm L}=P_{11}$. 
On the other hand, the SPT bispectrum at one-loop order has four terms as given in  Eqs.~(\ref{eq:bk_tree}) and (\ref{eq:bk_1loop}). From Eq.~(\ref{eq:def_bkred_SPT}), these are analytically expressed as follows: 
\begin{align}
    B_{112}^{\rm(S)}(\bfk_1,\,\bfk_2,\,\bfk_3) &= 2\,Z_2(\bfk_1,\,\bfk_2)Z_1(\bfk_1)Z_1(\bfk_2)
\nonumber
\\
&\times  P_{\rm L}(k_1)P_{\rm L}(k_2),
\label{eq:bkred_112}    
\\
   B_{123}^{\rm(S)}(\bfk_1,\,\bfk_2,\,\bfk_3) &=6 \,Z_1(\bfk_1)P_{\rm L}(k_1)
  \int \frac{d^3\bfp}{(2\pi)^3}Z_2(\bfp,\,\bfk_2-\bfp)
\nonumber
\\
&\times  Z_3(-\bfk_1,\,-\bfp,\,-\bfk_2+\bfp)
\nonumber
\\
& \times  P_{\rm L}(p)P_{\rm L}(|\bfk_2-\bfp|)
\nonumber
\\
& +  6\,Z_1(\bfk_1)Z_2(\bfk_1,\,\bfk_2)P_{\rm L}(k_1)P_{\rm L}(k_2) 
\nonumber
\\
&\times \int\frac{d^3\bfp}{(2\pi)^3} Z_3(\bfk_2,\,\bfp,\,-\bfp)P_{\rm L}(p),
\label{eq:bkred_123}    
\\
   B_{114}^{\rm(S)}(\bfk_1,\,\bfk_2,\,\bfk_3) &=   12\,Z_1(\bfk_1)Z_1(\bfk_2)P_{\rm L}(k_1)P_{\rm L}(k_2) 
\nonumber
\\
 & \times  \int \frac{d^3\bfp}{(2\pi)^3}Z_4(-\bfk_1,\,-\bfk_2,\,\bfp,\,-\bfp)P_{\rm L}(p),
\label{eq:bkred_114}
\\
   B_{222}^{\rm(S)}(\bfk_1,\,\bfk_2,\,\bfk_3) &=  8 \int\frac{d^3\bfp}{(2\pi)^3}\,Z_2(\bfk_1-\bfp,\,\bfp)
\nonumber
\\
 & \times   Z_2(\bfk_2+\bfp,\,-\bfp) Z_2(-\bfk_2-\bfp,\,-\bfk_1+\bfp)
\nonumber
\\
 & \times  P_{11}(|\bfk_1-\bfp|)P_{\rm L}(p)P_{\rm L}(|\bfk_2+\bfp).
\label{eq:bkred_222}
\end{align}

Finally, we note that the analytical SPT results presented in this paper are the multipole moments of the power spectrum and bispectrum. Thus, on top of the loop integrals as shown above, one has to also evaluate the integrals over the angles [see Eqs.~(\ref{eq:pk_multipoles}) and (\ref{eq:bk_multipoles})]. As a result, the six- and five-dimensional integrals have to be evaluated at the highest for the power spectrum and bispectrum, respectively\footnote{In the power spectrum case, one can use the rotational symmetry with respect to the line-of-sight direction to partly reduce the loop integrals. }. In order to deal with these multi-dimensional integrals, we adopt the Monte Carlo integration technique, and use specifically the quasi-random sampling in the CUBA library \cite{Hahn:2004fe} to directly compute them.

\begin{figure*}[tb]
\begin{center}
\includegraphics[width=18cm,angle=0]{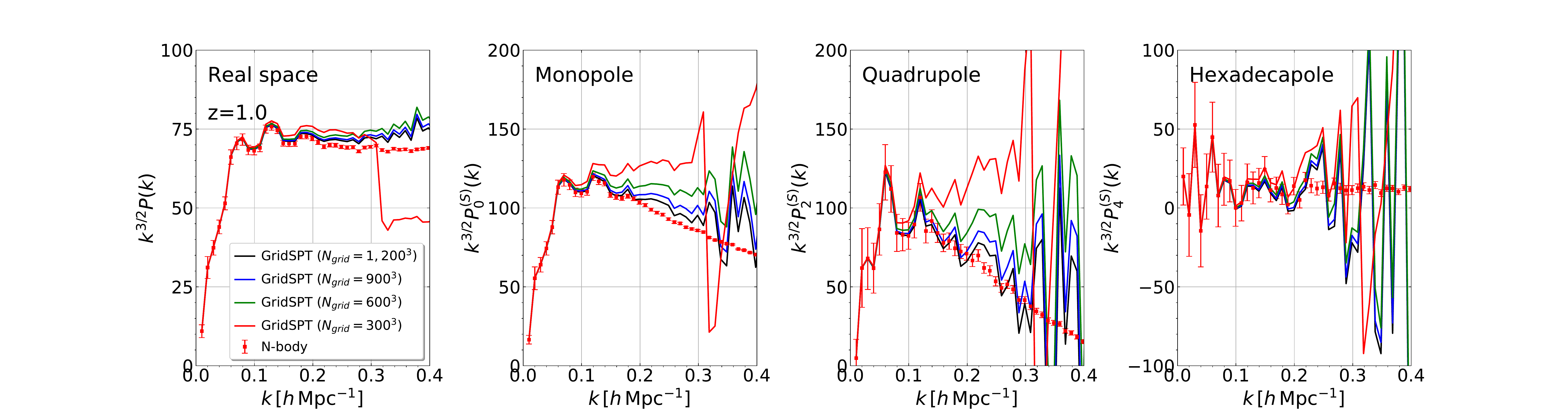}

\vspace*{-0.5cm}

\includegraphics[width=18cm,angle=0]{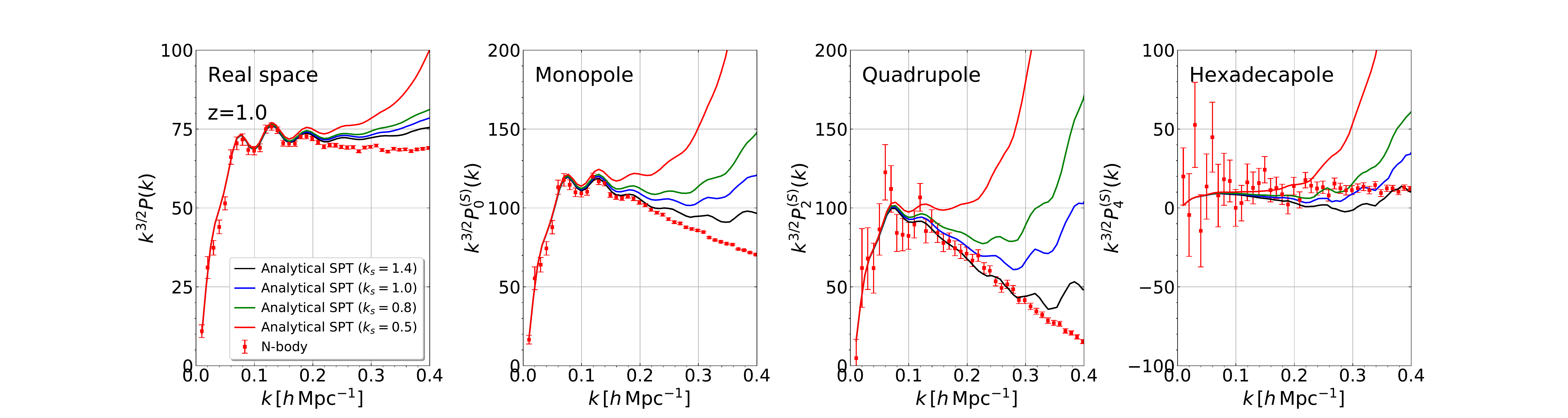}
\end{center}

\vspace*{-0.5cm}
\caption{Sensitivity of the power spectrum predictions to the high$-k$ cutoff in real and redshift space. The results at $z=1$ are plotted. In the upper panels, varying the number of grids $N_{\rm grid}$, the \gridspt\, results of the two-loop power spectra at $z=1$ are shown (solid). In all cases,  the box size of the \gridspt\, calculations is held fixed to $L_{\rm box}=1,000\,h^{-1}$\,Mpc. On the other hand, lower panels plot the analytical SPT predictions varying the high-$k$ cutoff in the linear power spectrum. Again, in all cases, the low-$k$ cutoff in the linear spectrum is set to $k_{\rm min}=2\pi/L_{\rm box}\simeq6.28\times10^{-3}\,h$\,Mpc$^{-1}$.  For reference, the $N$-body results are also shown in each panel, depicted as red crosses.  
\label{fig:pk_SPT_resolution}
}
\vspace*{-0.5cm}
\begin{center}
\includegraphics[width=18cm,angle=0]{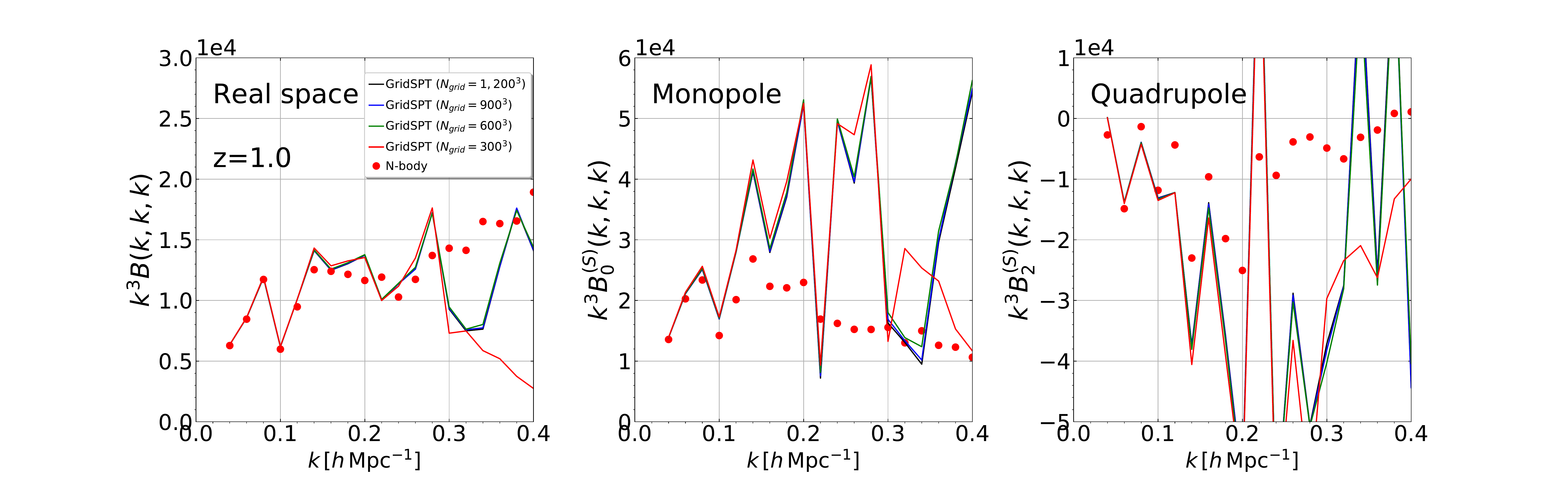}

\vspace*{-0.5cm}

\includegraphics[width=18cm,angle=0]{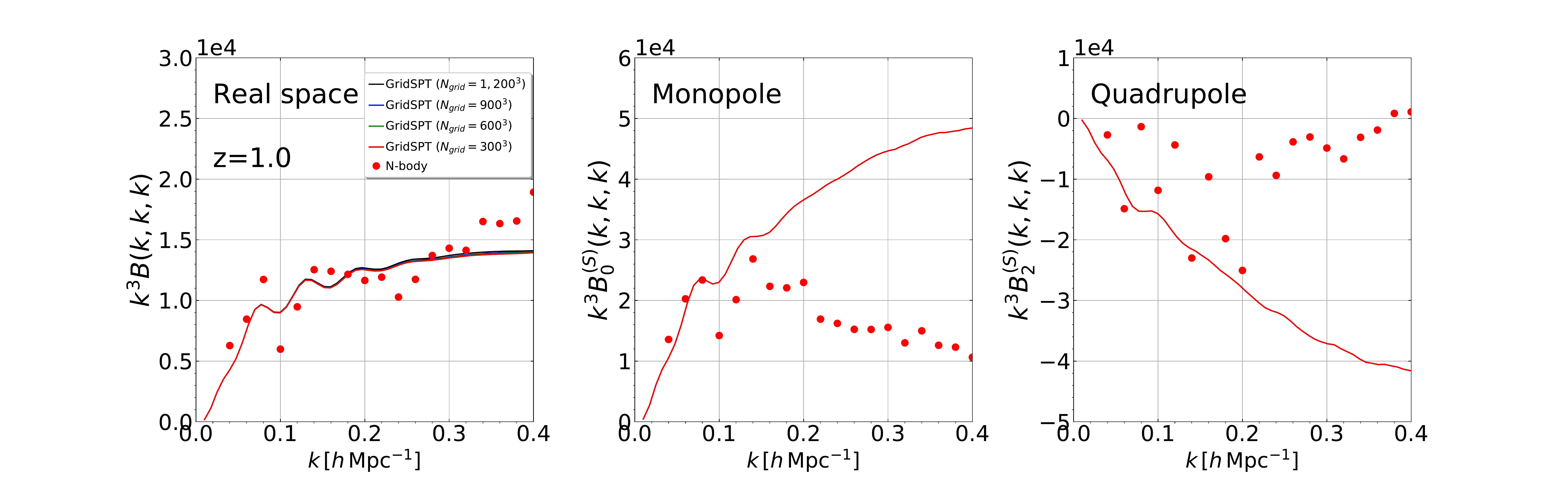}
\end{center}

\vspace*{-0.5cm}
\caption{Same as Fig.~\ref{fig:pk_SPT_resolution}, but the results of the one-loop bispectra spectra at $z=1$ are shown. 
\label{fig:bk_SPT_resolution}
}
\end{figure*}

\section{UV sensitivity of SPT calculations in redshift-space}
\label{sec:UV_sensitivity}

In this Appendix, we discuss the UV sensitivity of the SPT calculations, and examine the cutoff dependence of the predicted power spectra and bispectra. 

In SPT, higher-order PT corrections generally involve 
multi-dimensional loop integrals, and the support of their 
integrands gets wider for higher-loop integrals (e.g., \cite{Blas:2013aba,Bernardeau:2012ux}). That is, as we go to higher order, the result of the loop corrections becomes more sensitive to the cutoff of the integral. While the Galilean invariance of the SPT calculations ensures a cancellation of the IR-divergence and hence the IR sensitivity can become ignorable for a sufficiently small cutoff wavenumber, such a cancellation does not occur for the UV-sensitive behaviors, and a care must be taken for the choice of the UV cutoff (e.g., see Ref.~\cite{Taruya:2012ut,Konstandin_etal2019} for an explicit demonstration).

In Figs.~\ref{fig:pk_SPT_resolution} and \ref{fig:bk_SPT_resolution}, we respectively plot the SPT predictions of two-loop power spectra and one-loop bispectra at $z=1$. In each case, the upper panels show the \gridspt\, results varying the number of grids $N_{\rm grid}$, fixing the box size to $L_{\rm box}=1,000\,h^{-1}$\,Mpc. On the other hand, the lower panels plot the analytical SPT predictions varying the high-$k$ cutoff in the linear power spectrum.  In all analytical SPT results, the low-$k$ cutoff of $k_{\rm min}=2\pi/L_{\rm box}\simeq6.28\times10^{-3}\,h$\,Mpc$^{-1}$ is adopted. The high-$k$ cutoff scales in the analytical SPT results are taken to be slightly larger than the de-aliasing filter scales ($k_{\rm crit}=k_{\rm Nyq}/3$) for the \gridspt\, calculations\footnote{For reference, the de-aliasing filter scales shown in the upper panels are estimated as follows: $k_{\rm crit}=1.26\,h$\,Mpc$^{-1}$ (black, $N_{\rm grid}=1,200^3$), $0.94\,h$\,Mpc$^{-1}$ (blue $N_{\rm grid}=900^3$), $0.63\,h$\,Mpc$^{-1}$ (green, $N_{\rm grid}=600^3$), and $0.31\,h$\,Mpc$^{-1}$ (red, $N_{\rm grid}=300^3$).}, but we find a reasonable agreement between the two predictions. 

In Fig.~\ref{fig:pk_SPT_resolution}, we see that decreasing the high-$k$ cutoff or the number of grids enhances the power spectrum at small scales. While these trends have been known in real space (e.g., Ref.~\cite{Taruya:2012ut,Konstandin_etal2019}), a notable point is that the redshift-space power spectra exhibit a strong scale-dependent enhancement, not only in \gridspt\, but also in analytical SPT calculations. In real space, as increasing the cutoff scale or number of grids, the predicted amplitude of the power spectrum tends to converge. In redshift space, however, we still see a sizable change in the amplitude, especially at $k\gtrsim0.3\,h$\,Mpc$^{-1}$, indicating that the UV sensitivity is more serious in redshift space. This is perhaps due to the increasing number of PT corrections at higher order, arising from the line-of-sight velocity contributions [see Eq.~(\ref{eq:deltas_expansion})]. Thus, in redshift space, a careful choice of high-$k$ cutoff is necessary  for the two-loop SPT predictions of power spectrum.

On the other hand, the one-loop predictions of the bispectrum, shown in Fig.~\ref{fig:bk_SPT_resolution}, do not have a strong UV sensitivity in both real and redshift space, and the analytical SPT and \gridspt\, results with different high-$k$ cutoff or number of grids $N_{\rm grid}$ almost coincide with each other. These trends are qualitatively similar to those in the one-loop power spectrum, for which we checked to be rather insensitive to the high-$k$ cutoff.

To sum up, the cutoff dependence of the SPT prediction is significant in the power spectrum calculation at two-loop order, and in redshift space, even with a large UV cutoff, the convergence of the power spectrum result seems to be slow. Thus, the \gridspt\, prediction in redshift space suffers from  a rather strong UV sensitivity. However, this is indeed consistent with the analytical SPT calculations, and as long as we consider the PT calculations at fifth order, it is only the case for the power spectrum.

\section{Coefficients of Pad\'e approximations}
\label{sec:Pade_coefficients}

In this Appendix, we present the explicit form of the coefficients $a_n$ and $b_n$ for the density fields $\delta_{\rm Pade}^{\rm (S)}$ given in  Eq.~(\ref{eq:pade_approx}), which are expressed in terms of the redshift-space SPT density fields, $\delta_n^{\rm (S)}$. 

To derive the explicit expressions, we first introduce a book-keeping parameter $\epsilon$, and rewrite the coefficients as $a_n\to\,\epsilon^n\,a_n$, $b_n\to\,\epsilon^n\,b_n$ and $c_n\to\,\epsilon^n\,a_c$. We then equate the SPT density field up to $(M+N)$-th order to the rational form of Pad\'e $(M,\,N)$, $\delta_{\rm Pade}^{\rm(S)}$, given in Eq.~(\ref{eq:pade_approx}). We have
\begin{align}
 \Bigl\{1+\sum_{n=1}^N \epsilon^n\,b_n\Bigr\}\sum_{n=1}^{M+N}\epsilon^n\,c_n=\sum_{m=1}^M\,\epsilon^m\,a_m
\end{align}
Sorting the above expression with the power of expansion parameter $\epsilon$, the order-by-order comparison between both sides yields the equations for $a_n$ and $b_n$, involving also the coefficient $c_n$. Solving these equations for a given set of numbers $(M,\,N)$, the coefficients $a_n$ and $b_n$ are determined uniquely, and are expressed in terms of $c_n$.  Recalling that $c_n$ is written as $c_n=\delta_n^{\rm (S)}$, we obtain the explicit expressions for the coefficients $a_n$ and $b_n$ as follows:

\smallskip

\noindent
\underline{Pad\'e $(2,\,1)$}
\begin{align}
 a_1&=\delta_{\rm 1}^{\rm(S)}, 
\\
 a_2&=\frac{\{\delta_{\rm 2}^{\rm(S)}\}^2-\delta_{\rm 1}^{\rm(S)}\delta_{\rm 3}^{\rm(S)}}{\delta_{\rm 2}^{\rm(S)}}, 
\\
 b_1&=-\frac{\delta_{\rm 3}^{\rm(S)}}{\delta_{\rm 2}^{\rm(S)}}.
\end{align}

\noindent
\underline{Pad\'e $(2,\,2)$}
\begin{align}
 a_1&=\delta_{\rm 1}^{\rm(S)},
\\
a_2&=\frac{\{\delta_{\rm 2}^{\rm(S)}\}^3-2\,\delta_{\rm 1}^{\rm(S)}\delta_2^{\rm(S)}\delta_3^{\rm(S)}+\{\delta_1^{\rm(S)}\}^2\delta_4^{\rm(S)}}{
\{\delta_{\rm 2}^{\rm(S)}\}^2-\delta_{\rm 1}^{\rm(S)}\delta_{\rm 3}^{\rm(S)}},
\\
 b_1&=\frac{\delta_{\rm 1}^{\rm(S)}\delta_{\rm 4}^{\rm(S)} 
-\delta_{\rm 2}^{\rm(S)}\delta_{\rm 3}^{\rm(S)}}{
\{\delta_{\rm 2}^{\rm(S)}\}^2-\delta_{\rm 1}^{\rm(S)}\delta_{\rm 3}^{\rm(S)}},
\\
b_2&=\frac{\{\delta_{\rm 3}^{\rm(S)}\}^2-\delta_2^{\rm(S)}\delta_4^{\rm(S)}}{
\{\delta_{\rm 2}^{\rm(S)}\}^2-\delta_{\rm 1}^{\rm(S)}\delta_{\rm 3}^{\rm(S)}}.
\end{align}

\noindent
\underline{Pad\'e $(3,\,2)$}
\begin{widetext}
\begin{align}
a_1&=\delta_1^{\rm(S)} ,
\\
a_2&=\frac{\delta_2^{\rm(S)}\bigl[\{\delta_3^{\rm(S)}\}^2 -\delta_2^{\rm(S)}\delta_4^{\rm(S)}\bigr]-\delta_1^{\rm(S)}\bigl\{\delta_3^{\rm(S)}\delta_4^{\rm(S)}-\delta_2^{\rm(S)}\delta_5^{\rm(5)}\bigr\}
}{\{\delta_3^{\rm(S)}\}^2-\delta_2^{\rm(S)}\delta_4^{\rm(S)}} ,
\\
a_3&=\frac{\delta_3^{\rm(S)}\bigl[\{\delta_3^{\rm(S)}\}^2-2\delta_2^{\rm(S)}\delta_4^{\rm(S)}-\delta_1^{\rm(S)}\delta_5^{\rm(S)}\bigr]+\delta_1^{\rm(S)}\{\delta_4^{\rm(S)}\}^2+\{\delta_2^{\rm(S)}\}^2\delta_5^{\rm(S)}
}{\{\delta_3^{\rm(S)}\}^2-\delta_2^{\rm(S)}\delta_4^{\rm(S)}} ,
\\
b_1&=\frac{\delta_2^{\rm(S)}\delta_5^{\rm(S)}-\delta_3^{\rm(S)}\delta_4^{\rm(S)}}{\{\delta_3^{\rm(S)}\}^2-\delta_2^{\rm(S)}\delta_4^{\rm(S)}} ,
\\
b_2&=\frac{\{\delta_4^{\rm(S)}\}^2-\delta_3^{\rm(S)}\delta_5^{\rm(S)}}{\{\delta_3^{\rm(S)}\}^2-\delta_2^{\rm(S)}\delta_4^{\rm(S)}} .
\end{align}
\end{widetext}

\bibliographystyle{apsrev4-2}

\begin{thebibliography}{90}%
\makeatletter
\providecommand \@ifxundefined [1]{%
 \@ifx{#1\undefined}
}%
\providecommand \@ifnum [1]{%
 \ifnum #1\expandafter \@firstoftwo
 \else \expandafter \@secondoftwo
 \fi
}%
\providecommand \@ifx [1]{%
 \ifx #1\expandafter \@firstoftwo
 \else \expandafter \@secondoftwo
 \fi
}%
\providecommand \natexlab [1]{#1}%
\providecommand \enquote  [1]{``#1''}%
\providecommand \bibnamefont  [1]{#1}%
\providecommand \bibfnamefont [1]{#1}%
\providecommand \citenamefont [1]{#1}%
\providecommand \href@noop [0]{\@secondoftwo}%
\providecommand \href [0]{\begingroup \@sanitize@url \@href}%
\providecommand \@href[1]{\@@startlink{#1}\@@href}%
\providecommand \@@href[1]{\endgroup#1\@@endlink}%
\providecommand \@sanitize@url [0]{\catcode `\\12\catcode `\$12\catcode
  `\&12\catcode `\#12\catcode `\^12\catcode `\_12\catcode `\%12\relax}%
\providecommand \@@startlink[1]{}%
\providecommand \@@endlink[0]{}%
\providecommand \url  [0]{\begingroup\@sanitize@url \@url }%
\providecommand \@url [1]{\endgroup\@href {#1}{\urlprefix }}%
\providecommand \urlprefix  [0]{URL }%
\providecommand \Eprint [0]{\href }%
\providecommand \doibase [0]{https://doi.org/}%
\providecommand \selectlanguage [0]{\@gobble}%
\providecommand \bibinfo  [0]{\@secondoftwo}%
\providecommand \bibfield  [0]{\@secondoftwo}%
\providecommand \translation [1]{[#1]}%
\providecommand \BibitemOpen [0]{}%
\providecommand \bibitemStop [0]{}%
\providecommand \bibitemNoStop [0]{.\EOS\space}%
\providecommand \EOS [0]{\spacefactor3000\relax}%
\providecommand \BibitemShut  [1]{\csname bibitem#1\endcsname}%
\let\auto@bib@innerbib\@empty
\bibitem [{\citenamefont {{Desjacques}}\ \emph {et~al.}(2018)\citenamefont
  {{Desjacques}}, \citenamefont {{Jeong}},\ and\ \citenamefont
  {{Schmidt}}}]{Desjacques_Jeong_Schmidt2018}%
  \BibitemOpen
  \bibfield  {author} {\bibinfo {author} {\bibfnamefont {V.}~\bibnamefont
  {{Desjacques}}}, \bibinfo {author} {\bibfnamefont {D.}~\bibnamefont
  {{Jeong}}},\ and\ \bibinfo {author} {\bibfnamefont {F.}~\bibnamefont
  {{Schmidt}}},\ }\href {https://doi.org/10.1016/j.physrep.2017.12.002}
  {\bibfield  {journal} {\bibinfo  {journal} {\physrep}\ }\textbf {\bibinfo
  {volume} {733}},\ \bibinfo {pages} {1} (\bibinfo {year} {2018})},\ \Eprint
  {https://arxiv.org/abs/1611.09787} {arXiv:1611.09787 [astro-ph.CO]}
  \BibitemShut {NoStop}%
\bibitem [{\citenamefont {{Gebhardt}}(2021)}]{HETDEX}%
  \BibitemOpen
  \bibfield  {author} {\bibinfo {author} {\bibfnamefont {K.}~\bibnamefont
  {{Gebhardt}}},\ }\href@noop {} {\bibfield  {journal} {\bibinfo  {journal} {To
  appear in ApJ}\ } (\bibinfo {year} {2021})}\BibitemShut {NoStop}%
\bibitem [{\citenamefont {{Takada}}\ \emph {et~al.}(2014)\citenamefont
  {{Takada}}, \citenamefont {{Ellis}}, \citenamefont {{Chiba}}, \citenamefont
  {{Greene}}, \citenamefont {{Aihara}}, \citenamefont {{Arimoto}},
  \citenamefont {{Bundy}}, \citenamefont {{Cohen}}, \citenamefont {{Dor{\'e}}},
  \citenamefont {{Graves}}, \citenamefont {{Gunn}}, \citenamefont {{Heckman}},
  \citenamefont {{Hirata}}, \citenamefont {{Ho}}, \citenamefont {{Kneib}},
  \citenamefont {{Le F{\`e}vre}}, \citenamefont {{Lin}}, \citenamefont
  {{More}}, \citenamefont {{Murayama}}, \citenamefont {{Nagao}}, \citenamefont
  {{Ouchi}}, \citenamefont {{Seiffert}}, \citenamefont {{Silverman}},
  \citenamefont {{Sodr{\'e}}}, \citenamefont {{Spergel}}, \citenamefont
  {{Strauss}}, \citenamefont {{Sugai}}, \citenamefont {{Suto}}, \citenamefont
  {{Takami}},\ and\ \citenamefont {{Wyse}}}]{Subaru_PFS2014}%
  \BibitemOpen
  \bibfield  {author} {\bibinfo {author} {\bibfnamefont {M.}~\bibnamefont
  {{Takada}}}, \bibinfo {author} {\bibfnamefont {R.~S.}\ \bibnamefont
  {{Ellis}}}, \bibinfo {author} {\bibfnamefont {M.}~\bibnamefont {{Chiba}}},
  \bibinfo {author} {\bibfnamefont {J.~E.}\ \bibnamefont {{Greene}}}, \bibinfo
  {author} {\bibfnamefont {H.}~\bibnamefont {{Aihara}}}, \bibinfo {author}
  {\bibfnamefont {N.}~\bibnamefont {{Arimoto}}}, \bibinfo {author}
  {\bibfnamefont {K.}~\bibnamefont {{Bundy}}}, \bibinfo {author} {\bibfnamefont
  {J.}~\bibnamefont {{Cohen}}}, \bibinfo {author} {\bibfnamefont
  {O.}~\bibnamefont {{Dor{\'e}}}}, \bibinfo {author} {\bibfnamefont
  {G.}~\bibnamefont {{Graves}}}, \bibinfo {author} {\bibfnamefont {J.~E.}\
  \bibnamefont {{Gunn}}}, \bibinfo {author} {\bibfnamefont {T.}~\bibnamefont
  {{Heckman}}}, \bibinfo {author} {\bibfnamefont {C.~M.}\ \bibnamefont
  {{Hirata}}}, \bibinfo {author} {\bibfnamefont {P.}~\bibnamefont {{Ho}}},
  \bibinfo {author} {\bibfnamefont {J.-P.}\ \bibnamefont {{Kneib}}}, \bibinfo
  {author} {\bibfnamefont {O.}~\bibnamefont {{Le F{\`e}vre}}}, \bibinfo
  {author} {\bibfnamefont {L.}~\bibnamefont {{Lin}}}, \bibinfo {author}
  {\bibfnamefont {S.}~\bibnamefont {{More}}}, \bibinfo {author} {\bibfnamefont
  {H.}~\bibnamefont {{Murayama}}}, \bibinfo {author} {\bibfnamefont
  {T.}~\bibnamefont {{Nagao}}}, \bibinfo {author} {\bibfnamefont
  {M.}~\bibnamefont {{Ouchi}}}, \bibinfo {author} {\bibfnamefont
  {M.}~\bibnamefont {{Seiffert}}}, \bibinfo {author} {\bibfnamefont {J.~D.}\
  \bibnamefont {{Silverman}}}, \bibinfo {author} {\bibfnamefont
  {L.}~\bibnamefont {{Sodr{\'e}}}}, \bibinfo {author} {\bibfnamefont {D.~N.}\
  \bibnamefont {{Spergel}}}, \bibinfo {author} {\bibfnamefont {M.~A.}\
  \bibnamefont {{Strauss}}}, \bibinfo {author} {\bibfnamefont {H.}~\bibnamefont
  {{Sugai}}}, \bibinfo {author} {\bibfnamefont {Y.}~\bibnamefont {{Suto}}},
  \bibinfo {author} {\bibfnamefont {H.}~\bibnamefont {{Takami}}},\ and\
  \bibinfo {author} {\bibfnamefont {R.}~\bibnamefont {{Wyse}}},\ }\href
  {https://doi.org/10.1093/pasj/pst019} {\bibfield  {journal} {\bibinfo
  {journal} {\pasj}\ }\textbf {\bibinfo {volume} {66}},\ \bibinfo {eid} {R1}
  (\bibinfo {year} {2014})},\ \Eprint {https://arxiv.org/abs/1206.0737}
  {arXiv:1206.0737 [astro-ph.CO]} \BibitemShut {NoStop}%
\bibitem [{\citenamefont {{DESI Collaboration}}(2016)}]{DESI_ScienceBook2016}%
  \BibitemOpen
  \bibfield  {author} {\bibinfo {author} {\bibnamefont {{DESI
  Collaboration}}},\ }\href@noop {} {\bibfield  {journal} {\bibinfo  {journal}
  {arXiv e-prints}\ ,\ \bibinfo {eid} {arXiv:1611.00036}} (\bibinfo {year}
  {2016})},\ \Eprint {https://arxiv.org/abs/1611.00036} {arXiv:1611.00036
  [astro-ph.IM]} \BibitemShut {NoStop}%
\bibitem [{\citenamefont {{Laureijs}}\ \emph {et~al.}(2011)\citenamefont
  {{Laureijs}} \emph {et~al.}}]{Euclid2011}%
  \BibitemOpen
  \bibfield  {author} {\bibinfo {author} {\bibfnamefont {R.}~\bibnamefont
  {{Laureijs}}} \emph {et~al.},\ }\href@noop {} {\bibfield  {journal} {\bibinfo
   {journal} {arXiv e-prints}\ ,\ \bibinfo {eid} {arXiv:1110.3193}} (\bibinfo
  {year} {2011})},\ \Eprint {https://arxiv.org/abs/1110.3193} {arXiv:1110.3193
  [astro-ph.CO]} \BibitemShut {NoStop}%
\bibitem [{\citenamefont {{Green}}\ \emph {et~al.}(2012)\citenamefont {{Green}}
  \emph {et~al.}}]{WFIRST2012}%
  \BibitemOpen
  \bibfield  {author} {\bibinfo {author} {\bibfnamefont {J.}~\bibnamefont
  {{Green}}} \emph {et~al.},\ }\href@noop {} {\bibfield  {journal} {\bibinfo
  {journal} {arXiv e-prints}\ ,\ \bibinfo {eid} {arXiv:1208.4012}} (\bibinfo
  {year} {2012})},\ \Eprint {https://arxiv.org/abs/1208.4012} {arXiv:1208.4012
  [astro-ph.IM]} \BibitemShut {NoStop}%
\bibitem [{\citenamefont {{Dor{\'e}}}\ \emph {et~al.}(2018)\citenamefont
  {{Dor{\'e}}} \emph {et~al.}}]{SPHEREx2018}%
  \BibitemOpen
  \bibfield  {author} {\bibinfo {author} {\bibfnamefont {O.}~\bibnamefont
  {{Dor{\'e}}}} \emph {et~al.},\ }\href@noop {} {\bibfield  {journal} {\bibinfo
   {journal} {arXiv e-prints}\ ,\ \bibinfo {eid} {arXiv:1805.05489}} (\bibinfo
  {year} {2018})},\ \Eprint {https://arxiv.org/abs/1805.05489}
  {arXiv:1805.05489 [astro-ph.IM]} \BibitemShut {NoStop}%
\bibitem [{\citenamefont {{Weinberg}}\ \emph {et~al.}(2013)\citenamefont
  {{Weinberg}}, \citenamefont {{Mortonson}}, \citenamefont {{Eisenstein}},
  \citenamefont {{Hirata}}, \citenamefont {{Riess}},\ and\ \citenamefont
  {{Rozo}}}]{Weinberg_etal2013}%
  \BibitemOpen
  \bibfield  {author} {\bibinfo {author} {\bibfnamefont {D.~H.}\ \bibnamefont
  {{Weinberg}}}, \bibinfo {author} {\bibfnamefont {M.~J.}\ \bibnamefont
  {{Mortonson}}}, \bibinfo {author} {\bibfnamefont {D.~J.}\ \bibnamefont
  {{Eisenstein}}}, \bibinfo {author} {\bibfnamefont {C.}~\bibnamefont
  {{Hirata}}}, \bibinfo {author} {\bibfnamefont {A.~G.}\ \bibnamefont
  {{Riess}}},\ and\ \bibinfo {author} {\bibfnamefont {E.}~\bibnamefont
  {{Rozo}}},\ }\href {https://doi.org/10.1016/j.physrep.2013.05.001} {\bibfield
   {journal} {\bibinfo  {journal} {Phys. Rept.}\ }\textbf {\bibinfo {volume}
  {530}},\ \bibinfo {pages} {87} (\bibinfo {year} {2013})},\ \Eprint
  {https://arxiv.org/abs/1201.2434} {arXiv:1201.2434 [astro-ph.CO]}
  \BibitemShut {NoStop}%
\bibitem [{\citenamefont {Crocce}\ and\ \citenamefont
  {Scoccimarro}(2006)}]{Crocce:2005xy}%
  \BibitemOpen
  \bibfield  {author} {\bibinfo {author} {\bibfnamefont {M.}~\bibnamefont
  {Crocce}}\ and\ \bibinfo {author} {\bibfnamefont {R.}~\bibnamefont
  {Scoccimarro}},\ }\href {https://doi.org/10.1103/PhysRevD.73.063519}
  {\bibfield  {journal} {\bibinfo  {journal} {\prd}\ }\textbf {\bibinfo
  {volume} {73}},\ \bibinfo {pages} {063519} (\bibinfo {year} {2006})},\
  \Eprint {https://arxiv.org/abs/astro-ph/0509418} {arXiv:astro-ph/0509418}
  \BibitemShut {NoStop}%
\bibitem [{\citenamefont {Jeong}\ and\ \citenamefont
  {Komatsu}(2006)}]{Jeong:2006xd}%
  \BibitemOpen
  \bibfield  {author} {\bibinfo {author} {\bibfnamefont {D.}~\bibnamefont
  {Jeong}}\ and\ \bibinfo {author} {\bibfnamefont {E.}~\bibnamefont
  {Komatsu}},\ }\href {https://doi.org/10.1086/507781} {\bibfield  {journal}
  {\bibinfo  {journal} {Astrophys. J.}\ }\textbf {\bibinfo {volume} {651}},\
  \bibinfo {pages} {619} (\bibinfo {year} {2006})},\ \Eprint
  {https://arxiv.org/abs/astro-ph/0604075} {arXiv:astro-ph/0604075}
  \BibitemShut {NoStop}%
\bibitem [{\citenamefont {Jeong}\ and\ \citenamefont
  {Komatsu}(2009)}]{Jeong:2008rj}%
  \BibitemOpen
  \bibfield  {author} {\bibinfo {author} {\bibfnamefont {D.}~\bibnamefont
  {Jeong}}\ and\ \bibinfo {author} {\bibfnamefont {E.}~\bibnamefont
  {Komatsu}},\ }\href {https://doi.org/10.1088/0004-637X/691/1/569} {\bibfield
  {journal} {\bibinfo  {journal} {Astrophys. J.}\ }\textbf {\bibinfo {volume}
  {691}},\ \bibinfo {pages} {569} (\bibinfo {year} {2009})},\ \Eprint
  {https://arxiv.org/abs/0805.2632} {arXiv:0805.2632 [astro-ph]} \BibitemShut
  {NoStop}%
\bibitem [{\citenamefont {Crocce}\ and\ \citenamefont
  {Scoccimarro}(2008)}]{Crocce:2007dt}%
  \BibitemOpen
  \bibfield  {author} {\bibinfo {author} {\bibfnamefont {M.}~\bibnamefont
  {Crocce}}\ and\ \bibinfo {author} {\bibfnamefont {R.}~\bibnamefont
  {Scoccimarro}},\ }\href {https://doi.org/10.1103/PhysRevD.77.023533}
  {\bibfield  {journal} {\bibinfo  {journal} {\prd}\ }\textbf {\bibinfo
  {volume} {77}},\ \bibinfo {pages} {023533} (\bibinfo {year} {2008})},\
  \Eprint {https://arxiv.org/abs/0704.2783} {arXiv:0704.2783 [astro-ph]}
  \BibitemShut {NoStop}%
\bibitem [{\citenamefont {{Taruya}}\ and\ \citenamefont
  {{Hiramatsu}}(2008)}]{Taruya:2007xy}%
  \BibitemOpen
  \bibfield  {author} {\bibinfo {author} {\bibfnamefont {A.}~\bibnamefont
  {{Taruya}}}\ and\ \bibinfo {author} {\bibfnamefont {T.}~\bibnamefont
  {{Hiramatsu}}},\ }\href {https://doi.org/10.1086/526515} {\bibfield
  {journal} {\bibinfo  {journal} {\apj}\ }\textbf {\bibinfo {volume} {674}},\
  \bibinfo {eid} {617-635} (\bibinfo {year} {2008})},\ \Eprint
  {https://arxiv.org/abs/0708.1367} {arXiv:0708.1367} \BibitemShut {NoStop}%
\bibitem [{\citenamefont {Matsubara}(2008)}]{Matsubara2008a}%
  \BibitemOpen
  \bibfield  {author} {\bibinfo {author} {\bibfnamefont {T.}~\bibnamefont
  {Matsubara}},\ }\href {https://doi.org/10.1103/PhysRevD.77.063530} {\bibfield
   {journal} {\bibinfo  {journal} {\prd}\ }\textbf {\bibinfo {volume} {77}},\
  \bibinfo {pages} {063530} (\bibinfo {year} {2008})},\ \Eprint
  {https://arxiv.org/abs/0711.2521} {arXiv:0711.2521 [astro-ph]} \BibitemShut
  {NoStop}%
\bibitem [{\citenamefont {Bernardeau}\ \emph {et~al.}(2008)\citenamefont
  {Bernardeau}, \citenamefont {Crocce},\ and\ \citenamefont
  {Scoccimarro}}]{Bernardeau:2008fa}%
  \BibitemOpen
  \bibfield  {author} {\bibinfo {author} {\bibfnamefont {F.}~\bibnamefont
  {Bernardeau}}, \bibinfo {author} {\bibfnamefont {M.}~\bibnamefont {Crocce}},\
  and\ \bibinfo {author} {\bibfnamefont {R.}~\bibnamefont {Scoccimarro}},\
  }\href {https://doi.org/10.1103/PhysRevD.78.103521} {\bibfield  {journal}
  {\bibinfo  {journal} {\prd}\ }\textbf {\bibinfo {volume} {78}},\ \bibinfo
  {pages} {103521} (\bibinfo {year} {2008})},\ \Eprint
  {https://arxiv.org/abs/0806.2334} {arXiv:0806.2334 [astro-ph]} \BibitemShut
  {NoStop}%
\bibitem [{\citenamefont {Nishimichi}\ \emph {et~al.}(2009)\citenamefont
  {Nishimichi} \emph {et~al.}}]{Nishimichi:2008ry}%
  \BibitemOpen
  \bibfield  {author} {\bibinfo {author} {\bibfnamefont {T.}~\bibnamefont
  {Nishimichi}} \emph {et~al.},\ }\href@noop {} {\bibfield  {journal} {\bibinfo
   {journal} {Publ. Astron. Soc. Jap.}\ }\textbf {\bibinfo {volume} {61}},\
  \bibinfo {pages} {321} (\bibinfo {year} {2009})},\ \Eprint
  {https://arxiv.org/abs/0810.0813} {arXiv:0810.0813 [astro-ph]} \BibitemShut
  {NoStop}%
\bibitem [{\citenamefont {Lawrence}\ \emph {et~al.}(2010)\citenamefont
  {Lawrence} \emph {et~al.}}]{Lawrence:2009uk}%
  \BibitemOpen
  \bibfield  {author} {\bibinfo {author} {\bibfnamefont {E.}~\bibnamefont
  {Lawrence}} \emph {et~al.},\ }\href
  {https://doi.org/10.1088/0004-637X/713/2/1322} {\bibfield  {journal}
  {\bibinfo  {journal} {Astrophys. J.}\ }\textbf {\bibinfo {volume} {713}},\
  \bibinfo {pages} {1322} (\bibinfo {year} {2010})},\ \Eprint
  {https://arxiv.org/abs/0912.4490} {arXiv:0912.4490 [astro-ph.CO]}
  \BibitemShut {NoStop}%
\bibitem [{\citenamefont {Taruya}\ \emph {et~al.}(2010)\citenamefont {Taruya},
  \citenamefont {Nishimichi},\ and\ \citenamefont {Saito}}]{Taruya:2010mx}%
  \BibitemOpen
  \bibfield  {author} {\bibinfo {author} {\bibfnamefont {A.}~\bibnamefont
  {Taruya}}, \bibinfo {author} {\bibfnamefont {T.}~\bibnamefont {Nishimichi}},\
  and\ \bibinfo {author} {\bibfnamefont {S.}~\bibnamefont {Saito}},\ }\href
  {https://doi.org/10.1103/PhysRevD.82.063522} {\bibfield  {journal} {\bibinfo
  {journal} {\prd}\ }\textbf {\bibinfo {volume} {82}},\ \bibinfo {pages}
  {063522} (\bibinfo {year} {2010})},\ \Eprint
  {https://arxiv.org/abs/1006.0699} {arXiv:1006.0699 [astro-ph.CO]}
  \BibitemShut {NoStop}%
\bibitem [{\citenamefont {Nishimichi}\ and\ \citenamefont
  {Taruya}(2011)}]{Nishimichi:2011jm}%
  \BibitemOpen
  \bibfield  {author} {\bibinfo {author} {\bibfnamefont {T.}~\bibnamefont
  {Nishimichi}}\ and\ \bibinfo {author} {\bibfnamefont {A.}~\bibnamefont
  {Taruya}},\ }\href {https://doi.org/10.1103/PhysRevD.84.043526} {\bibfield
  {journal} {\bibinfo  {journal} {\prd}\ }\textbf {\bibinfo {volume} {84}},\
  \bibinfo {pages} {043526} (\bibinfo {year} {2011})},\ \Eprint
  {https://arxiv.org/abs/1106.4562} {arXiv:1106.4562 [astro-ph.CO]}
  \BibitemShut {NoStop}%
\bibitem [{\citenamefont {{Baumann}}\ \emph {et~al.}(2012)\citenamefont
  {{Baumann}}, \citenamefont {{Nicolis}}, \citenamefont {{Senatore}},\ and\
  \citenamefont {{Zaldarriaga}}}]{2012JCAP...07..051B}%
  \BibitemOpen
  \bibfield  {author} {\bibinfo {author} {\bibfnamefont {D.}~\bibnamefont
  {{Baumann}}}, \bibinfo {author} {\bibfnamefont {A.}~\bibnamefont
  {{Nicolis}}}, \bibinfo {author} {\bibfnamefont {L.}~\bibnamefont
  {{Senatore}}},\ and\ \bibinfo {author} {\bibfnamefont {M.}~\bibnamefont
  {{Zaldarriaga}}},\ }\href {https://doi.org/10.1088/1475-7516/2012/07/051}
  {\bibfield  {journal} {\bibinfo  {journal} {\jcap}\ }\textbf {\bibinfo
  {volume} {7}},\ \bibinfo {eid} {051} (\bibinfo {year} {2012})},\ \Eprint
  {https://arxiv.org/abs/1004.2488} {arXiv:1004.2488 [astro-ph.CO]}
  \BibitemShut {NoStop}%
\bibitem [{\citenamefont {{Senatore}}(2015)}]{Senatore2015}%
  \BibitemOpen
  \bibfield  {author} {\bibinfo {author} {\bibfnamefont {L.}~\bibnamefont
  {{Senatore}}},\ }\href {https://doi.org/10.1088/1475-7516/2015/11/007}
  {\bibfield  {journal} {\bibinfo  {journal} {\jcap}\ }\textbf {\bibinfo
  {volume} {2015}},\ \bibinfo {eid} {007} (\bibinfo {year} {2015})},\ \Eprint
  {https://arxiv.org/abs/1406.7843} {arXiv:1406.7843 [astro-ph.CO]}
  \BibitemShut {NoStop}%
\bibitem [{\citenamefont {{Mirbabayi}}\ \emph {et~al.}(2015)\citenamefont
  {{Mirbabayi}}, \citenamefont {{Schmidt}},\ and\ \citenamefont
  {{Zaldarriaga}}}]{Mirbabayi_Schmidt_Zaldarriaga2015}%
  \BibitemOpen
  \bibfield  {author} {\bibinfo {author} {\bibfnamefont {M.}~\bibnamefont
  {{Mirbabayi}}}, \bibinfo {author} {\bibfnamefont {F.}~\bibnamefont
  {{Schmidt}}},\ and\ \bibinfo {author} {\bibfnamefont {M.}~\bibnamefont
  {{Zaldarriaga}}},\ }\href {https://doi.org/10.1088/1475-7516/2015/07/030}
  {\bibfield  {journal} {\bibinfo  {journal} {\jcap}\ }\textbf {\bibinfo
  {volume} {2015}},\ \bibinfo {eid} {030} (\bibinfo {year} {2015})},\ \Eprint
  {https://arxiv.org/abs/1412.5169} {arXiv:1412.5169 [astro-ph.CO]}
  \BibitemShut {NoStop}%
\bibitem [{\citenamefont {{Nishimichi}}\ \emph {et~al.}(2017)\citenamefont
  {{Nishimichi}}, \citenamefont {{Bernardeau}},\ and\ \citenamefont
  {{Taruya}}}]{Nishimichi_etal2017}%
  \BibitemOpen
  \bibfield  {author} {\bibinfo {author} {\bibfnamefont {T.}~\bibnamefont
  {{Nishimichi}}}, \bibinfo {author} {\bibfnamefont {F.}~\bibnamefont
  {{Bernardeau}}},\ and\ \bibinfo {author} {\bibfnamefont {A.}~\bibnamefont
  {{Taruya}}},\ }\href {https://doi.org/10.1103/PhysRevD.96.123515} {\bibfield
  {journal} {\bibinfo  {journal} {\prd}\ }\textbf {\bibinfo {volume} {96}},\
  \bibinfo {eid} {123515} (\bibinfo {year} {2017})},\ \Eprint
  {https://arxiv.org/abs/1708.08946} {arXiv:1708.08946} \BibitemShut {NoStop}%
\bibitem [{\citenamefont {{Nishimichi}}\ \emph {et~al.}(2019)\citenamefont
  {{Nishimichi}}, \citenamefont {{Takada}}, \citenamefont {{Takahashi}},
  \citenamefont {{Osato}}, \citenamefont {{Shirasaki}}, \citenamefont {{Oogi}},
  \citenamefont {{Miyatake}}, \citenamefont {{Oguri}}, \citenamefont
  {{Murata}}, \citenamefont {{Kobayashi}},\ and\ \citenamefont
  {{Yoshida}}}]{Nishimichi_etal2019_DQ_I}%
  \BibitemOpen
  \bibfield  {author} {\bibinfo {author} {\bibfnamefont {T.}~\bibnamefont
  {{Nishimichi}}}, \bibinfo {author} {\bibfnamefont {M.}~\bibnamefont
  {{Takada}}}, \bibinfo {author} {\bibfnamefont {R.}~\bibnamefont
  {{Takahashi}}}, \bibinfo {author} {\bibfnamefont {K.}~\bibnamefont
  {{Osato}}}, \bibinfo {author} {\bibfnamefont {M.}~\bibnamefont
  {{Shirasaki}}}, \bibinfo {author} {\bibfnamefont {T.}~\bibnamefont {{Oogi}}},
  \bibinfo {author} {\bibfnamefont {H.}~\bibnamefont {{Miyatake}}}, \bibinfo
  {author} {\bibfnamefont {M.}~\bibnamefont {{Oguri}}}, \bibinfo {author}
  {\bibfnamefont {R.}~\bibnamefont {{Murata}}}, \bibinfo {author}
  {\bibfnamefont {Y.}~\bibnamefont {{Kobayashi}}},\ and\ \bibinfo {author}
  {\bibfnamefont {N.}~\bibnamefont {{Yoshida}}},\ }\href
  {https://doi.org/10.3847/1538-4357/ab3719} {\bibfield  {journal} {\bibinfo
  {journal} {\apj}\ }\textbf {\bibinfo {volume} {884}},\ \bibinfo {eid} {29}
  (\bibinfo {year} {2019})},\ \Eprint {https://arxiv.org/abs/1811.09504}
  {arXiv:1811.09504 [astro-ph.CO]} \BibitemShut {NoStop}%
\bibitem [{\citenamefont {Bernardeau}\ \emph {et~al.}(2002)\citenamefont
  {Bernardeau}, \citenamefont {Colombi}, \citenamefont {Gaztanaga},\ and\
  \citenamefont {Scoccimarro}}]{Bernardeau:2001qr}%
  \BibitemOpen
  \bibfield  {author} {\bibinfo {author} {\bibfnamefont {F.}~\bibnamefont
  {Bernardeau}}, \bibinfo {author} {\bibfnamefont {S.}~\bibnamefont {Colombi}},
  \bibinfo {author} {\bibfnamefont {E.}~\bibnamefont {Gaztanaga}},\ and\
  \bibinfo {author} {\bibfnamefont {R.}~\bibnamefont {Scoccimarro}},\ }\href
  {https://doi.org/10.1016/S0370-1573(02)00135-7} {\bibfield  {journal}
  {\bibinfo  {journal} {Phys. Rept.}\ }\textbf {\bibinfo {volume} {367}},\
  \bibinfo {pages} {1} (\bibinfo {year} {2002})},\ \Eprint
  {https://arxiv.org/abs/astro-ph/0112551} {arXiv:astro-ph/0112551}
  \BibitemShut {NoStop}%
\bibitem [{\citenamefont {{Taruya}}\ \emph {et~al.}(2018)\citenamefont
  {{Taruya}}, \citenamefont {{Nishimichi}},\ and\ \citenamefont
  {{Jeong}}}]{Taruya_Nishimichi_Jeong2018}%
  \BibitemOpen
  \bibfield  {author} {\bibinfo {author} {\bibfnamefont {A.}~\bibnamefont
  {{Taruya}}}, \bibinfo {author} {\bibfnamefont {T.}~\bibnamefont
  {{Nishimichi}}},\ and\ \bibinfo {author} {\bibfnamefont {D.}~\bibnamefont
  {{Jeong}}},\ }\href {https://doi.org/10.1103/PhysRevD.98.103532} {\bibfield
  {journal} {\bibinfo  {journal} {\prd}\ }\textbf {\bibinfo {volume} {98}},\
  \bibinfo {eid} {103532} (\bibinfo {year} {2018})},\ \Eprint
  {https://arxiv.org/abs/1807.04215} {arXiv:1807.04215 [astro-ph.CO]}
  \BibitemShut {NoStop}%
\bibitem [{\citenamefont {{Roth}}\ and\ \citenamefont
  {{Porciani}}(2011)}]{Roth_Porciani2011}%
  \BibitemOpen
  \bibfield  {author} {\bibinfo {author} {\bibfnamefont {N.}~\bibnamefont
  {{Roth}}}\ and\ \bibinfo {author} {\bibfnamefont {C.}~\bibnamefont
  {{Porciani}}},\ }\href {https://doi.org/10.1111/j.1365-2966.2011.18768.x}
  {\bibfield  {journal} {\bibinfo  {journal} {\mnras}\ }\textbf {\bibinfo
  {volume} {415}},\ \bibinfo {pages} {829} (\bibinfo {year} {2011})},\ \Eprint
  {https://arxiv.org/abs/1101.1520} {arXiv:1101.1520 [astro-ph.CO]}
  \BibitemShut {NoStop}%
\bibitem [{\citenamefont {{Tassev}}(2014)}]{Tassev2014}%
  \BibitemOpen
  \bibfield  {author} {\bibinfo {author} {\bibfnamefont {S.}~\bibnamefont
  {{Tassev}}},\ }\href {https://doi.org/10.1088/1475-7516/2014/06/008}
  {\bibfield  {journal} {\bibinfo  {journal} {\jcap}\ }\textbf {\bibinfo
  {volume} {2014}},\ \bibinfo {eid} {008} (\bibinfo {year} {2014})},\ \Eprint
  {https://arxiv.org/abs/1311.4884} {arXiv:1311.4884 [astro-ph.CO]}
  \BibitemShut {NoStop}%
\bibitem [{\citenamefont {{Taruya}}\ \emph {et~al.}(2021)\citenamefont
  {{Taruya}}, \citenamefont {{Nishimichi}},\ and\ \citenamefont
  {{Jeong}}}]{Taruya_Nishimichi_Jeong2021}%
  \BibitemOpen
  \bibfield  {author} {\bibinfo {author} {\bibfnamefont {A.}~\bibnamefont
  {{Taruya}}}, \bibinfo {author} {\bibfnamefont {T.}~\bibnamefont
  {{Nishimichi}}},\ and\ \bibinfo {author} {\bibfnamefont {D.}~\bibnamefont
  {{Jeong}}},\ }\href {https://doi.org/10.1103/PhysRevD.103.023501} {\bibfield
  {journal} {\bibinfo  {journal} {\prd}\ }\textbf {\bibinfo {volume} {103}},\
  \bibinfo {eid} {023501} (\bibinfo {year} {2021})},\ \Eprint
  {https://arxiv.org/abs/2007.05504} {arXiv:2007.05504 [astro-ph.CO]}
  \BibitemShut {NoStop}%
\bibitem [{\citenamefont {{Peebles}}(1980)}]{Peebles:1980}%
  \BibitemOpen
  \bibfield  {author} {\bibinfo {author} {\bibfnamefont {P.~J.~E.}\
  \bibnamefont {{Peebles}}},\ }\href@noop {} {\emph {\bibinfo {title} {{The
  large-scale structure of the universe}}}}\ (\bibinfo  {publisher} {Princeton
  University Press},\ \bibinfo {year} {1980})\BibitemShut {NoStop}%
\bibitem [{\citenamefont {{Hamilton}}(1998)}]{Hamilton_RSD_review1998}%
  \BibitemOpen
  \bibfield  {author} {\bibinfo {author} {\bibfnamefont {A.~J.~S.}\
  \bibnamefont {{Hamilton}}},\ }\bibinfo {title} {{Linear Redshift Distortions:
  a Review}},\ in\ \href {https://doi.org/10.1007/978-94-011-4960-0\_17} {\emph
  {\bibinfo {booktitle} {The Evolving Universe}}},\ \bibinfo {series}
  {Astrophysics and Space Science Library}, Vol.\ \bibinfo {volume} {231},\
  \bibinfo {editor} {edited by\ \bibinfo {editor} {\bibfnamefont
  {D.}~\bibnamefont {{Hamilton}}}}\ (\bibinfo  {publisher} {Kluwer Academic
  Publishers},\ \bibinfo {year} {1998})\ p.\ \bibinfo {pages} {185}\BibitemShut
  {NoStop}%
\bibitem [{\citenamefont {Scoccimarro}(2004)}]{Scoccimarro:2004tg}%
  \BibitemOpen
  \bibfield  {author} {\bibinfo {author} {\bibfnamefont {R.}~\bibnamefont
  {Scoccimarro}},\ }\href {https://doi.org/10.1103/PhysRevD.70.083007}
  {\bibfield  {journal} {\bibinfo  {journal} {\prd}\ }\textbf {\bibinfo
  {volume} {70}},\ \bibinfo {pages} {083007} (\bibinfo {year} {2004})},\
  \Eprint {https://arxiv.org/abs/astro-ph/0407214} {arXiv:astro-ph/0407214}
  \BibitemShut {NoStop}%
\bibitem [{\citenamefont {Vlah}\ \emph {et~al.}(2012)\citenamefont {Vlah},
  \citenamefont {Seljak}, \citenamefont {McDonald}, \citenamefont {Okumura},\
  and\ \citenamefont {Baldauf}}]{Vlah:2012ni}%
  \BibitemOpen
  \bibfield  {author} {\bibinfo {author} {\bibfnamefont {Z.}~\bibnamefont
  {Vlah}}, \bibinfo {author} {\bibfnamefont {U.}~\bibnamefont {Seljak}},
  \bibinfo {author} {\bibfnamefont {P.}~\bibnamefont {McDonald}}, \bibinfo
  {author} {\bibfnamefont {T.}~\bibnamefont {Okumura}},\ and\ \bibinfo {author}
  {\bibfnamefont {T.}~\bibnamefont {Baldauf}},\ }\href
  {https://doi.org/10.1088/1475-7516/2012/11/009} {\bibfield  {journal}
  {\bibinfo  {journal} {\jcap}\ }\textbf {\bibinfo {volume} {1211}},\ \bibinfo
  {pages} {009} (\bibinfo {year} {2012})},\ \Eprint
  {https://arxiv.org/abs/1207.0839} {arXiv:1207.0839 [astro-ph.CO]}
  \BibitemShut {NoStop}%
\bibitem [{\citenamefont {{Taruya}}\ \emph {et~al.}(2013)\citenamefont
  {{Taruya}}, \citenamefont {{Nishimichi}},\ and\ \citenamefont
  {{Bernardeau}}}]{Taruya_Nishimichi_Bernardeau2013}%
  \BibitemOpen
  \bibfield  {author} {\bibinfo {author} {\bibfnamefont {A.}~\bibnamefont
  {{Taruya}}}, \bibinfo {author} {\bibfnamefont {T.}~\bibnamefont
  {{Nishimichi}}},\ and\ \bibinfo {author} {\bibfnamefont {F.}~\bibnamefont
  {{Bernardeau}}},\ }\href {https://doi.org/10.1103/PhysRevD.87.083509}
  {\bibfield  {journal} {\bibinfo  {journal} {\prd}\ }\textbf {\bibinfo
  {volume} {87}},\ \bibinfo {eid} {083509} (\bibinfo {year} {2013})},\ \Eprint
  {https://arxiv.org/abs/1301.3624} {arXiv:1301.3624 [astro-ph.CO]}
  \BibitemShut {NoStop}%
\bibitem [{\citenamefont {{Vlah}}\ \emph {et~al.}(2013)\citenamefont {{Vlah}},
  \citenamefont {{Seljak}}, \citenamefont {{Okumura}},\ and\ \citenamefont
  {{Desjacques}}}]{Vlah_etal2013}%
  \BibitemOpen
  \bibfield  {author} {\bibinfo {author} {\bibfnamefont {Z.}~\bibnamefont
  {{Vlah}}}, \bibinfo {author} {\bibfnamefont {U.}~\bibnamefont {{Seljak}}},
  \bibinfo {author} {\bibfnamefont {T.}~\bibnamefont {{Okumura}}},\ and\
  \bibinfo {author} {\bibfnamefont {V.}~\bibnamefont {{Desjacques}}},\ }\href
  {https://doi.org/10.1088/1475-7516/2013/10/053} {\bibfield  {journal}
  {\bibinfo  {journal} {\jcap}\ }\textbf {\bibinfo {volume} {2013}},\ \bibinfo
  {eid} {053} (\bibinfo {year} {2013})},\ \Eprint
  {https://arxiv.org/abs/1308.6294} {arXiv:1308.6294 [astro-ph.CO]}
  \BibitemShut {NoStop}%
\bibitem [{\citenamefont {{Carlson}}\ \emph {et~al.}(2013)\citenamefont
  {{Carlson}}, \citenamefont {{Reid}},\ and\ \citenamefont
  {{White}}}]{Carlson_Reid_White2013}%
  \BibitemOpen
  \bibfield  {author} {\bibinfo {author} {\bibfnamefont {J.}~\bibnamefont
  {{Carlson}}}, \bibinfo {author} {\bibfnamefont {B.}~\bibnamefont {{Reid}}},\
  and\ \bibinfo {author} {\bibfnamefont {M.}~\bibnamefont {{White}}},\ }\href
  {https://doi.org/10.1093/mnras/sts457} {\bibfield  {journal} {\bibinfo
  {journal} {\mnras}\ }\textbf {\bibinfo {volume} {429}},\ \bibinfo {pages}
  {1674} (\bibinfo {year} {2013})},\ \Eprint {https://arxiv.org/abs/1209.0780}
  {arXiv:1209.0780 [astro-ph.CO]} \BibitemShut {NoStop}%
\bibitem [{\citenamefont {Wang}\ \emph {et~al.}(2013)\citenamefont {Wang},
  \citenamefont {Reid},\ and\ \citenamefont {White}}]{Wang:2013hwa}%
  \BibitemOpen
  \bibfield  {author} {\bibinfo {author} {\bibfnamefont {L.}~\bibnamefont
  {Wang}}, \bibinfo {author} {\bibfnamefont {B.}~\bibnamefont {Reid}},\ and\
  \bibinfo {author} {\bibfnamefont {M.}~\bibnamefont {White}},\ }\href@noop {}
  {\bibfield  {journal} {\bibinfo  {journal} {\mnras}\ } (\bibinfo {year}
  {2013})},\ \Eprint {https://arxiv.org/abs/1306.1804} {arXiv:1306.1804
  [astro-ph.CO]} \BibitemShut {NoStop}%
\bibitem [{\citenamefont {{Matsubara}}(2014)}]{Matsubara2014}%
  \BibitemOpen
  \bibfield  {author} {\bibinfo {author} {\bibfnamefont {T.}~\bibnamefont
  {{Matsubara}}},\ }\href {https://doi.org/10.1103/PhysRevD.90.043537}
  {\bibfield  {journal} {\bibinfo  {journal} {\prd}\ }\textbf {\bibinfo
  {volume} {90}},\ \bibinfo {eid} {043537} (\bibinfo {year} {2014})},\ \Eprint
  {https://arxiv.org/abs/1304.4226} {arXiv:1304.4226 [astro-ph.CO]}
  \BibitemShut {NoStop}%
\bibitem [{\citenamefont {{Hand}}\ \emph {et~al.}(2017)\citenamefont {{Hand}},
  \citenamefont {{Seljak}}, \citenamefont {{Beutler}},\ and\ \citenamefont
  {{Vlah}}}]{Hand_etal2017}%
  \BibitemOpen
  \bibfield  {author} {\bibinfo {author} {\bibfnamefont {N.}~\bibnamefont
  {{Hand}}}, \bibinfo {author} {\bibfnamefont {U.}~\bibnamefont {{Seljak}}},
  \bibinfo {author} {\bibfnamefont {F.}~\bibnamefont {{Beutler}}},\ and\
  \bibinfo {author} {\bibfnamefont {Z.}~\bibnamefont {{Vlah}}},\ }\href
  {https://doi.org/10.1088/1475-7516/2017/10/009} {\bibfield  {journal}
  {\bibinfo  {journal} {\jcap}\ }\textbf {\bibinfo {volume} {2017}},\ \bibinfo
  {eid} {009} (\bibinfo {year} {2017})},\ \Eprint
  {https://arxiv.org/abs/1706.02362} {arXiv:1706.02362 [astro-ph.CO]}
  \BibitemShut {NoStop}%
\bibitem [{\citenamefont {{Vlah}}\ and\ \citenamefont
  {{White}}(2019)}]{Vlah_White2019}%
  \BibitemOpen
  \bibfield  {author} {\bibinfo {author} {\bibfnamefont {Z.}~\bibnamefont
  {{Vlah}}}\ and\ \bibinfo {author} {\bibfnamefont {M.}~\bibnamefont
  {{White}}},\ }\href {https://doi.org/10.1088/1475-7516/2019/03/007}
  {\bibfield  {journal} {\bibinfo  {journal} {\jcap}\ }\textbf {\bibinfo
  {volume} {2019}},\ \bibinfo {eid} {007} (\bibinfo {year} {2019})},\ \Eprint
  {https://arxiv.org/abs/1812.02775} {arXiv:1812.02775 [astro-ph.CO]}
  \BibitemShut {NoStop}%
\bibitem [{\citenamefont {{Chen}}\ \emph {et~al.}(2021)\citenamefont {{Chen}},
  \citenamefont {{Vlah}}, \citenamefont {{Castorina}},\ and\ \citenamefont
  {{White}}}]{Chen_etal2021}%
  \BibitemOpen
  \bibfield  {author} {\bibinfo {author} {\bibfnamefont {S.-F.}\ \bibnamefont
  {{Chen}}}, \bibinfo {author} {\bibfnamefont {Z.}~\bibnamefont {{Vlah}}},
  \bibinfo {author} {\bibfnamefont {E.}~\bibnamefont {{Castorina}}},\ and\
  \bibinfo {author} {\bibfnamefont {M.}~\bibnamefont {{White}}},\ }\href
  {https://doi.org/10.1088/1475-7516/2021/03/100} {\bibfield  {journal}
  {\bibinfo  {journal} {\jcap}\ }\textbf {\bibinfo {volume} {2021}},\ \bibinfo
  {eid} {100} (\bibinfo {year} {2021})},\ \Eprint
  {https://arxiv.org/abs/2012.04636} {arXiv:2012.04636 [astro-ph.CO]}
  \BibitemShut {NoStop}%
\bibitem [{\citenamefont {{Carrasco}}\ \emph {et~al.}(2012)\citenamefont
  {{Carrasco}}, \citenamefont {{Hertzberg}},\ and\ \citenamefont
  {{Senatore}}}]{2012JHEP...09..082C}%
  \BibitemOpen
  \bibfield  {author} {\bibinfo {author} {\bibfnamefont {J.~J.~M.}\
  \bibnamefont {{Carrasco}}}, \bibinfo {author} {\bibfnamefont {M.~P.}\
  \bibnamefont {{Hertzberg}}},\ and\ \bibinfo {author} {\bibfnamefont
  {L.}~\bibnamefont {{Senatore}}},\ }\href
  {https://doi.org/10.1007/JHEP09(2012)082} {\bibfield  {journal} {\bibinfo
  {journal} {Journal of High Energy Physics}\ }\textbf {\bibinfo {volume}
  {9}},\ \bibinfo {eid} {82} (\bibinfo {year} {2012})},\ \Eprint
  {https://arxiv.org/abs/1206.2926} {arXiv:1206.2926 [astro-ph.CO]}
  \BibitemShut {NoStop}%
\bibitem [{\citenamefont {Baldauf}\ \emph {et~al.}(2015)\citenamefont
  {Baldauf}, \citenamefont {Mercolli},\ and\ \citenamefont
  {Zaldarriaga}}]{Baldauf:2015aha}%
  \BibitemOpen
  \bibfield  {author} {\bibinfo {author} {\bibfnamefont {T.}~\bibnamefont
  {Baldauf}}, \bibinfo {author} {\bibfnamefont {L.}~\bibnamefont {Mercolli}},\
  and\ \bibinfo {author} {\bibfnamefont {M.}~\bibnamefont {Zaldarriaga}},\
  }\href {https://doi.org/10.1103/PhysRevD.92.123007} {\bibfield  {journal}
  {\bibinfo  {journal} {\prd}\ }\textbf {\bibinfo {volume} {92}},\ \bibinfo
  {pages} {123007} (\bibinfo {year} {2015})},\ \Eprint
  {https://arxiv.org/abs/1507.02256} {arXiv:1507.02256 [astro-ph.CO]}
  \BibitemShut {NoStop}%
\bibitem [{\citenamefont {{Nishimichi}}\ \emph {et~al.}(2020)\citenamefont
  {{Nishimichi}}, \citenamefont {{D'Amico}}, \citenamefont {{Ivanov}},
  \citenamefont {{Senatore}}, \citenamefont {{Simonovi{\'c}}}, \citenamefont
  {{Takada}}, \citenamefont {{Zaldarriaga}},\ and\ \citenamefont
  {{Zhang}}}]{Nishimichi_etal2020}%
  \BibitemOpen
  \bibfield  {author} {\bibinfo {author} {\bibfnamefont {T.}~\bibnamefont
  {{Nishimichi}}}, \bibinfo {author} {\bibfnamefont {G.}~\bibnamefont
  {{D'Amico}}}, \bibinfo {author} {\bibfnamefont {M.~M.}\ \bibnamefont
  {{Ivanov}}}, \bibinfo {author} {\bibfnamefont {L.}~\bibnamefont
  {{Senatore}}}, \bibinfo {author} {\bibfnamefont {M.}~\bibnamefont
  {{Simonovi{\'c}}}}, \bibinfo {author} {\bibfnamefont {M.}~\bibnamefont
  {{Takada}}}, \bibinfo {author} {\bibfnamefont {M.}~\bibnamefont
  {{Zaldarriaga}}},\ and\ \bibinfo {author} {\bibfnamefont {P.}~\bibnamefont
  {{Zhang}}},\ }\href@noop {} {\bibfield  {journal} {\bibinfo  {journal} {arXiv
  e-prints}\ ,\ \bibinfo {eid} {arXiv:2003.08277}} (\bibinfo {year} {2020})},\
  \Eprint {https://arxiv.org/abs/2003.08277} {arXiv:2003.08277 [astro-ph.CO]}
  \BibitemShut {NoStop}%
\bibitem [{\citenamefont {{Steele}}\ and\ \citenamefont
  {{Baldauf}}(2021{\natexlab{a}})}]{Steele_Baldauf2021a}%
  \BibitemOpen
  \bibfield  {author} {\bibinfo {author} {\bibfnamefont {T.}~\bibnamefont
  {{Steele}}}\ and\ \bibinfo {author} {\bibfnamefont {T.}~\bibnamefont
  {{Baldauf}}},\ }\href {https://doi.org/10.1103/PhysRevD.103.023520}
  {\bibfield  {journal} {\bibinfo  {journal} {\prd}\ }\textbf {\bibinfo
  {volume} {103}},\ \bibinfo {eid} {023520} (\bibinfo {year}
  {2021}{\natexlab{a}})},\ \Eprint {https://arxiv.org/abs/2009.01200}
  {arXiv:2009.01200 [astro-ph.CO]} \BibitemShut {NoStop}%
\bibitem [{\citenamefont {{Steele}}\ and\ \citenamefont
  {{Baldauf}}(2021{\natexlab{b}})}]{Steele_Baldauf2021b}%
  \BibitemOpen
  \bibfield  {author} {\bibinfo {author} {\bibfnamefont {T.}~\bibnamefont
  {{Steele}}}\ and\ \bibinfo {author} {\bibfnamefont {T.}~\bibnamefont
  {{Baldauf}}},\ }\href {https://doi.org/10.1103/PhysRevD.103.103518}
  {\bibfield  {journal} {\bibinfo  {journal} {\prd}\ }\textbf {\bibinfo
  {volume} {103}},\ \bibinfo {eid} {103518} (\bibinfo {year}
  {2021}{\natexlab{b}})},\ \Eprint {https://arxiv.org/abs/2101.10289}
  {arXiv:2101.10289 [astro-ph.CO]} \BibitemShut {NoStop}%
\bibitem [{\citenamefont {Pietroni}(2008)}]{Pietroni:2008jx}%
  \BibitemOpen
  \bibfield  {author} {\bibinfo {author} {\bibfnamefont {M.}~\bibnamefont
  {Pietroni}},\ }\href {https://doi.org/10.1088/1475-7516/2008/10/036}
  {\bibfield  {journal} {\bibinfo  {journal} {\jcap}\ }\textbf {\bibinfo
  {volume} {0810}},\ \bibinfo {pages} {036} (\bibinfo {year} {2008})},\ \Eprint
  {https://arxiv.org/abs/0806.0971} {arXiv:0806.0971 [astro-ph]} \BibitemShut
  {NoStop}%
\bibitem [{\citenamefont {{Takahashi}}(2008)}]{Takahashi2008}%
  \BibitemOpen
  \bibfield  {author} {\bibinfo {author} {\bibfnamefont {R.}~\bibnamefont
  {{Takahashi}}},\ }\href {https://doi.org/10.1143/PTP.120.549} {\bibfield
  {journal} {\bibinfo  {journal} {Progress of Theoretical Physics}\ }\textbf
  {\bibinfo {volume} {120}},\ \bibinfo {pages} {549} (\bibinfo {year}
  {2008})},\ \Eprint {https://arxiv.org/abs/0806.1437} {arXiv:0806.1437
  [astro-ph]} \BibitemShut {NoStop}%
\bibitem [{\citenamefont {{Hiramatsu}}\ and\ \citenamefont
  {{Taruya}}(2009)}]{Hiramatsu_Taruya2009}%
  \BibitemOpen
  \bibfield  {author} {\bibinfo {author} {\bibfnamefont {T.}~\bibnamefont
  {{Hiramatsu}}}\ and\ \bibinfo {author} {\bibfnamefont {A.}~\bibnamefont
  {{Taruya}}},\ }\href {https://doi.org/10.1103/PhysRevD.79.103526} {\bibfield
  {journal} {\bibinfo  {journal} {\prd}\ }\textbf {\bibinfo {volume} {79}},\
  \bibinfo {eid} {103526} (\bibinfo {year} {2009})},\ \Eprint
  {https://arxiv.org/abs/0902.3772} {arXiv:0902.3772 [astro-ph.CO]}
  \BibitemShut {NoStop}%
\bibitem [{\citenamefont {Blas}\ \emph {et~al.}(2014)\citenamefont {Blas},
  \citenamefont {Garny},\ and\ \citenamefont {Konstandin}}]{Blas:2013aba}%
  \BibitemOpen
  \bibfield  {author} {\bibinfo {author} {\bibfnamefont {D.}~\bibnamefont
  {Blas}}, \bibinfo {author} {\bibfnamefont {M.}~\bibnamefont {Garny}},\ and\
  \bibinfo {author} {\bibfnamefont {T.}~\bibnamefont {Konstandin}},\ }\href
  {https://doi.org/10.1088/1475-7516/2014/01/010} {\bibfield  {journal}
  {\bibinfo  {journal} {\jcap}\ }\textbf {\bibinfo {volume} {1401}},\ \bibinfo
  {pages} {010} (\bibinfo {year} {2014})},\ \Eprint
  {https://arxiv.org/abs/1309.3308} {arXiv:1309.3308 [astro-ph.CO]}
  \BibitemShut {NoStop}%
\bibitem [{\citenamefont {Bernardeau}\ \emph {et~al.}(2014)\citenamefont
  {Bernardeau}, \citenamefont {Taruya},\ and\ \citenamefont
  {Nishimichi}}]{Bernardeau:2012ux}%
  \BibitemOpen
  \bibfield  {author} {\bibinfo {author} {\bibfnamefont {F.}~\bibnamefont
  {Bernardeau}}, \bibinfo {author} {\bibfnamefont {A.}~\bibnamefont {Taruya}},\
  and\ \bibinfo {author} {\bibfnamefont {T.}~\bibnamefont {Nishimichi}},\
  }\href {https://doi.org/10.1103/PhysRevD.89.023502} {\bibfield  {journal}
  {\bibinfo  {journal} {\prd}\ }\textbf {\bibinfo {volume} {89}},\ \bibinfo
  {pages} {023502} (\bibinfo {year} {2014})},\ \Eprint
  {https://arxiv.org/abs/1211.1571} {arXiv:1211.1571 [astro-ph.CO]}
  \BibitemShut {NoStop}%
\bibitem [{\citenamefont {Nishimichi}\ \emph {et~al.}(2016)\citenamefont
  {Nishimichi}, \citenamefont {Bernardeau},\ and\ \citenamefont
  {Taruya}}]{Nishimichi:2014rra}%
  \BibitemOpen
  \bibfield  {author} {\bibinfo {author} {\bibfnamefont {T.}~\bibnamefont
  {Nishimichi}}, \bibinfo {author} {\bibfnamefont {F.}~\bibnamefont
  {Bernardeau}},\ and\ \bibinfo {author} {\bibfnamefont {A.}~\bibnamefont
  {Taruya}},\ }\href {https://doi.org/10.1016/j.physletb.2016.09.035}
  {\bibfield  {journal} {\bibinfo  {journal} {Phys. Lett.}\ }\textbf {\bibinfo
  {volume} {B762}},\ \bibinfo {pages} {247} (\bibinfo {year} {2016})},\ \Eprint
  {https://arxiv.org/abs/1411.2970} {arXiv:1411.2970 [astro-ph.CO]}
  \BibitemShut {NoStop}%
\bibitem [{\citenamefont {{Kaiser}}(1987)}]{Kaiser1987}%
  \BibitemOpen
  \bibfield  {author} {\bibinfo {author} {\bibfnamefont {N.}~\bibnamefont
  {{Kaiser}}},\ }\href {https://doi.org/10.1093/mnras/227.1.1} {\bibfield
  {journal} {\bibinfo  {journal} {\mnras}\ }\textbf {\bibinfo {volume} {227}},\
  \bibinfo {pages} {1} (\bibinfo {year} {1987})}\BibitemShut {NoStop}%
\bibitem [{\citenamefont {Cole}\ \emph {et~al.}(1994)\citenamefont {Cole},
  \citenamefont {Fisher},\ and\ \citenamefont {Weinberg}}]{Cole:1993kh}%
  \BibitemOpen
  \bibfield  {author} {\bibinfo {author} {\bibfnamefont {S.}~\bibnamefont
  {Cole}}, \bibinfo {author} {\bibfnamefont {K.~B.}\ \bibnamefont {Fisher}},\
  and\ \bibinfo {author} {\bibfnamefont {D.~H.}\ \bibnamefont {Weinberg}},\
  }\href@noop {} {\bibfield  {journal} {\bibinfo  {journal} {\mnras}\ }\textbf
  {\bibinfo {volume} {267}},\ \bibinfo {pages} {785} (\bibinfo {year}
  {1994})},\ \Eprint {https://arxiv.org/abs/astro-ph/9308003}
  {arXiv:astro-ph/9308003} \BibitemShut {NoStop}%
\bibitem [{\citenamefont {{Raccanelli}}\ \emph {et~al.}(2018)\citenamefont
  {{Raccanelli}}, \citenamefont {{Bertacca}}, \citenamefont {{Jeong}},
  \citenamefont {{Neyrinck}},\ and\ \citenamefont
  {{Szalay}}}]{Raccanelli_etal2018}%
  \BibitemOpen
  \bibfield  {author} {\bibinfo {author} {\bibfnamefont {A.}~\bibnamefont
  {{Raccanelli}}}, \bibinfo {author} {\bibfnamefont {D.}~\bibnamefont
  {{Bertacca}}}, \bibinfo {author} {\bibfnamefont {D.}~\bibnamefont {{Jeong}}},
  \bibinfo {author} {\bibfnamefont {M.~C.}\ \bibnamefont {{Neyrinck}}},\ and\
  \bibinfo {author} {\bibfnamefont {A.~S.}\ \bibnamefont {{Szalay}}},\ }\href
  {https://doi.org/10.1016/j.dark.2017.12.003} {\bibfield  {journal} {\bibinfo
  {journal} {Physics of the Dark Universe}\ }\textbf {\bibinfo {volume} {19}},\
  \bibinfo {pages} {109} (\bibinfo {year} {2018})},\ \Eprint
  {https://arxiv.org/abs/1602.03186} {arXiv:1602.03186 [astro-ph.CO]}
  \BibitemShut {NoStop}%
\bibitem [{\citenamefont {Springel}(2005)}]{Springel:2005mi}%
  \BibitemOpen
  \bibfield  {author} {\bibinfo {author} {\bibfnamefont {V.}~\bibnamefont
  {Springel}},\ }\href {https://doi.org/10.1111/j.1365-2966.2005.09655.x}
  {\bibfield  {journal} {\bibinfo  {journal} {\mnras}\ }\textbf {\bibinfo
  {volume} {364}},\ \bibinfo {pages} {1105} (\bibinfo {year} {2005})},\ \Eprint
  {https://arxiv.org/abs/astro-ph/0505010} {arXiv:astro-ph/0505010}
  \BibitemShut {NoStop}%
\bibitem [{\citenamefont {Crocce}\ \emph {et~al.}(2006)\citenamefont {Crocce},
  \citenamefont {Pueblas},\ and\ \citenamefont {Scoccimarro}}]{Crocce:2006ve}%
  \BibitemOpen
  \bibfield  {author} {\bibinfo {author} {\bibfnamefont {M.}~\bibnamefont
  {Crocce}}, \bibinfo {author} {\bibfnamefont {S.}~\bibnamefont {Pueblas}},\
  and\ \bibinfo {author} {\bibfnamefont {R.}~\bibnamefont {Scoccimarro}},\
  }\href {https://doi.org/10.1111/j.1365-2966.2006.11040.x} {\bibfield
  {journal} {\bibinfo  {journal} {\mnras}\ }\textbf {\bibinfo {volume} {373}},\
  \bibinfo {pages} {369} (\bibinfo {year} {2006})},\ \Eprint
  {https://arxiv.org/abs/astro-ph/0606505} {arXiv:astro-ph/0606505}
  \BibitemShut {NoStop}%
\bibitem [{\citenamefont {{Jeong}}\ \emph {et~al.}(2015)\citenamefont
  {{Jeong}}, \citenamefont {{Dai}}, \citenamefont {{Kamionkowski}},\ and\
  \citenamefont {{Szalay}}}]{Jeong/etal:2015}%
  \BibitemOpen
  \bibfield  {author} {\bibinfo {author} {\bibfnamefont {D.}~\bibnamefont
  {{Jeong}}}, \bibinfo {author} {\bibfnamefont {L.}~\bibnamefont {{Dai}}},
  \bibinfo {author} {\bibfnamefont {M.}~\bibnamefont {{Kamionkowski}}},\ and\
  \bibinfo {author} {\bibfnamefont {A.~S.}\ \bibnamefont {{Szalay}}},\ }\href
  {https://doi.org/10.1093/mnras/stv525} {\bibfield  {journal} {\bibinfo
  {journal} {\mnras}\ }\textbf {\bibinfo {volume} {449}},\ \bibinfo {pages}
  {3312} (\bibinfo {year} {2015})},\ \Eprint {https://arxiv.org/abs/1408.4648}
  {arXiv:1408.4648 [astro-ph.CO]} \BibitemShut {NoStop}%
\bibitem [{\citenamefont {{Tanaka}}\ \emph {et~al.}(2017)\citenamefont
  {{Tanaka}}, \citenamefont {{Yoshikawa}}, \citenamefont {{Minoshima}},\ and\
  \citenamefont {{Yoshida}}}]{Tanaka_Satoshi_etal2017}%
  \BibitemOpen
  \bibfield  {author} {\bibinfo {author} {\bibfnamefont {S.}~\bibnamefont
  {{Tanaka}}}, \bibinfo {author} {\bibfnamefont {K.}~\bibnamefont
  {{Yoshikawa}}}, \bibinfo {author} {\bibfnamefont {T.}~\bibnamefont
  {{Minoshima}}},\ and\ \bibinfo {author} {\bibfnamefont {N.}~\bibnamefont
  {{Yoshida}}},\ }\href {https://doi.org/10.3847/1538-4357/aa901f} {\bibfield
  {journal} {\bibinfo  {journal} {\apj}\ }\textbf {\bibinfo {volume} {849}},\
  \bibinfo {eid} {76} (\bibinfo {year} {2017})},\ \Eprint
  {https://arxiv.org/abs/1702.08521} {arXiv:1702.08521 [physics.comp-ph]}
  \BibitemShut {NoStop}%
\bibitem [{\citenamefont {Sefusatti}\ \emph {et~al.}(2016)\citenamefont
  {Sefusatti}, \citenamefont {Crocce}, \citenamefont {Scoccimarro},\ and\
  \citenamefont {Couchman}}]{sefusatti2016}%
  \BibitemOpen
  \bibfield  {author} {\bibinfo {author} {\bibfnamefont {E.}~\bibnamefont
  {Sefusatti}}, \bibinfo {author} {\bibfnamefont {M.}~\bibnamefont {Crocce}},
  \bibinfo {author} {\bibfnamefont {R.}~\bibnamefont {Scoccimarro}},\ and\
  \bibinfo {author} {\bibfnamefont {H.}~\bibnamefont {Couchman}},\ }\href
  {https://doi.org/10.1093/mnras/stw1229} {\bibfield  {journal} {\bibinfo
  {journal} {Mon. Not. R. Astron. Soc.}\ }\textbf {\bibinfo {volume} {460}},\
  \bibinfo {pages} {3624} (\bibinfo {year} {2016})},\ \Eprint
  {https://arxiv.org/abs/1512.07295} {arXiv:1512.07295} \BibitemShut {NoStop}%
\bibitem [{\citenamefont {{Hashimoto}}\ \emph {et~al.}(2017)\citenamefont
  {{Hashimoto}}, \citenamefont {{Rasera}},\ and\ \citenamefont
  {{Taruya}}}]{Hashimoto_Rasera_Taruya2017}%
  \BibitemOpen
  \bibfield  {author} {\bibinfo {author} {\bibfnamefont {I.}~\bibnamefont
  {{Hashimoto}}}, \bibinfo {author} {\bibfnamefont {Y.}~\bibnamefont
  {{Rasera}}},\ and\ \bibinfo {author} {\bibfnamefont {A.}~\bibnamefont
  {{Taruya}}},\ }\href {https://doi.org/10.1103/PhysRevD.96.043526} {\bibfield
  {journal} {\bibinfo  {journal} {\prd}\ }\textbf {\bibinfo {volume} {96}},\
  \bibinfo {eid} {043526} (\bibinfo {year} {2017})},\ \Eprint
  {https://arxiv.org/abs/1705.02574} {arXiv:1705.02574 [astro-ph.CO]}
  \BibitemShut {NoStop}%
\bibitem [{\citenamefont {{Scoccimarro}}\ \emph {et~al.}(1999)\citenamefont
  {{Scoccimarro}}, \citenamefont {{Couchman}},\ and\ \citenamefont
  {{Frieman}}}]{Scoccimarro_etal1999}%
  \BibitemOpen
  \bibfield  {author} {\bibinfo {author} {\bibfnamefont {R.}~\bibnamefont
  {{Scoccimarro}}}, \bibinfo {author} {\bibfnamefont {H.~M.~P.}\ \bibnamefont
  {{Couchman}}},\ and\ \bibinfo {author} {\bibfnamefont {J.~A.}\ \bibnamefont
  {{Frieman}}},\ }\href {https://doi.org/10.1086/307220} {\bibfield  {journal}
  {\bibinfo  {journal} {\apj}\ }\textbf {\bibinfo {volume} {517}},\ \bibinfo
  {pages} {531} (\bibinfo {year} {1999})},\ \Eprint
  {https://arxiv.org/abs/astro-ph/9808305} {arXiv:astro-ph/9808305 [astro-ph]}
  \BibitemShut {NoStop}%
\bibitem [{\citenamefont {Scoccimarro}(2015)}]{Scoccimarro2015}%
  \BibitemOpen
  \bibfield  {author} {\bibinfo {author} {\bibfnamefont {R.}~\bibnamefont
  {Scoccimarro}},\ }\href {https://doi.org/10.1103/PhysRevD.92.083532}
  {\bibfield  {journal} {\bibinfo  {journal} {Physical Review D}\ }\textbf
  {\bibinfo {volume} {92}},\ \bibinfo {pages} {1} (\bibinfo {year} {2015})},\
  \Eprint {https://arxiv.org/abs/1506.02729} {arXiv:1506.02729} \BibitemShut
  {NoStop}%
\bibitem [{\citenamefont {{Gagrani}}\ and\ \citenamefont
  {{Samushia}}(2017)}]{Gagrani_Samushia2017}%
  \BibitemOpen
  \bibfield  {author} {\bibinfo {author} {\bibfnamefont {P.}~\bibnamefont
  {{Gagrani}}}\ and\ \bibinfo {author} {\bibfnamefont {L.}~\bibnamefont
  {{Samushia}}},\ }\href {https://doi.org/10.1093/mnras/stx135} {\bibfield
  {journal} {\bibinfo  {journal} {\mnras}\ }\textbf {\bibinfo {volume} {467}},\
  \bibinfo {pages} {928} (\bibinfo {year} {2017})},\ \Eprint
  {https://arxiv.org/abs/1610.03488} {arXiv:1610.03488 [astro-ph.CO]}
  \BibitemShut {NoStop}%
\bibitem [{\citenamefont {{Yamamoto}}\ \emph {et~al.}(2017)\citenamefont
  {{Yamamoto}}, \citenamefont {{Nan}},\ and\ \citenamefont
  {{Hikage}}}]{Yamamoto_Nan_Hikage2017}%
  \BibitemOpen
  \bibfield  {author} {\bibinfo {author} {\bibfnamefont {K.}~\bibnamefont
  {{Yamamoto}}}, \bibinfo {author} {\bibfnamefont {Y.}~\bibnamefont {{Nan}}},\
  and\ \bibinfo {author} {\bibfnamefont {C.}~\bibnamefont {{Hikage}}},\ }\href
  {https://doi.org/10.1103/PhysRevD.95.043528} {\bibfield  {journal} {\bibinfo
  {journal} {\prd}\ }\textbf {\bibinfo {volume} {95}},\ \bibinfo {eid} {043528}
  (\bibinfo {year} {2017})},\ \Eprint {https://arxiv.org/abs/1610.03665}
  {arXiv:1610.03665 [astro-ph.CO]} \BibitemShut {NoStop}%
\bibitem [{\citenamefont {{Sugiyama}}\ \emph {et~al.}(2019)\citenamefont
  {{Sugiyama}}, \citenamefont {{Saito}}, \citenamefont {{Beutler}},\ and\
  \citenamefont {{Seo}}}]{Sugiyama_etal2019}%
  \BibitemOpen
  \bibfield  {author} {\bibinfo {author} {\bibfnamefont {N.~S.}\ \bibnamefont
  {{Sugiyama}}}, \bibinfo {author} {\bibfnamefont {S.}~\bibnamefont {{Saito}}},
  \bibinfo {author} {\bibfnamefont {F.}~\bibnamefont {{Beutler}}},\ and\
  \bibinfo {author} {\bibfnamefont {H.-J.}\ \bibnamefont {{Seo}}},\ }\href
  {https://doi.org/10.1093/mnras/sty3249} {\bibfield  {journal} {\bibinfo
  {journal} {\mnras}\ }\textbf {\bibinfo {volume} {484}},\ \bibinfo {pages}
  {364} (\bibinfo {year} {2019})},\ \Eprint {https://arxiv.org/abs/1803.02132}
  {arXiv:1803.02132 [astro-ph.CO]} \BibitemShut {NoStop}%
\bibitem [{\citenamefont {{Baldauf}}\ \emph {et~al.}(2015)\citenamefont
  {{Baldauf}}, \citenamefont {{Mercolli}}, \citenamefont {{Mirbabayi}},\ and\
  \citenamefont {{Pajer}}}]{Baldauf_etal2015}%
  \BibitemOpen
  \bibfield  {author} {\bibinfo {author} {\bibfnamefont {T.}~\bibnamefont
  {{Baldauf}}}, \bibinfo {author} {\bibfnamefont {L.}~\bibnamefont
  {{Mercolli}}}, \bibinfo {author} {\bibfnamefont {M.}~\bibnamefont
  {{Mirbabayi}}},\ and\ \bibinfo {author} {\bibfnamefont {E.}~\bibnamefont
  {{Pajer}}},\ }\href {https://doi.org/10.1088/1475-7516/2015/05/007}
  {\bibfield  {journal} {\bibinfo  {journal} {\jcap}\ }\textbf {\bibinfo
  {volume} {2015}},\ \bibinfo {eid} {007} (\bibinfo {year} {2015})},\ \Eprint
  {https://arxiv.org/abs/1406.4135} {arXiv:1406.4135 [astro-ph.CO]}
  \BibitemShut {NoStop}%
\bibitem [{\citenamefont {Bianchi}\ \emph {et~al.}(2015)\citenamefont
  {Bianchi}, \citenamefont {Gil-Mar{\'{i}}n}, \citenamefont {Ruggeri},\ and\
  \citenamefont {Percival}}]{Bianchi_etal2015}%
  \BibitemOpen
  \bibfield  {author} {\bibinfo {author} {\bibfnamefont {D.}~\bibnamefont
  {Bianchi}}, \bibinfo {author} {\bibfnamefont {H.}~\bibnamefont
  {Gil-Mar{\'{i}}n}}, \bibinfo {author} {\bibfnamefont {R.}~\bibnamefont
  {Ruggeri}},\ and\ \bibinfo {author} {\bibfnamefont {W.~J.}\ \bibnamefont
  {Percival}},\ }\href {https://doi.org/10.1093/mnrasl/slv090} {\bibfield
  {journal} {\bibinfo  {journal} {Monthly Notices of the Royal Astronomical
  Society: Letters}\ }\textbf {\bibinfo {volume} {453}},\ \bibinfo {pages}
  {L11} (\bibinfo {year} {2015})}\BibitemShut {NoStop}%
\bibitem [{\citenamefont {{Taruya}}\ \emph {et~al.}(2012)\citenamefont
  {{Taruya}}, \citenamefont {{Bernardeau}}, \citenamefont {{Nishimichi}},\ and\
  \citenamefont {{Codis}}}]{Taruya:2012ut}%
  \BibitemOpen
  \bibfield  {author} {\bibinfo {author} {\bibfnamefont {A.}~\bibnamefont
  {{Taruya}}}, \bibinfo {author} {\bibfnamefont {F.}~\bibnamefont
  {{Bernardeau}}}, \bibinfo {author} {\bibfnamefont {T.}~\bibnamefont
  {{Nishimichi}}},\ and\ \bibinfo {author} {\bibfnamefont {S.}~\bibnamefont
  {{Codis}}},\ }\href {https://doi.org/10.1103/PhysRevD.86.103528} {\bibfield
  {journal} {\bibinfo  {journal} {\prd}\ }\textbf {\bibinfo {volume} {86}},\
  \bibinfo {eid} {103528} (\bibinfo {year} {2012})},\ \Eprint
  {https://arxiv.org/abs/1208.1191} {arXiv:1208.1191 [astro-ph.CO]}
  \BibitemShut {NoStop}%
\bibitem [{\citenamefont {{Davis}}\ and\ \citenamefont
  {{Peebles}}(1983)}]{1983ApJ...267..465D}%
  \BibitemOpen
  \bibfield  {author} {\bibinfo {author} {\bibfnamefont {M.}~\bibnamefont
  {{Davis}}}\ and\ \bibinfo {author} {\bibfnamefont {P.~J.~E.}\ \bibnamefont
  {{Peebles}}},\ }\href {https://doi.org/10.1086/160884} {\bibfield  {journal}
  {\bibinfo  {journal} {Astrophys. J.}\ }\textbf {\bibinfo {volume} {267}},\
  \bibinfo {pages} {465} (\bibinfo {year} {1983})}\BibitemShut {NoStop}%
\bibitem [{\citenamefont {{Jackson}}(1972)}]{1972MNRAS.156P...1J}%
  \BibitemOpen
  \bibfield  {author} {\bibinfo {author} {\bibfnamefont {J.~C.}\ \bibnamefont
  {{Jackson}}},\ }\href@noop {} {\bibfield  {journal} {\bibinfo  {journal}
  {\mnras}\ }\textbf {\bibinfo {volume} {156}},\ \bibinfo {pages} {1P}
  (\bibinfo {year} {1972})}\BibitemShut {NoStop}%
\bibitem [{\citenamefont {Okumura}\ \emph {et~al.}(2012)\citenamefont
  {Okumura}, \citenamefont {Seljak}, \citenamefont {McDonald},\ and\
  \citenamefont {Desjacques}}]{Okumura:2011pb}%
  \BibitemOpen
  \bibfield  {author} {\bibinfo {author} {\bibfnamefont {T.}~\bibnamefont
  {Okumura}}, \bibinfo {author} {\bibfnamefont {U.}~\bibnamefont {Seljak}},
  \bibinfo {author} {\bibfnamefont {P.}~\bibnamefont {McDonald}},\ and\
  \bibinfo {author} {\bibfnamefont {V.}~\bibnamefont {Desjacques}},\ }\href
  {https://doi.org/10.1088/1475-7516/2012/02/010} {\bibfield  {journal}
  {\bibinfo  {journal} {\jcap}\ }\textbf {\bibinfo {volume} {1202}},\ \bibinfo
  {pages} {010} (\bibinfo {year} {2012})},\ \Eprint
  {https://arxiv.org/abs/1109.1609} {arXiv:1109.1609 [astro-ph.CO]}
  \BibitemShut {NoStop}%
\bibitem [{\citenamefont {{Yoshisato}}\ \emph {et~al.}(1998)\citenamefont
  {{Yoshisato}}, \citenamefont {{Matsubara}},\ and\ \citenamefont
  {{Morikawa}}}]{Yoshisato_Matsubara_Morikawa1998}%
  \BibitemOpen
  \bibfield  {author} {\bibinfo {author} {\bibfnamefont {A.}~\bibnamefont
  {{Yoshisato}}}, \bibinfo {author} {\bibfnamefont {T.}~\bibnamefont
  {{Matsubara}}},\ and\ \bibinfo {author} {\bibfnamefont {M.}~\bibnamefont
  {{Morikawa}}},\ }\href {https://doi.org/10.1086/305534} {\bibfield  {journal}
  {\bibinfo  {journal} {\apj}\ }\textbf {\bibinfo {volume} {498}},\ \bibinfo
  {pages} {48} (\bibinfo {year} {1998})},\ \Eprint
  {https://arxiv.org/abs/astro-ph/9707296} {arXiv:astro-ph/9707296 [astro-ph]}
  \BibitemShut {NoStop}%
\bibitem [{\citenamefont {{Matsubara}}\ \emph {et~al.}(1998)\citenamefont
  {{Matsubara}}, \citenamefont {{Yoshisato}},\ and\ \citenamefont
  {{Morikawa}}}]{Matsubara_Yoshisato_Moriakwa1998}%
  \BibitemOpen
  \bibfield  {author} {\bibinfo {author} {\bibfnamefont {T.}~\bibnamefont
  {{Matsubara}}}, \bibinfo {author} {\bibfnamefont {A.}~\bibnamefont
  {{Yoshisato}}},\ and\ \bibinfo {author} {\bibfnamefont {M.}~\bibnamefont
  {{Morikawa}}},\ }\href {https://doi.org/10.1086/306085} {\bibfield  {journal}
  {\bibinfo  {journal} {\apj}\ }\textbf {\bibinfo {volume} {504}},\ \bibinfo
  {pages} {7} (\bibinfo {year} {1998})},\ \Eprint
  {https://arxiv.org/abs/astro-ph/9708154} {arXiv:astro-ph/9708154 [astro-ph]}
  \BibitemShut {NoStop}%
\bibitem [{\citenamefont {{Tatekawa}}(2007)}]{Tatekawa2007}%
  \BibitemOpen
  \bibfield  {author} {\bibinfo {author} {\bibfnamefont {T.}~\bibnamefont
  {{Tatekawa}}},\ }\href {https://doi.org/10.1103/PhysRevD.75.044028}
  {\bibfield  {journal} {\bibinfo  {journal} {\prd}\ }\textbf {\bibinfo
  {volume} {75}},\ \bibinfo {eid} {044028} (\bibinfo {year} {2007})},\ \Eprint
  {https://arxiv.org/abs/astro-ph/0605250} {arXiv:astro-ph/0605250 [astro-ph]}
  \BibitemShut {NoStop}%
\bibitem [{\citenamefont {{Hinch}}(1991)}]{Hinch1991}%
  \BibitemOpen
  \bibfield  {author} {\bibinfo {author} {\bibfnamefont {E.~J.}\ \bibnamefont
  {{Hinch}}},\ }\href@noop {} {\emph {\bibinfo {title} {{Perturbation
  methods}}}}\ (\bibinfo  {publisher} {Cambridge University Press},\ \bibinfo
  {year} {1991})\BibitemShut {NoStop}%
\bibitem [{\citenamefont {{McDonald}}\ and\ \citenamefont
  {{Roy}}(2009)}]{McDonald_Roy2009}%
  \BibitemOpen
  \bibfield  {author} {\bibinfo {author} {\bibfnamefont {P.}~\bibnamefont
  {{McDonald}}}\ and\ \bibinfo {author} {\bibfnamefont {A.}~\bibnamefont
  {{Roy}}},\ }\href {https://doi.org/10.1088/1475-7516/2009/08/020} {\bibfield
  {journal} {\bibinfo  {journal} {\jcap}\ }\textbf {\bibinfo {volume} {2009}},\
  \bibinfo {eid} {020} (\bibinfo {year} {2009})},\ \Eprint
  {https://arxiv.org/abs/0902.0991} {arXiv:0902.0991 [astro-ph.CO]}
  \BibitemShut {NoStop}%
\bibitem [{\citenamefont {{Chan}}\ \emph {et~al.}(2012)\citenamefont {{Chan}},
  \citenamefont {{Scoccimarro}},\ and\ \citenamefont
  {{Sheth}}}]{Chan_Scoccimarro_Sheth2012}%
  \BibitemOpen
  \bibfield  {author} {\bibinfo {author} {\bibfnamefont {K.~C.}\ \bibnamefont
  {{Chan}}}, \bibinfo {author} {\bibfnamefont {R.}~\bibnamefont
  {{Scoccimarro}}},\ and\ \bibinfo {author} {\bibfnamefont {R.~K.}\
  \bibnamefont {{Sheth}}},\ }\href {https://doi.org/10.1103/PhysRevD.85.083509}
  {\bibfield  {journal} {\bibinfo  {journal} {\prd}\ }\textbf {\bibinfo
  {volume} {85}},\ \bibinfo {eid} {083509} (\bibinfo {year} {2012})},\ \Eprint
  {https://arxiv.org/abs/1201.3614} {arXiv:1201.3614 [astro-ph.CO]}
  \BibitemShut {NoStop}%
\bibitem [{\citenamefont {{Saito}}\ \emph {et~al.}(2014)\citenamefont
  {{Saito}}, \citenamefont {{Baldauf}}, \citenamefont {{Vlah}}, \citenamefont
  {{Seljak}}, \citenamefont {{Okumura}},\ and\ \citenamefont
  {{McDonald}}}]{Saito_etal2014}%
  \BibitemOpen
  \bibfield  {author} {\bibinfo {author} {\bibfnamefont {S.}~\bibnamefont
  {{Saito}}}, \bibinfo {author} {\bibfnamefont {T.}~\bibnamefont {{Baldauf}}},
  \bibinfo {author} {\bibfnamefont {Z.}~\bibnamefont {{Vlah}}}, \bibinfo
  {author} {\bibfnamefont {U.}~\bibnamefont {{Seljak}}}, \bibinfo {author}
  {\bibfnamefont {T.}~\bibnamefont {{Okumura}}},\ and\ \bibinfo {author}
  {\bibfnamefont {P.}~\bibnamefont {{McDonald}}},\ }\href
  {https://doi.org/10.1103/PhysRevD.90.123522} {\bibfield  {journal} {\bibinfo
  {journal} {\prd}\ }\textbf {\bibinfo {volume} {90}},\ \bibinfo {eid} {123522}
  (\bibinfo {year} {2014})},\ \Eprint {https://arxiv.org/abs/1405.1447}
  {arXiv:1405.1447 [astro-ph.CO]} \BibitemShut {NoStop}%
\bibitem [{\citenamefont {{Fujita}}\ and\ \citenamefont
  {{Vlah}}(2020)}]{Fujita_Vlah2020}%
  \BibitemOpen
  \bibfield  {author} {\bibinfo {author} {\bibfnamefont {T.}~\bibnamefont
  {{Fujita}}}\ and\ \bibinfo {author} {\bibfnamefont {Z.}~\bibnamefont
  {{Vlah}}},\ }\href {https://doi.org/10.1088/1475-7516/2020/10/059} {\bibfield
   {journal} {\bibinfo  {journal} {\jcap}\ }\textbf {\bibinfo {volume}
  {2020}},\ \bibinfo {eid} {059} (\bibinfo {year} {2020})},\ \Eprint
  {https://arxiv.org/abs/2003.10114} {arXiv:2003.10114 [astro-ph.CO]}
  \BibitemShut {NoStop}%
\bibitem [{\citenamefont {{Kitaura}}\ and\ \citenamefont
  {{En{\ss}lin}}(2008)}]{Kitaura_Enslin2008}%
  \BibitemOpen
  \bibfield  {author} {\bibinfo {author} {\bibfnamefont {F.~S.}\ \bibnamefont
  {{Kitaura}}}\ and\ \bibinfo {author} {\bibfnamefont {T.~A.}\ \bibnamefont
  {{En{\ss}lin}}},\ }\href {https://doi.org/10.1111/j.1365-2966.2008.13341.x}
  {\bibfield  {journal} {\bibinfo  {journal} {\mnras}\ }\textbf {\bibinfo
  {volume} {389}},\ \bibinfo {pages} {497} (\bibinfo {year} {2008})},\ \Eprint
  {https://arxiv.org/abs/0705.0429} {arXiv:0705.0429 [astro-ph]} \BibitemShut
  {NoStop}%
\bibitem [{\citenamefont {{Jasche}}\ and\ \citenamefont
  {{Wandelt}}(2013)}]{Jasche_Wandelt2013}%
  \BibitemOpen
  \bibfield  {author} {\bibinfo {author} {\bibfnamefont {J.}~\bibnamefont
  {{Jasche}}}\ and\ \bibinfo {author} {\bibfnamefont {B.~D.}\ \bibnamefont
  {{Wandelt}}},\ }\href {https://doi.org/10.1093/mnras/stt449} {\bibfield
  {journal} {\bibinfo  {journal} {\mnras}\ }\textbf {\bibinfo {volume} {432}},\
  \bibinfo {pages} {894} (\bibinfo {year} {2013})},\ \Eprint
  {https://arxiv.org/abs/1203.3639} {arXiv:1203.3639 [astro-ph.CO]}
  \BibitemShut {NoStop}%
\bibitem [{\citenamefont {{Schmidt}}\ \emph {et~al.}(2019)\citenamefont
  {{Schmidt}}, \citenamefont {{Elsner}}, \citenamefont {{Jasche}},
  \citenamefont {{Nguyen}},\ and\ \citenamefont {{Lavaux}}}]{Schmidt_etal2019}%
  \BibitemOpen
  \bibfield  {author} {\bibinfo {author} {\bibfnamefont {F.}~\bibnamefont
  {{Schmidt}}}, \bibinfo {author} {\bibfnamefont {F.}~\bibnamefont {{Elsner}}},
  \bibinfo {author} {\bibfnamefont {J.}~\bibnamefont {{Jasche}}}, \bibinfo
  {author} {\bibfnamefont {N.~M.}\ \bibnamefont {{Nguyen}}},\ and\ \bibinfo
  {author} {\bibfnamefont {G.}~\bibnamefont {{Lavaux}}},\ }\href
  {https://doi.org/10.1088/1475-7516/2019/01/042} {\bibfield  {journal}
  {\bibinfo  {journal} {\jcap}\ }\textbf {\bibinfo {volume} {2019}},\ \bibinfo
  {eid} {042} (\bibinfo {year} {2019})},\ \Eprint
  {https://arxiv.org/abs/1808.02002} {arXiv:1808.02002 [astro-ph.CO]}
  \BibitemShut {NoStop}%
\bibitem [{\citenamefont {{Orszag}}(1971)}]{Orszag1971a}%
  \BibitemOpen
  \bibfield  {author} {\bibinfo {author} {\bibfnamefont {S.~A.}\ \bibnamefont
  {{Orszag}}},\ }\href
  {https://doi.org/10.1175/1520-0469(1971)028<1074:OTEOAI>2.0.CO;2} {\bibfield
  {journal} {\bibinfo  {journal} {Journal of Atmospheric Sciences}\ }\textbf
  {\bibinfo {volume} {28}},\ \bibinfo {pages} {1074} (\bibinfo {year}
  {1971})}\BibitemShut {NoStop}%
\bibitem [{\citenamefont {Takahashi}\ \emph {et~al.}(2008)\citenamefont
  {Takahashi} \emph {et~al.}}]{Takahashi:2008wn}%
  \BibitemOpen
  \bibfield  {author} {\bibinfo {author} {\bibfnamefont {R.}~\bibnamefont
  {Takahashi}} \emph {et~al.},\ }\href@noop {} {\bibfield  {journal} {\bibinfo
  {journal} {\mnras}\ }\textbf {\bibinfo {volume} {389}},\ \bibinfo {pages}
  {1675} (\bibinfo {year} {2008})},\ \Eprint {https://arxiv.org/abs/0802.1808}
  {arXiv:0802.1808 [astro-ph]} \BibitemShut {NoStop}%
\bibitem [{\citenamefont {Goroff}\ \emph {et~al.}(1986)\citenamefont {Goroff},
  \citenamefont {Grinstein}, \citenamefont {Rey},\ and\ \citenamefont
  {Wise}}]{Goroff:1986ep}%
  \BibitemOpen
  \bibfield  {author} {\bibinfo {author} {\bibfnamefont {M.~H.}\ \bibnamefont
  {Goroff}}, \bibinfo {author} {\bibfnamefont {B.}~\bibnamefont {Grinstein}},
  \bibinfo {author} {\bibfnamefont {S.~J.}\ \bibnamefont {Rey}},\ and\ \bibinfo
  {author} {\bibfnamefont {M.~B.}\ \bibnamefont {Wise}},\ }\href
  {https://doi.org/10.1086/164749} {\bibfield  {journal} {\bibinfo  {journal}
  {Astrophys. J.}\ }\textbf {\bibinfo {volume} {311}},\ \bibinfo {pages} {6}
  (\bibinfo {year} {1986})}\BibitemShut {NoStop}%
\bibitem [{\citenamefont {{Jain}}\ and\ \citenamefont
  {{Bertschinger}}(1994)}]{Jain_Betschinger1994}%
  \BibitemOpen
  \bibfield  {author} {\bibinfo {author} {\bibfnamefont {B.}~\bibnamefont
  {{Jain}}}\ and\ \bibinfo {author} {\bibfnamefont {E.}~\bibnamefont
  {{Bertschinger}}},\ }\href {https://doi.org/10.1086/174502} {\bibfield
  {journal} {\bibinfo  {journal} {\apj}\ }\textbf {\bibinfo {volume} {431}},\
  \bibinfo {pages} {495} (\bibinfo {year} {1994})},\ \Eprint
  {https://arxiv.org/abs/astro-ph/9311070} {arXiv:astro-ph/9311070 [astro-ph]}
  \BibitemShut {NoStop}%
\bibitem [{\citenamefont {Scoccimarro}\ \emph {et~al.}(1999)\citenamefont
  {Scoccimarro}, \citenamefont {Zaldarriaga},\ and\ \citenamefont
  {Hui}}]{Scoccimarro1999}%
  \BibitemOpen
  \bibfield  {author} {\bibinfo {author} {\bibfnamefont {R.}~\bibnamefont
  {Scoccimarro}}, \bibinfo {author} {\bibfnamefont {M.}~\bibnamefont
  {Zaldarriaga}},\ and\ \bibinfo {author} {\bibfnamefont {L.}~\bibnamefont
  {Hui}},\ }\href {https://doi.org/10.1086/308059} {\bibfield  {journal}
  {\bibinfo  {journal} {Astrophys. J.}\ }\textbf {\bibinfo {volume} {527}},\
  \bibinfo {pages} {1} (\bibinfo {year} {1999})},\ \Eprint
  {https://arxiv.org/abs/astro-ph/9901099} {arXiv:astro-ph/9901099}
  \BibitemShut {NoStop}%
\bibitem [{\citenamefont {Hahn}(2005)}]{Hahn:2004fe}%
  \BibitemOpen
  \bibfield  {author} {\bibinfo {author} {\bibfnamefont {T.}~\bibnamefont
  {Hahn}},\ }\href {https://doi.org/10.1016/j.cpc.2005.01.010} {\bibfield
  {journal} {\bibinfo  {journal} {Comput. Phys. Commun.}\ }\textbf {\bibinfo
  {volume} {168}},\ \bibinfo {pages} {78} (\bibinfo {year} {2005})},\ \Eprint
  {https://arxiv.org/abs/hep-ph/0404043} {arXiv:hep-ph/0404043} \BibitemShut
  {NoStop}%
\bibitem [{\citenamefont {{Konstandin}}\ \emph {et~al.}(2019)\citenamefont
  {{Konstandin}}, \citenamefont {{Porto}},\ and\ \citenamefont
  {{Rubira}}}]{Konstandin_etal2019}%
  \BibitemOpen
  \bibfield  {author} {\bibinfo {author} {\bibfnamefont {T.}~\bibnamefont
  {{Konstandin}}}, \bibinfo {author} {\bibfnamefont {R.~A.}\ \bibnamefont
  {{Porto}}},\ and\ \bibinfo {author} {\bibfnamefont {H.}~\bibnamefont
  {{Rubira}}},\ }\href {https://doi.org/10.1088/1475-7516/2019/11/027}
  {\bibfield  {journal} {\bibinfo  {journal} {\jcap}\ }\textbf {\bibinfo
  {volume} {2019}},\ \bibinfo {eid} {027} (\bibinfo {year} {2019})},\ \Eprint
  {https://arxiv.org/abs/1906.00997} {arXiv:1906.00997 [astro-ph.CO]}
  \BibitemShut {NoStop}%
\end{thebibliography}
%


\end{document}